\documentclass[aps,prb,twocolumn,superscriptaddress,longbibliography]{revtex4-2}
\usepackage{graphicx}
\usepackage{dcolumn}   % needed for some tables
\usepackage{bm}        % for math
\usepackage{amssymb}   % for math
\usepackage{amsmath}
\usepackage{url}
\usepackage{layout}
\usepackage{color}
\usepackage{epsfig}
\usepackage{graphicx}
\usepackage{mathrsfs}\usepackage{bm}\usepackage{appendix}\usepackage{color}

\usepackage[thinlines]{easytable}
\usepackage{makecell}
\usepackage{tikz,pgf}
\usepackage{booktabs}
\usepackage{float}
\usepackage{subfigure}
\usepackage{hyperref}
\hypersetup{colorlinks,allcolors=blue}

\begin{document}

\title{Variation Monte Carlo Study on the bilayer $t-J_{\parallel}-J_{\perp}$ model for La$_3$Ni$_2$O$_7$}
\author{Zeyu Chen}

\affiliation{School of Physics, Beijing Institute of Technology, Beijing 100081, China}

\author{Yu-Bo Liu}
\affiliation{Institute of Theoretical Physics, Chinese Academy of Sciences, Beijing 100190, China}

\author{Fan Yang}
\email{yangfan\_blg@bit.edu.cn}
\affiliation{School of Physics, Beijing Institute of Technology, Beijing 100081, China}

\begin{abstract}
The discovery of high-temperature superconductivity (HTSC) in La$_3$Ni$_2$O$_7$ has aroused significant interest in exploring the pairing mechanism. Previous studies have proposed an effective d$_{x^2-y^2}$-orbital bilayer $t-J_{\parallel}-J_{\perp}$ model, in which the electrons of the d$_{x^2-y^2}$ orbital are charge carriers, which are subject to the intralayer antiferromagnetic (AFM) superexchange $J_{\parallel}$ and the large interlayer AFM superexchange $J_{\perp}\approx 2J_{\parallel}$, with the latter transferred from the nearly half-filled and hence localized $d_{z^2}$ orbital through the strong Hund's rule coupling. Here we study this model by the variational Monte Carlo (VMC) simulation and find a dominant interlayer s-wave pairing, in which the SC order parameters have a drastic improvement compared with those of the mean field (MF) type of theories. In real materials, the Hund's coupling is finite, leading to reduced $J_{\perp}$, dictating that the MF-type theories have difficulty explaining the HTSC. However, our VMC calculations find that even for effective $J_{\perp}$ as weak as $J_{\perp}=J_{\parallel}$, the interlayer pairing is still considerably large and can be compared with the $T_c$ observed in experiments, which is very weak in MF-type theories.  This result indicates the important role of the Gutzwiller projection in improving the $T_c$, which is ignored in the MF-type theories. In addition, our results show that suppressed interlayer hopping can promote interlayer pairing, which is consistent with the fact that the interlayer hopping of the d$_{x^2-y^2}$ orbital in La$_3$Ni$_2$O$_7$ is very weak. Our research offers a new perspective for understanding the pairing mechanism of bilayer nickelates and provides a reference for recent ultra-cold atom experiments in mixed-dimensional systems.

\end{abstract}

\maketitle

\section{Introduction}
The discovery of high-temperature superconductivity (HTSC) of  nickelate family La$_{n+1}$Ni$_n$O$_{3n+1}$ of Ruddlesden-Popper (RP) phase has aroused a great upsurge\cite{Wang2023LNO,YuanHQ2023LNO,Wang2023LNOb,wang2023LNOpoly,10.1063/5.0247684,zhang2023pressure,puphal2024unconven,wang2023structure,li2024pressure,Dong2024vis,PhysRevB.111.075140,huang2024signature,du2024correlated,cui2023strain,li2024distinguishing,zhou2024revealing,fan2024tunn,li2024la3,zhu2024superconductivity,Li2023trilayer,zhang2023superconductivity,wang2024bulk,ko2024signatures,liu2025,Zhou2025}, representatives of them are bilayer nickelate La$_3$Ni$_2$O$_7$ with  critical temperature $T_{c}\approx80$K above 14GPa \cite{Wang2023LNO,YuanHQ2023LNO,Wang2023LNOb,wang2023LNOpoly,10.1063/5.0247684,zhang2023pressure,puphal2024unconven,wang2023structure,li2024pressure,Dong2024vis} and trilayer nickelate La$_4$Ni$_3$O$_{10}$ with maximum $T_{c}\approx30$K at 69.0 GPa\cite{zhu2024superconductivity,Li2023trilayer,zhang2023superconductivity}. The element substitution compound La$_2$PrNi$_2$O$_{7-\delta}$ has $T_{c}^{onset}=82.5$K and $T_{c}^{zero}=60$K at 18-20 GPa\cite{wang2024bulk}. Recently, new progress has been made that thin film La$_3$Ni$_2$O$_7$\cite{ko2024signatures}, La$_2$PrNi$_2$O$_7$ film\cite{liu2025} and (La,Pr)$_3$Ni$_2$O$_7$ films\cite{Zhou2025}  were found to host SC at ambient pressure, greatly expanding the characterization techniques and potential application of the nickelate superconductor. 

La$_3$Ni$_2$O$_7$ has a quasi-two-dimensional structure that is formed by two NiO$_6$ octahedral layers connected by vertical Ni-O-Ni bond. In bulk La$_3$Ni$_2$O$_7$, SC arises with a structure transition at 14 GPa from the Amam phase to the Fmmm phase, through which the vertical Ni-O-Ni bond angle changes from 168$^{\circ}$ to 180$^{\circ}$ \cite{Wang2023LNO}. Similarly, recent experiments on the superconducting thin film of La$_3$Ni$_2$O$_7$ also reveal that the vertical bond angle is close to 180$^{\circ}$\cite{10.1093/nsr/nwaf253,bhatt2025}. These experiments suggest that the vertical bond angle of 180$^{\circ}$, which corresponds to improved interlayer coupling, is beneficial to SC. In La$_3$Ni$_2$O$_7$, Ni$^{2.5+}$ has a configuration of 3d$^{7.5}$. Density functional theory (DFT) calculations reveal that the Ni-d$_{3z^2-r^2}$ and $d_{x^2 - y^2}$ orbitals hybridized with the O-p orbitals are near the Fermi surface~\cite{YaoDX2023,Dagotto2023,rhodes2023structural,Ouyang2024absence,Yi2024nature,labollita2024,zhang2024emergent,labollita2024assessing,wang2024electronic,PhysRevB.111.014515,geisler2024optical}, implying strong electron correlation. In this system, the $d_{z^2}$ orbital is nearly half filled and the $d_{x^2 - y^2}$ orbital is nearly quarter filled.  

The pairing mechanism of La$_3$Ni$_2$O$_7$ has been extensively studied, but remains a challenging problem. The ultrafast dynamics experiment reveals a relatively minor role for electron–phonon coupling at ambient pressure\cite{LI2025180} and DFT calculations find that electron-phonon coupling is insufficient to explain the high $T_{c}\approx$ 80K\cite{Ouyang2024absence,You2025}. It is indicated that La$_3$Ni$_2$O$_7$ is an unconventional superconductor and the electron-electron correlation is critical for HTSC. La$_3$Ni$_2$O$_7$ has charge and spin density wave order at ambient pressure\cite{Khasanov2025,chen2024evidence,Kakoi2024,Ren2025,dan2024spin,chen2024electronic,Wang2022LNO} that is close to the SC region in the phase diagram, which is similar to cuprates\cite{tsuei2000pairing,lee2006doping,xiang2022d} and iron-based SC\cite{Fernandes2022}, and it is believed that the pairing is closely related to magnetic interaction. 

Based on the experiments, many theoretical and numerical analyses have been proposed to explore the magnetic properties and the pairing mechanism of the nickelate SC\cite{lu2023bilayertJ,oh2023type2,liao2023electron,Yi_Feng2023,jiang2023high,huang2023impurity,lu2023sc,luo2023high,pan2023rno,yang2024decom,Lu2024interplay,wu2024deconfined,duan2025,
YangF2023,zhang2023structural,zhang2024electronic,zhang2024prediction,zhang2024s,zhang2023trends,chen2024tri,zhang2023la3ni2o6,Xia2025,botzel2024theory,Yubo_Liu2024,lin2024magnetic,
WangQH2023,HuJP2023,PhysRevLett.134.076001,Yang2024effective,
Kuroki2023,sakakibara2023La4Ni3O10,heier2023competing,
PhysRevB.110.235119,Leonov2024electronicc,lechermann2023,Werner2023,shilenko2023correlated,WuWei2023charge,cao2023flat,wang2024non,ryee2024quenched,PhysRevResearch.7.L012066,ouyang2023hund,tian2023correlation,PhysRevB.111.035108,
qin2023high,chang2023fermisurfacesymmetricmass,
qu2023bilayer,ZhangGM2023DMRG,zhang2023strong,lange2023mixedtj,PhysRevB.110.104517,lange2023feshbach,kaneko2023pair,Grusdt2023lno03349,qu2023roles,kakoi2023pair,ji2025,
chen2023iPEPS}. The picture of weak coupling emphasizes that the effective attraction of electrons leading to SC is caused by exchanging spin fluctuations, and most studies find that the symmetry is the s$^{\pm}$ wave\cite{YangF2023,zhang2023structural,zhang2023trends,WangQH2023,HuJP2023,PhysRevLett.134.076001,Kuroki2023}. However, a series of experiments indicate the strong correlation characteristic of La$_3$Ni$_2$O$_7$. For example, angle-resolved photoemission spectroscopy (ARPES) experiments show strong band renormalization\cite{yang2024orbital,Li2024ele}, suggesting strong electron correlation. Optical measurement declares that La$_3$Ni$_2$O$_7$ is near the Mott phase and the electron kinetic energy is significantly reduced
\cite{liu2024electronic}. The transport experiment finds that the normal state shows strange metal behavior that has linearly temperature-dependent resistivity\cite{YuanHQ2023LNO} and the ultrafast dynamics experiment also suggests non-Fermi liquid behavior\cite{LI2025180}. 

In the picture of strong coupling, some believe that the hole of the d$_{3z^2-r^2}$ orbital in the $\gamma$ pocket is crucial for SC, the strong interlayer antiferromagnetic (AFM) superexchange causes pairing and the hybridization of the $d_{z^2}$ and d$_{x^2-y^2}$ orbitals leads to phase coherence\cite{Yi_Feng2023,ZhangGM2023DMRG,qin2023high}. Recently, ARPES experiment on (La,Pr)$_3$Ni$_2$O$_7$ superconducting film claims that there is a $\gamma$ band pocket of d$_{z^2}$ orbital\cite{10.1093}. However, another ARPES experiment of the thin film superconductor La$_2$PrNi$_2$O$_7$\cite{wang2025electronic} shows no $\gamma$-pocket. These experiments seem to show that the $\gamma$ pocket is unnecessary for the SC of bilayer nickelates. Relatedly, some calculations of the two-orbital model\cite{qu2023roles,tian2023correlation,ji2025,chen2023iPEPS,Lu2024interplay} find that the $d_{x^2 - y^2}$ orbital dominates the SC in La$_3$Ni$_2$O$_7$, which does not rely on the presence of the $\gamma$ pocket. At first glance, it is difficult for the d$_{x^2-y^2}$ orbital with quarter filling to carry the HTSC. However, the strong Hund's rule coupling can transfer the interlayer AFM superexchange interaction of the $d_{z^2}$ orbital to the d$_{x^2-y^2}$ orbital, giving rise to an effective interlayer superexchange interaction $J_{\perp}$ of the latter. In combination with the intralayer superexchange interaction $J_{\parallel}$, an effective bilayer $t-J_{\parallel}-J_{\perp}$ model\cite{lu2023bilayertJ,oh2023type2,pan2023rno,qu2023bilayer,zhang2023strong,Grusdt2023lno03349,yang2025} of the $d_{x^2 - y^2}$ orbital is acquired, which is used to understand the HTSC in La$_3$Ni$_2$O$_7$.

 The bilayer $t-J_{\parallel}-J_{\perp}$ model has been studied using the mean-field (MF) and the slave-boson MF (SBMF) theory\cite{lu2023bilayertJ,oh2023type2}. In the strong limit of Hund's coupling $J_{H}\rightarrow\infty$, the interlayer AFM superexchange of the $d_{z^2}$ orbital $J_{z}$ is completely transferred to the d$_{x^2-y^2}$ orbital, which makes effective $J_{\perp}=J_{z}\approx2J_{\parallel}$. Under such a strong $J_{\perp}$, the $T_{c}$ obtained by the SBMF theory is high and can be compared with experiments. However, in real materials, the Hund's coupling is finite, which makes a transfer ratio of the interlayer superexchange interaction from the $d_{z^2}$ orbital to the d$_{x^2-y^2}$ orbital, dictating that $J_{\perp}<J_{z}$. With suppressed $J_{\perp}$, $T_c$ would be significantly suppressed. For example, in the MF and SBMF theory, $T_c$ for $J_{\perp}=J_{\parallel}$ is too low to be comparable to experiments. Therefore, an investigation of this model beyond the MF type of studies is necessary currently. In addition, the density matrix renormalization group (DMRG) calculations of this model\cite{qu2023bilayer,zhang2023strong,yang2025} have been performed, which however adopt a chain with width $L_{y}=1,2$ to get a power-law-decay pairing as a quasi-1D system, and the competition between the s-wave and the d-wave SC is difficult to discern.

In this paper, we adopt the variation Monte Carlo (VMC) approach to study the $d_{x^2 - y^2}$ - orbital bilayer $t-J_{\parallel}-J_{\perp}$ model, with a focus on adjusting $J_{\perp}$ to simulate different transfer ratios of the interlayer superexchange interaction between the two Ni-3d orbitals driven by finite Hund's rule coupling. Our central finding here is that the interlayer pairing dominates the intralayer pairing for realistic superexchange parameters, with the pairing strength at least an order of magnitude stronger than that obtained in the MF and the SBMF calculation\cite{lu2023bilayertJ}, indicating that the Gutzwiller projection is crucial for the HTSC. Especially at effective $J_{\perp}= J_{\parallel}$, there is still considerable interlayer pairing with its amplitude comparable with experimental $T_c$, which means that HTSC can exist even under about 50\% transfer ratio of the interlayer superexchange. The interlayer pairing can be further promoted by strong effective $J_{\perp}$ and electron filling. Another important result here is that suppressed interlayer hopping can promote interlayer pairing, which is consistent with the fact that the interlayer hopping of the $d_{x^2 - y^2}$ orbital in La$_3$Ni$_2$O$_7$ is very weak. In addition, our result yields that under weak $J_{\perp}$, the d-wave intralayer SC becomes energetically more favorable than the s-wave interlayer SC. Our results help to understand the HTSC of La$_3$Ni$_2$O$_7$ beyond the MF level.

The organization of the remaining part of this article is as follows. Sect.~\ref{Model and approach} introduces the bilayer $t-J_{\parallel}-J_{\perp}$ model relevant to the HTSC of La$_3$Ni$_2$O$_7$, and the VMC approach. Sect.~\ref{results} introduces our main results obtained through the VMC calculations, where we shall investigate the parameter dependence of the pairing strength and pairing symmetry, in comparison with previous MF and SBMF results. Sect.~\ref{Discussion and Conclusions} provides the discussion and conclusions.

\section{Model and approach}
\label{Model and approach}
\begin{figure}[t!]
\centering
\includegraphics[width=0.8\linewidth]{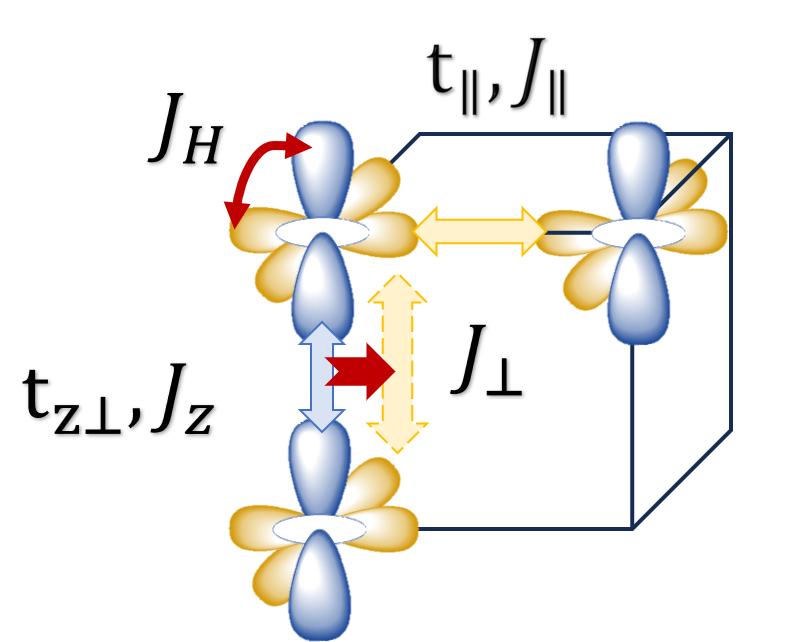}
\caption{Schematic diagram for the two orbital's $t-J-J_{H}$ model, the yellow orbital is $d_{x^2 - y^2}$ with  $t_{\parallel}$ and $J_{\parallel}$, and the blue one is d$_{z^2}$ orbital with $t_{z\perp}$ and $J_{z}$. The red arrow represents the Hund coupling which transfers $J_{z}$ to  $d_{x^2 - y^2}$ orbital and makes an effective $J_{\perp}$ of $d_{x^2 - y^2}$ orbital.}\label{structure}
\end{figure}

In La$_3$Ni$_2$O$_7$, the Ni - $d_{z^2}$ orbital and $d_{x^2 - y^2}$ orbital are near the Fermi surface. The $d_{z^2}$ orbital is nearly half-filling as the local spin. The $d_{x^2 - y^2}$ orbital is almost quarter-filling and plays the role of itinerant carriers in the SC state. As shown in Fig.~\ref{structure}, the $d_{z^2}$ orbital has interlayer hopping $t_{z\perp}$ and strong interlayer AFM superexchange $J_{z}=4t_{z\perp}^{2}/U$, and the $d_{x^2 - y^2}$ orbital has intralayer hopping $t_{\parallel}$ and intralayer AFM superexchange $J_{\parallel}=4t_{\parallel}^2/U$ ($U$ is onsite Coulomb repulsion). The Hund coupling between the two orbitals transfers $J_{z}$ to the $d_{x^2 - y^2}$ orbital which makes an effective $J_{\perp}$, so we consider the effective bilayer $t-J_{\parallel}-J_{\perp}$ model of the $d_{x^2 - y^2}$ orbital as follows:
\begin{equation}
\begin{aligned}
H=&-t_{\parallel}\sum_{\langle i,j \rangle,\alpha,\sigma}\mathcal{P}\left(c_{i\alpha\sigma}^{\dagger}c_{j\alpha\sigma}+h.c.\right)\mathcal{P}\\&+J_{\parallel}\sum_{\langle i,j\rangle,\alpha}\mathbf{S}_{i\alpha}\cdot\mathbf{S}_{j\alpha}+J_{\perp}\sum_{i}\mathbf{S}_{i1}\cdot\mathbf{S}_{i2}.
\end{aligned}
\end{equation}
Here $c_{i\alpha\sigma}^{\dagger}$ ($c_{i\alpha\sigma}$) is the creation (annihilation) operator of the $d_{x^2 - y^2}$ electron at the $i$ site and the $\alpha$ layer ($\alpha=1,2$) with the $\sigma$ spin. $\mathcal{P}$ is the projector to exclude double occupation, $\mathbf{S}_{i\alpha}$ is spin operator of the $d_{x^2 - y^2}$ orbital at the $i$ site and the $\alpha$ layer, $\langle i,j \rangle$ represents nearest-neighboring (NN) sites.

In the limit of strong coupling, the interlayer AFM interaction of the $d_{z^2}$ orbital compared to the intralayer AFM interaction of the $d_{x^2 - y^2}$ orbital makes $J_{z}/J_{\parallel}\approx2$. In the strong limit of Hund's coupling $J_{H}\rightarrow\infty$,  $J_{z}$ transfers the interlayer AFM interaction to $d_{x^2 - y^2}$ orbital completely, which makes $J_\perp=J_z$. However, at a finite Hund's coupling, there is $J_{\perp}/J_{z}=\eta\in(0,1)$. Here we adopt $J_{\perp}/J_{\parallel}=0.5, 0.75, 1, 2$ to simulate different transfer ratios of interlayer AFM interaction by Hund's coupling (corresponding to $\eta=25\%, 37.5\%, 50\%, 100\%)$.
We adopt $t_{\parallel}=1$, $J_{\parallel}=0.4$ and electron filling $n\in(0.25,0.35)$ to simulate hybridization
of the $d_{x^2 - y^2}$ and $d_{z^2}$ orbitals and the possibility of holes moving to the O 2p orbital. These parameters are consistent with the SBMF calculation\cite{lu2023bilayertJ}. 

We use a mean-field Hamiltonian:
\begin{equation}
\begin{aligned}
H_{mf}&=-t\sum_{\langle i,j\rangle,\alpha,\sigma}\left(c_{i\alpha\sigma}^{\dagger}c_{j\alpha\sigma}+h.c.\right)\\&-\chi_{\perp}\sum_{i\sigma}\left(c_{i1\sigma}^{\dagger}c_{i2\sigma}+h.c.\right)-\mu\sum_{i\alpha\sigma}c_{i\alpha\sigma}^{\dagger}c_{i\alpha\sigma}\\&+\sum_{\langle i,j\rangle,\alpha}\Delta_{\parallel}\left(c_{i\alpha\uparrow}^{\dagger}c_{j\alpha\downarrow}^{\dagger}-c_{i\alpha\downarrow}^{\dagger}c_{j\alpha\uparrow}^{\dagger} \right)+h.c.\\&+\sum_{i}\Delta_{\perp}\left(c_{i1\uparrow}^{\dagger}c_{i2\downarrow}^{\dagger}-c_{i1\downarrow}^{\dagger}c_{i2\uparrow}^{\dagger} \right)+h.c.
\end{aligned}
\end{equation}
to get a trial wavefunction 
\begin{equation}
\begin{aligned}
|\Psi\rangle=P_{G}|MF\rangle.
\end{aligned}
\end{equation}
Here, $P_{G}$ is the Gutzwiller projector to exclude double occupation, and $t$ is an energy unit. Because the energy band splitting caused by $t_{\perp}$ is very small, we consider both intraband pairing and interband pairing, the details can be seen in Appendix A. We use the steepest descent method to optimize the variational parameters and calculate the energy $E=E(\chi_{\perp},\Delta_{\parallel},\Delta_{\perp},\mu)$. The chemical potential is adjusted to satisfy the close shell condition. We adopt the $2\times10\times10$ lattice and the periodic and antiperiodic boundary conditions to avoid the singularity of the matrix in the VMC calculation. We calculate $5\times 10^6$ MC samples, which corresponds to $1\times 10^9$ Markov processes.

\section{Results}
\label{results}
\subsection{Dominant Strong Interlayer Pairing}
\label{interlayer s - wave SC}
In this part, we introduce our core results obtained by VMC calculation. Compared with the MF and SBMF solution, the pairing strength is drastically enhanced. The details are as follows.

In our calculation, we find that $\chi_{\perp}$ is almost zero within the error bar of VMC. The reason is that this bilayer system has U(1)$\times$U(1) symmetry and large $J_{\perp}$ inducing $\chi_{\perp}$ leads to interlayer hopping and breaks the symmetry to U(1). It is similar to the Stoner criterion
in the itinerant ferromagnetic phase when the interaction exceeds the critical value $U>U_{c}=1/N_{E_{f}}$. But here the density of state is not large enough to satisfy the effective interlayer superexchange $J_{\perp}$ above the critical value $J_{\perp c}$, the details can be seen in Appendix B.

Fig.~\ref{delta}(a) shows optimized variational parameters of interlayer and intralayer pairing with $J_{\perp}/J_{\parallel}=0.5, 0.75, 1, 2$ ($\eta=25\%, 37.5\%, 50\%, 100\%)$ obtained from our VMC calculations. At $J_{\perp}/ J_{\parallel}=0.75, 1, 2$, the interlayer pairing parameter $\Delta_{\perp}$ is enhanced by electron filling, which indicates that self-electron doping of the $d_{x^2 - y^2}$ orbital increases the density of state and benefits SC. Strong effective interlayer AFM superexchange also promotes SC, which comes from strong interlayer AFM superexchange of $d_{z^2}$ orbital and strong Hund coupling as a medium. 

Fig.~\ref{delta}(b) shows the SBMF\cite{lu2023bilayertJ} and MF results with the same parameters. 
Our results of VMC are qualitatively consistent with the SBMF calculation\cite{lu2023bilayertJ} in which the interlayer pairing is much stronger than the intralayer pairing.
\begin{figure*}[t]
\centering
\begin{minipage}{0.49\textwidth}{  (a)
\centering
\includegraphics[width=1\linewidth]{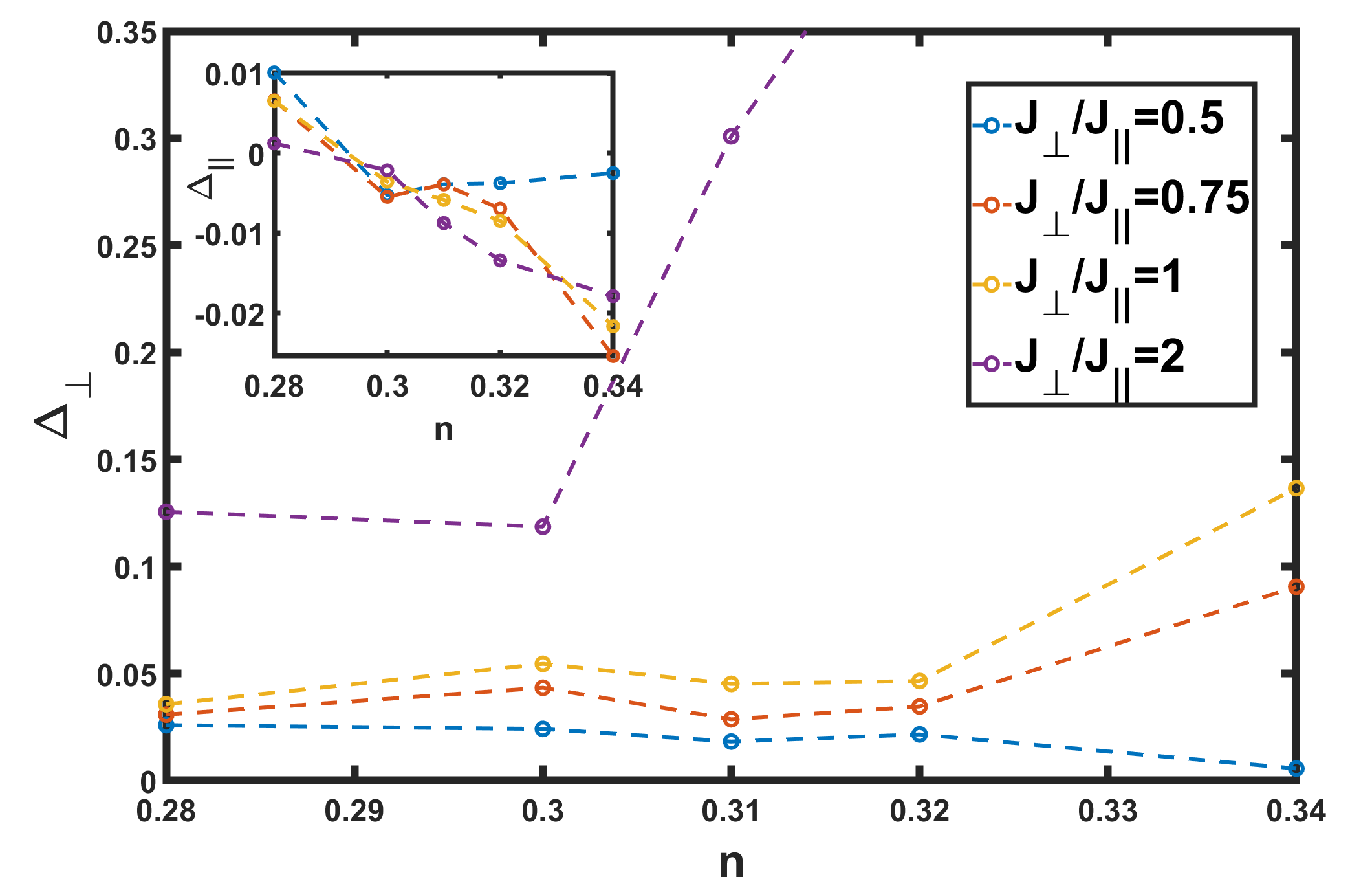} }
\label{delta_VMC}
 \end{minipage}
\hfill
\begin{minipage}{0.49\textwidth}{  (b)
\centering
\includegraphics[width=1\linewidth]{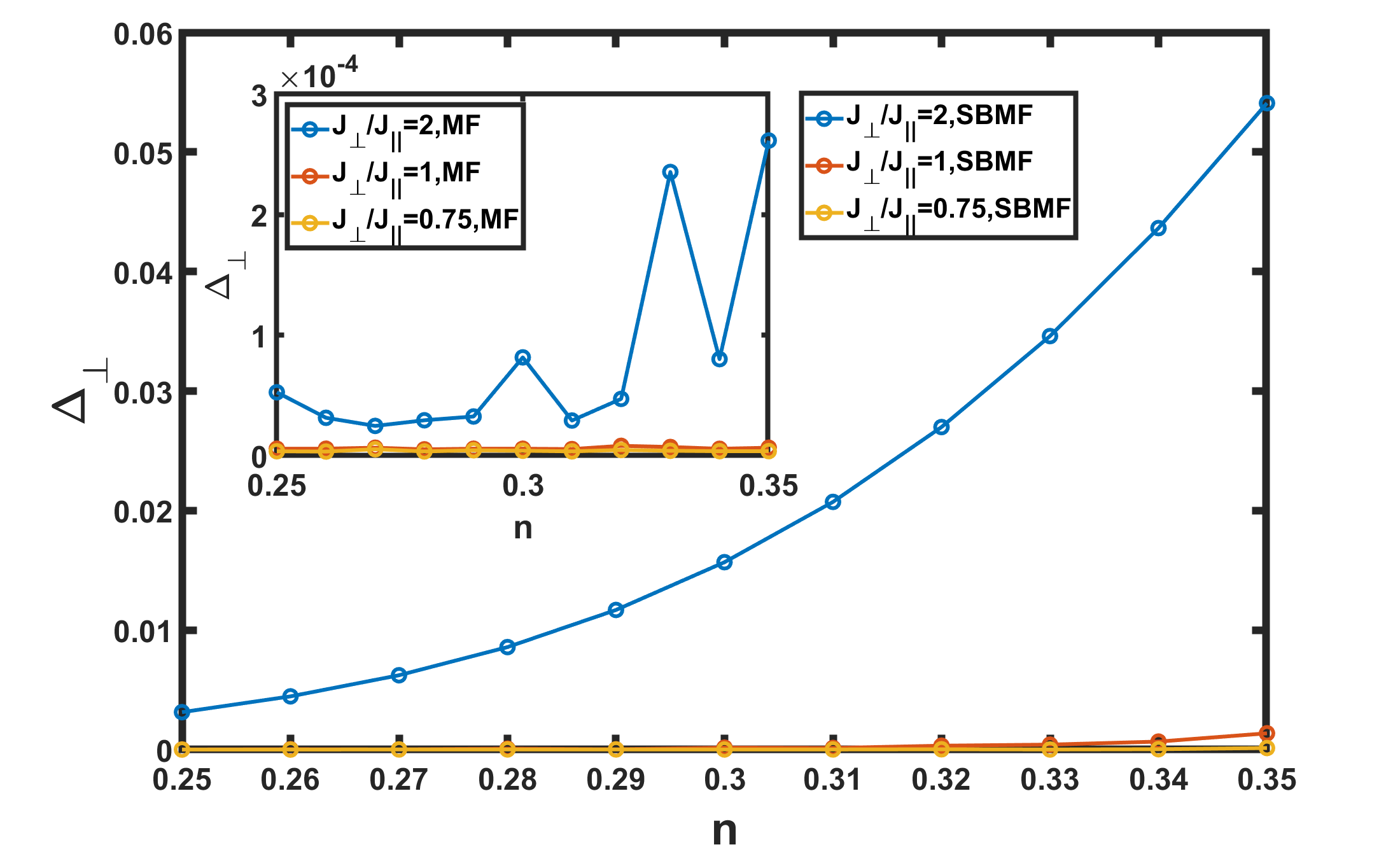} }
\end{minipage}
\caption{Pairing parameters calculated by VMC and SBMF. The picture (a) is the pairing parameters $\Delta_{\perp}$ versus electron filling for s-wave SC with $J_{\perp}/J_{\parallel}=0.5,0.75,1,2$ by VMC calculation, and insert shows pairing parameters $\Delta_{\parallel}$. The picture (b) is SBMF solution of $\Delta_{\perp}$ at $J_{\perp}/J_{\parallel}=0.75,1,2$\cite{lu2023bilayertJ}, and insert shows the MF solution of $\Delta_{\perp}$ at $J_{\perp}/J_{\parallel}=0.75,1,2$ as a comparison.}
\label{delta}
\end{figure*} 
It is noted that the pairing parameters at $J_{\perp}/J_{\parallel}=0.75,1,2$ are one to two orders of magnitude larger than those in the SBMF calculation\cite{lu2023bilayertJ} and about three to four orders of magnitude larger than those in the MF solution with the same filling, as shown in Fig.~\ref{delta}. Even with the
finite size effect, the pairing strength of the VMC calculation
is larger than that in the SBMF theory, the details can be seen in Appendix C. For example, at $n=0.3$ and $J_{\perp}/J_{\parallel}=2$, the interlayer pairing is $\Delta_{\perp}\approx0.1185$ in VMC, $\Delta_{\perp}\approx0.0157$ in the SBMF solution and $\Delta_{\perp}\approx8.2\times10^{-5}$ in the MF solution. At $J_{\perp}/J_{\parallel}=0.75$, the interlayer pairing is $\Delta_{\perp}\approx0.043$ by VMC, $\Delta_{\perp}\approx1.1\times10^{-5}$ by SBMF calculation, and $\Delta_{\perp}\approx4.1\times10^{-6}$ by MF calculation. The reason why the interlayer pairing of SBMF is much larger than that of MF is that in the BCS theory, $T_{c}\sim\Delta\sim e^{-1/\lambda}\sim e^{-1/(N_{E_{f}}J)}$, the increase in pairing can be seen as the renormalization of  the kinetic energy and the increase in the density of state near the Fermi surface. Furthermore, the VMC results are much larger than the results in the MF level because we take account of the influence of the exclusion of double occupancy in the microscopic quantum state in VMC calculation. This agrees with the conclusion of ref.\cite{zhang2023strong} that binding energy has a remarkable enhancement with $U/t_{\parallel}$ increasing, in which $U\rightarrow\infty$ is equal to the $t-J_{\parallel}-J_{\perp}$ model with projection of no double occupancy.

In addition, the interlayer pairing is still considerably large at $J_{\perp}/J_{\parallel}=1$ in the VMC calculation. Although we start from a strong Hund coupling limit to construct the model, we can get a large interlayer pairing with a finite Hund coupling, here the effective interlayer superexchange of the $d_{x^2 - y^2}$ orbital is half of that of the $d_{z^2}$ orbital which corresponds to the transfer ratio from the latter to the former by the finite Hund coupling $\eta=50\%$. This is very different from the SBMF\cite{lu2023bilayertJ} and the MF calculation in which the interlayer pairing is very weak at $J_{\perp}\le J_{\parallel}$. For example, at $J_{\perp}/J_{\parallel}=1$ and $n=0.3$ the interlayer pairing is $\Delta_{\perp}\approx5.7\times10^{-6}$ by the MF calculation, $\Delta_{\perp}\approx1.6\times10^{-4}$ by the SBMF calculation, but $\Delta_{\perp}\approx0.0545$ by the VMC calculation. In our calculation, $t_{\parallel}$ is the energy unit and $t_{\parallel}\approx0.5eV$ is calculated by DFT\cite{YaoDX2023} in the real material. Considering the suppression of the Gutzwiller factor in pairing, there is $\tilde{\Delta}=\delta\Delta_{\perp}$ in which $\delta=0.4$ is the density of holons. According to BCS theory: $2\Delta/T_{c}\approx3.52$, we can obtain our $T_{c}\approx\tilde{\Delta}\times t_{\parallel}/1.77=0.0062$eV = 62K in the VMC calculation, which can explain the experimental results of $T_c\approx80$K in La$_3$Ni$_2$O$_7$. However, the interlayer pairing at the MF level is too small, which contradicts the experimental results.
\begin{figure}[b]
\centering
\includegraphics[width=1\linewidth]{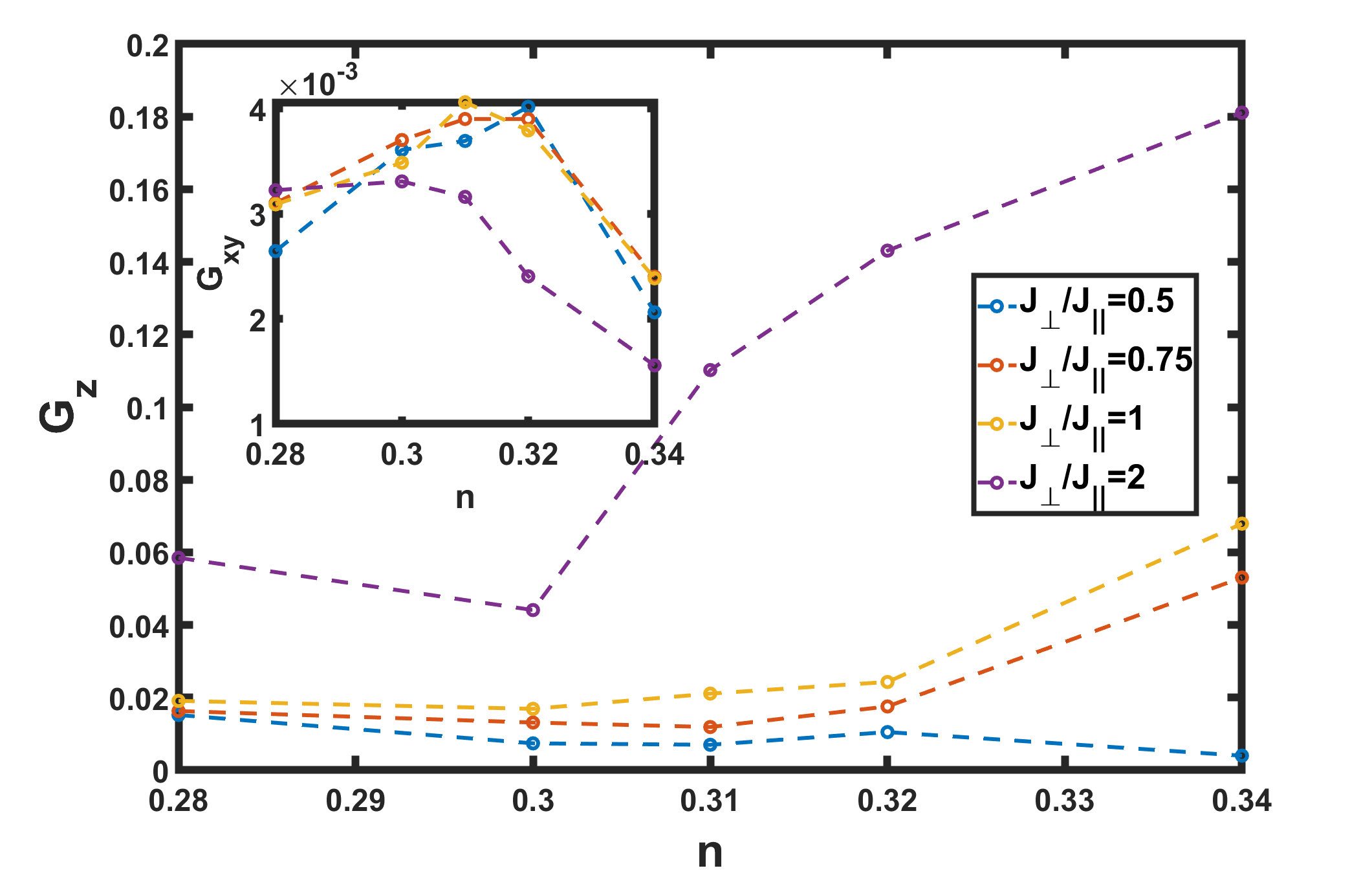}
\caption{Correlation functions $G_z$ and $G_{xy}$ changing with electron filling at $J_{\perp}/J_{\parallel}=0.5,0.75,1,2.$}
\label{correlation}
\end{figure}

We also calculate the correlation function of interlayer pairing $G_{z}$ and intralayer pairing $G_{xy}$:
\begin{equation}
\begin{aligned}
&G_{z}=\left\langle\frac{1}{N_{site}}\left(\sum_{i}c_{i\uparrow}^{\dagger}c_{i+z\downarrow}^{\dagger}\right)\left(\sum_{j}c_{j\downarrow}c_{j+z\uparrow}\right)\right\rangle^{\frac{1}{2}}\\
&G_{xy}=\left\langle\frac{1}{16N_{site}}\left(\sum_{i\tau}c_{i\uparrow}^{\dagger}c_{i+\tau\downarrow}^{\dagger}\right)\left(\sum_{j\tau'}c_{j\downarrow}c_{j+\tau'\uparrow}\right)\right\rangle^{\frac{1}{2}}.
\end{aligned}
\end{equation}
Here $i,j$ are the two farthest sites in the lattice ($L_{i,j}=5\sqrt{2}$) and $\tau(\tau')$ is NN intralayer sites of site $i(j)$. $5\times 10^6$ samples are calculated and the results are shown in Fig.~\ref{correlation}. The interlayer pairing correlation is much larger than intralayer pairing correlation, and increases with $J_{\perp}$ and electron filling increasing, showing the same behavior as pairing parameters changing with filling and $J_{\perp}$. We consider the nonmonotonicity of pairing parameter and correlation function possibly comes from finite-size effect of the system.
\begin{figure}[t]
    \centering
    \begin{subfigure}
        \centering
        \includegraphics[width=0.5\textwidth]{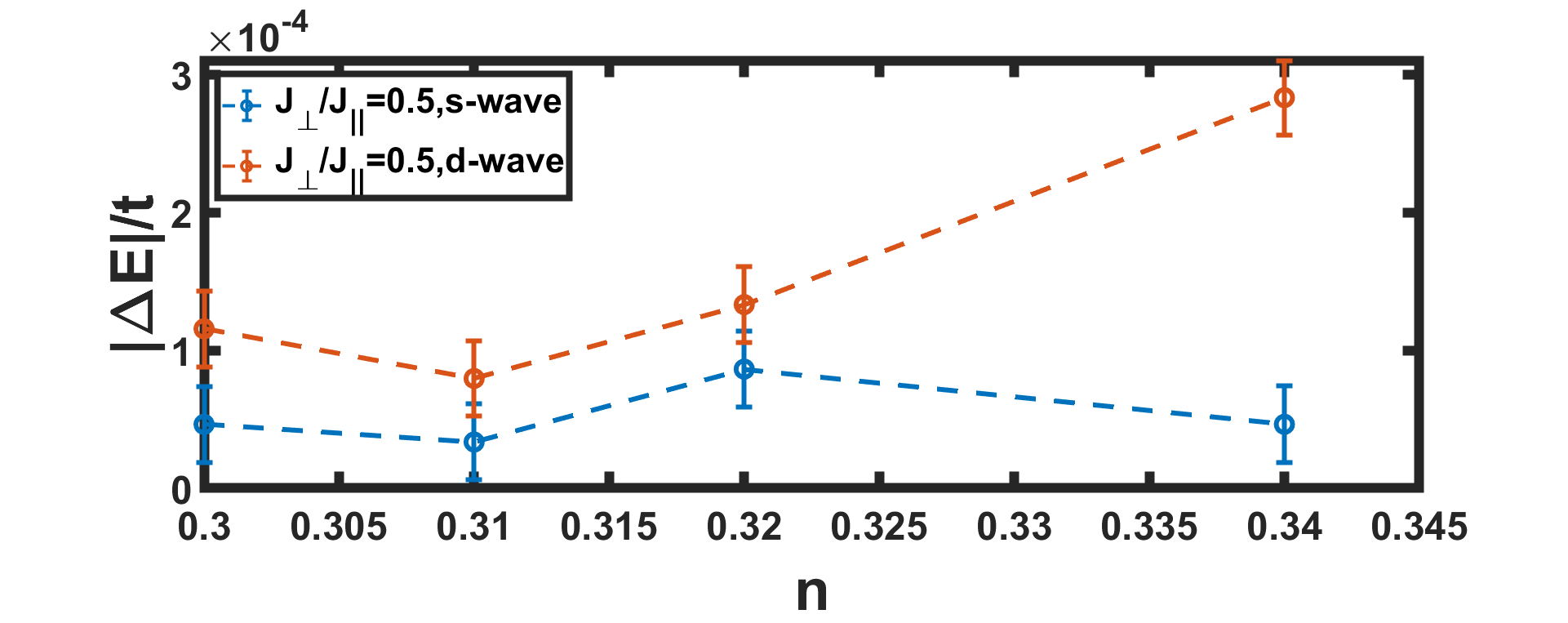}
        (a)
    \end{subfigure}
    \begin{subfigure}
        \centering 
        \includegraphics[width=0.5\textwidth]{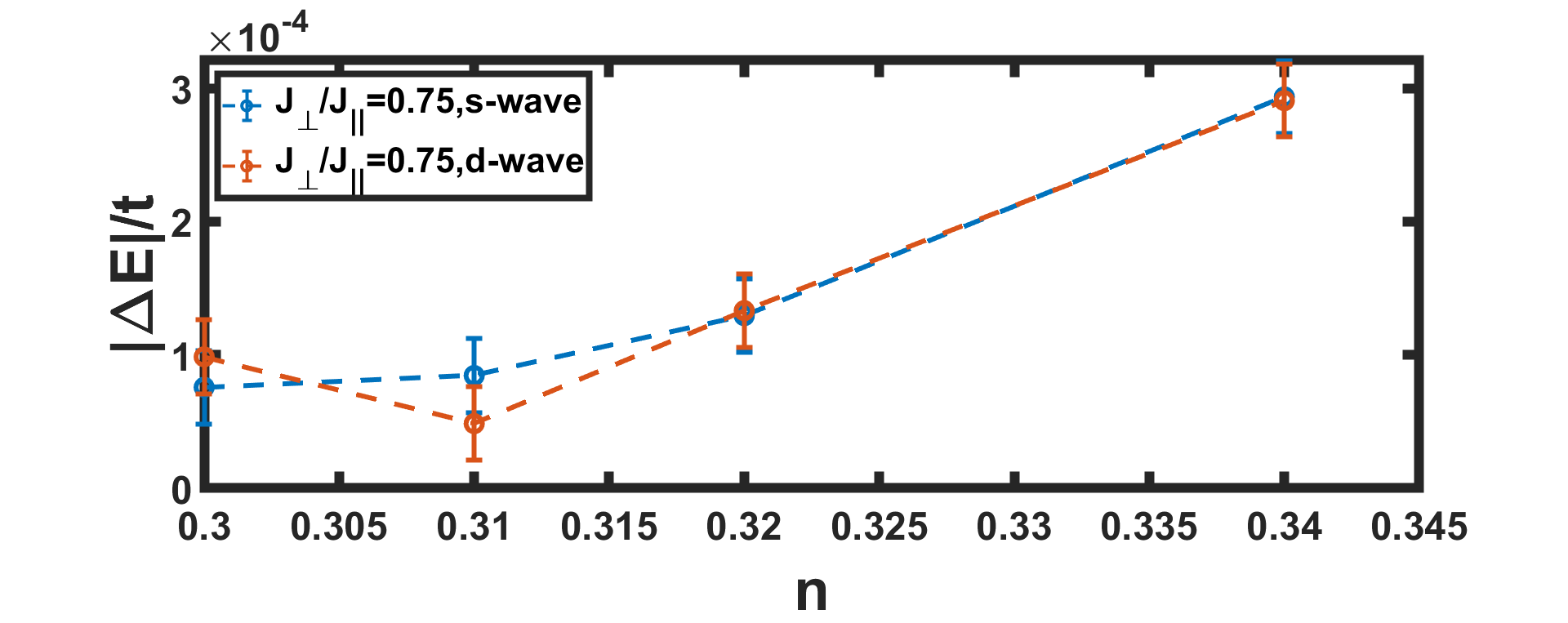}
        (b)
    \end{subfigure}
    \begin{subfigure}
        \centering  
        \includegraphics[width=0.5\textwidth]{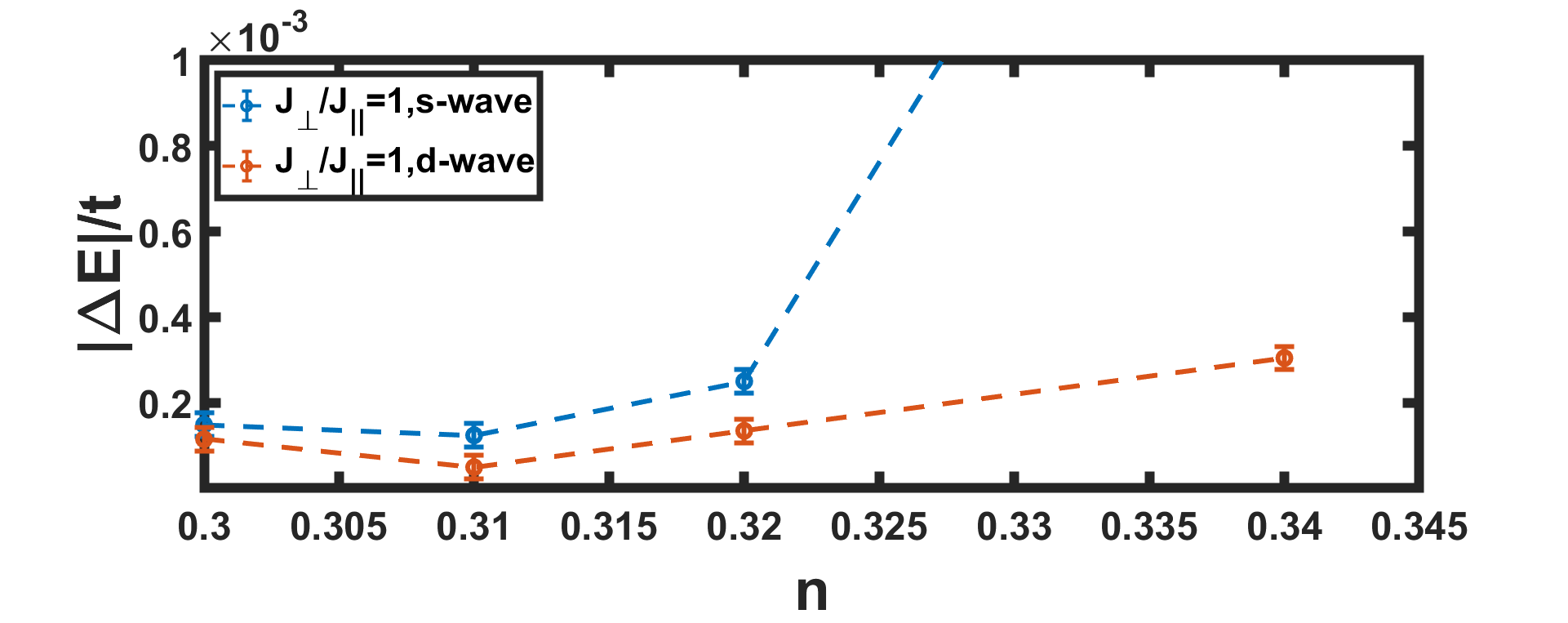}
        (c)
    \end{subfigure} 
    \caption{Energy gain per site ($E_{normal}-E_{SC}$) $|\Delta E|/t$ of s-wave and d-wave pairing versus electron filling at (a) $J_{\perp}/J_{\parallel}=0.5$, (b) $J_{\perp}/J_{\parallel}=0.75$ and (c) $J_{\perp}/J_{\parallel}=1$.}
  \label{s-d}
\end{figure}

\subsection{$J_{\perp}$-Dependence of The Pairing Symmetry}
\label{competition of s-wave and d-wave SC }
In this part we study how $J_{\perp}$ changes pairing symmetry. At $J_{\perp}/J_{\parallel}=0.5$, $\Delta_{\perp}$ shows a reduction trend with increasing electron filling, indicating that s-wave SC becomes more unfavorable with increasing filling. We check the d-wave trial wavefunction which $\Delta_{\perp}=0$ and $\Delta_{x}=-\Delta_{y}$ in (\ref{MF}), both the energy gains of s-wave and d-wave SC are shown in Fig.~\ref{s-d}. The s-wave SC becomes more favorable at $J_{\perp}/J_{\parallel}=1 (\eta=50\%)$ in Fig.~\ref{s-d}(c), and the energy gain of s-wave SC is almost the same as that of d-wave SC at $J_{\perp}/J_{\parallel}=0.75 (\eta=37.5\%)$ in Fig.~\ref{s-d}(b); however the d-wave SC becomes more favorable at $J_{\perp}/J_{\parallel}=0.5 (\eta=25\%)$ in Fig.~\ref{s-d}(a). These results show a $J_{\perp}$ dependence of the pairing symmetry that as $J_{\perp}$ increases, the intralayer d-wave SC transitions to the interlayer s-wave SC at $J_{\perp}/J_{\parallel}\approx0.75$. It is agreed with the SBMF calculation that there is a first-order phase transition from d-wave SC to s-wave SC at nearly $J_{\perp}/J_{\parallel}=0.8$\cite{oh2023type2}.
Besides, a recent DMRG calculation of $t-J_{\parallel}-J_{\perp}$ model\cite{yang2025} also finds robust interlayer SC at $J_{\perp}/J_{\parallel}\geq1$, but cannot confirm the d-wave intralayer SC at $J_{\perp}/J_{\parallel}<1$ due to the computational complexity caused by the size and long-range entanglement.

\begin{figure}[t!]
\centering
\includegraphics[width=1\linewidth]{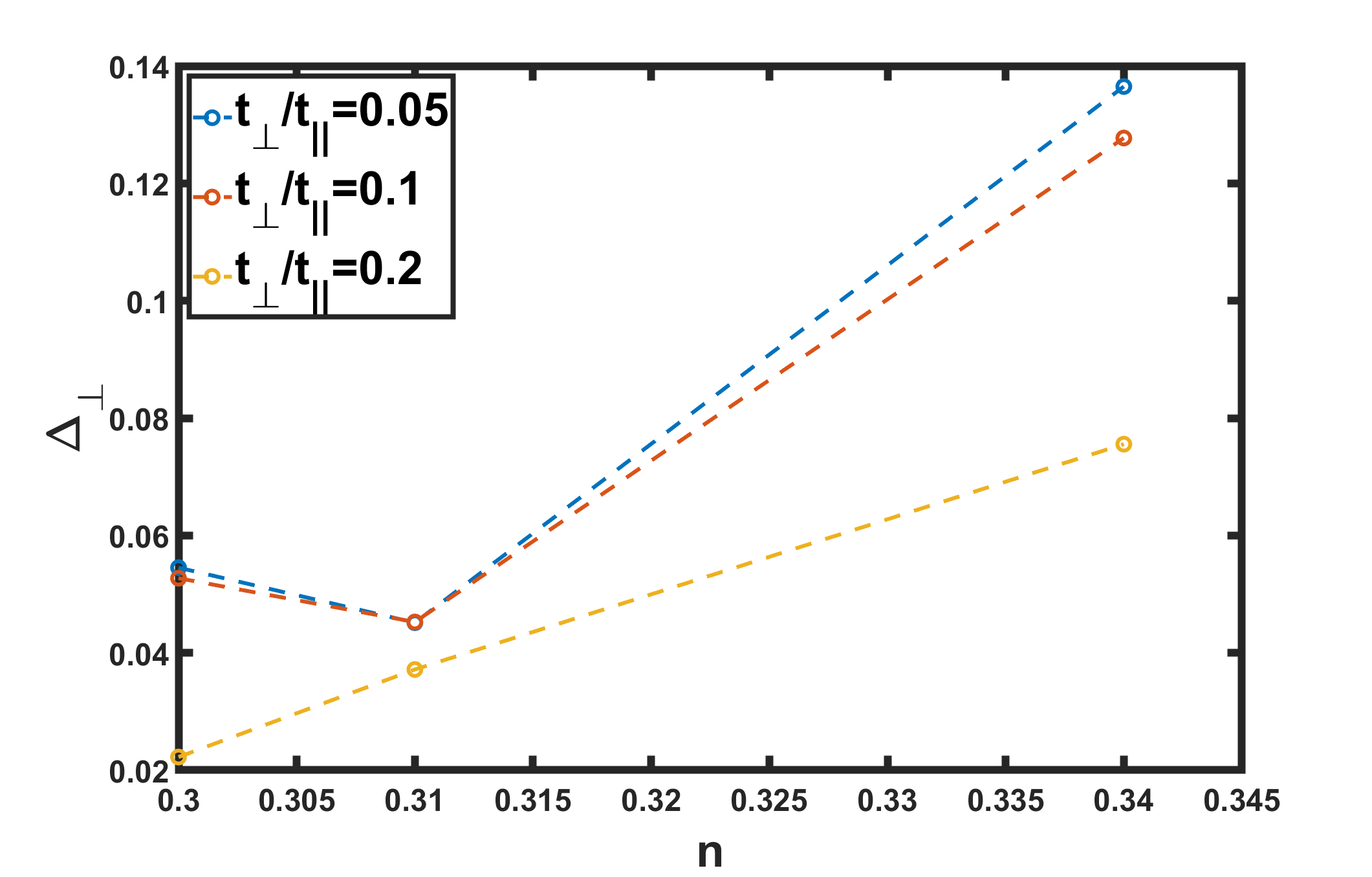}
\caption{Pairing parameters $\Delta_{\perp}$ versus electron filling at different interlayer hopping $t_{\perp}/t_{\parallel}=0.05,0.1,0.2$ at $J_{\perp}/J_{\parallel}=1$.}\label{interlayer hopping}
\end{figure}
\subsection{$t_\perp$-Dependence of the Pairing Strength}
\label{the interlayer hopping decreases interlayer SC}
In this part we study how the interlayer hopping of $d_{x^2 - y^2}$ orbital influences the interlayer pairing.

For the Hamiltonian:
\begin{equation}
\begin{aligned}
H=&-t_{\parallel}\sum_{\langle i,j \rangle,\alpha,\sigma}\mathcal{P}\left(c_{i\alpha\sigma}^{\dagger}c_{j\alpha\sigma}+h.c.\right)\mathcal{P}\\&-t_{\perp}\sum_{i\sigma}\mathcal{P}\left(c_{i1\sigma}^{\dagger}c_{i2\sigma}+h.c.\right)\mathcal{P}
\\&+J_{\parallel}\sum_{\langle i,j\rangle,\alpha}\mathbf{S}_{i\alpha}\cdot\mathbf{S}_{j\alpha}+J_{\perp}\sum_{i}\mathbf{S}_{i1}\cdot\mathbf{S}_{i2}.,\end{aligned}
\end{equation}
in which $t_{\perp}$ is the interlayer hopping of the $d_{x^2 - y^2}$ orbital. Fig.~\ref{interlayer hopping} shows different $t_{\perp}$ and the corresponding order parameters $\Delta_{\perp}$ by the VMC calculation at $J_{\perp}/J_{\parallel}=1$. As $t_{\perp}$ increases, $\Delta_{\perp}$ shows a decreasing trend with the same filling, indicating that $t_{\perp}$ is harmful to SC. This is consistent with the conclusion of\cite{oh2023type2,qu2023bilayer,chen2023iPEPS}, that interlayer hopping could introduce the Pauli blocking effect, which promotes repulsion between interlayer holes rather than forming interlayer spin-singlet pairs. Besides, strong $t_{\perp}$ may induce strong $\chi_{z}$ in the particle-hole channel and reduce $\Delta_{\perp}$ in the pairing channel. According to the DFT calculation\cite{YaoDX2023}, the interlayer hopping $t_{x\perp}=0.005$ is very weak compared to the intralayer hopping $t_{x\parallel}=-0.483$ in the $d_{x^2 - y^2}$ orbital, providing a favorable condition for HTSC in La$_3$Ni$_2$O$_7$. 

It should be noted that the HTSC in La$_3$Ni$_2$O$_7$ can not be described by any one orbital's $t-J$ model that is physically derived from a single orbital. As shown in Fig.~\ref{structure}, the $d_{x^2 - y^2}$ orbital has a small $t_{\perp}$ that leads to a small direct interlayer superexchange $J_{x\perp}$ according to $J_{x\perp}=4t_{\perp}^2/U$, although it has a large $t_{\parallel}$ and $J_{\parallel}$, the quarter-filling of $d_{x^2 - y^2}$  in La$_3$Ni$_2$O$_7$ which corresponds to a heavily overdoped region in the 2D $t-J$ model, leads to low carrier concentration and low $T_c$.
The d$_{z^2}$ orbital has strong interlayer superexchange, but its half-filling and small intralayer hopping lead to the lack of phase coherence. By introducing effective interlayer superexchange from the d$_{z^2}$ orbital by Hund coupling, both suppressed $t_{\perp}$ and large $J_{\perp}$ of the $d_{x^2 - y^2}$ orbital can be satisfied simultaneously which leads to the HTSC of La$_3$Ni$_2$O$_7$.

Interestingly, some ultracold atom experiments of mixed-dimensional systems\cite{Hirthe2023,Bohrdt2022} can be described by this $t-J_{\parallel}-J_{\perp}$ model. By controlling the potential gradient between two layers, the interlayer hopping is suppressed but $J_{\perp}$ still exists. The system undergoes a BEC–BCS crossover when $t_{\parallel}/J_{\perp}$ increases, and they find a (quasi-) long-range s-wave pairing order with $T_{c}\approx J_{\perp}/2$ at $t_{\parallel}/J_{\perp}=0.6$ in the 2D limit\cite{Grusdt2023lno03349,PRXQuantum.5.040341}. This leads to a possible way to realize HTSC in artificial systems. 

\section{Discussion and Conclusions}
\label{Discussion and Conclusions}

Recently, some thin-film nickelate experiments under ambient pressure have deepened our understanding of pairing mechanism. The ARPES experiments on (La,Pr)$_3$Ni$_2$O$_7$ film claim that there is a $\gamma$ band pocket of d$_{z^2}$ orbital\cite{10.1093} and a gap node is absent along the diagonal of the Brillouin zone\cite{shen2025}. However, another ARPES on the strained thin film La$_2$PrNi$_2$O$_7$ reports that the $\gamma$ band pocket stays $\sim$70meV below the Fermi surface\cite{wang2025electronic}, indicating that the $\gamma$ pocket is not a necessary condition for HTSC and d$_{z^2}$ orbital electrons are possibly not charge carriers. We ignore the details near the Fermi surface and start from the strong-coupling view to construct the effective $d_{x^2 - y^2}$ orbital's model, and our results are not in contradiction to the aforementioned experimental results. Besides recent STM measurement\cite{fan2025STM} shows a two-gap structure and fittings based on the Dynes model, which indicates that the dominant gap should have an anisotropic s-wave structure with a dominant interlayer pairing, although the V-shape spectrum should be considered more carefully in future studies, such as the two-orbital Hubbard model with Hund coupling\cite{YaoDX2023,wang2024electronic,liao2023electron,YangF2023,zhang2023structural,Xia2025,WangQH2023,HuJP2023,PhysRevLett.134.076001,Kuroki2023,heier2023competing,PhysRevB.110.235119,lechermann2023,ryee2024quenched,PhysRevResearch.7.L012066,tian2023correlation,PhysRevB.111.035108}.

In summary, we study a d$_{x^2-y^2}$ orbital's $t-J_{\parallel}-J_{\perp}$ model of La$_3$Ni$_2$O$_7$ by VMC simulation. We stress the significance of Hund coupling in the transfer interlayer AFM superexchange of d$_{z^2}$ orbital to d$_{x^2-y^2}$ orbital and find a dominant interlayer s-wave pairing at effective $J_{\perp}/ J_{\parallel}=1,2$, meanwhile the SC order parameter has a drastic improvement compared to those of the MF type of theories, indicating the important role of Gutzwiller projection to HTSC. The interlayer pairing is still considerably
large even at $J_{\perp}= J_{\parallel}$, which corresponds to a finite transfer ratio $\eta=50\%$ with a finite Hund coupling. Under this parameter, we obtain a reasonable $T_{c}$ by VMC calculation, which is close to the experimental result, but the HTSC cannot be explained by SBMF and MF solution with the same parameters. SC can be promoted by self-electron doping of the $d_{x^2 - y^2}$ orbital, and strong effective $J_{\perp}$. In addition, suppressed $d_{x^2 - y^2}$ interlayer hopping $t_{\perp}$ is beneficial in forming the interlayer spin-singlet state, which also promotes SC. This corresponds to the small interlayer hopping of d$_{x^2-y^2}$ orbital in the real materials. The interlayer s-wave SC changes to the intralayer d-wave SC as the effective $J_{\perp}$ decreases in this model. Our research offers a new perspective for understanding the
pairing mechanism of bilayer nickelates, and provides a reference for recent ultracold atom experiments in mixed-dimensional systems.

\section*{Acknowledgments}
The authors thank Chen Lu, Zhi-Yan Shao, and Jia-Heng Ji for helpful discussions. 
F. Y. is supported by the National Natural Science Foundation of China under Grant Nos. 12234016 and 12074031.

\subsection*{Appendix A: Details of VMC Calculation}

We consider both intraband pairing and interband pairing in the $t-J_{\parallel}-J_{\perp}$ model. By particle-hole transformation on the spin-down electrons:
\begin{equation}
\begin{aligned}
c_{-\pmb{k}\alpha\downarrow}^{\dagger}=d_{-\pmb{k}\alpha\downarrow}, c_{-\pmb{k}\alpha\downarrow}=d_{-\pmb{k}\alpha\downarrow}^{\dagger}
\end{aligned}
\end{equation}
\begin{equation}
\begin{aligned}
H_{mf}&=\sum_{\pmb{k}\alpha}\varepsilon_{\pmb{k}\alpha}\left( c_{\pmb{k}\alpha\uparrow}^{\dagger}c_{\pmb{k}\alpha\uparrow}+d_{-\pmb{k}\alpha\downarrow}d_{-\pmb{k}\alpha\downarrow}^{\dagger}\right)\\
&-t_{\perp}\sum_{\pmb{k}}\left( c_{\pmb{k}1\uparrow}^{\dagger}c_{\pmb{k}2\uparrow}+d_{-\pmb{k}1\downarrow}d_{-\pmb{k}2\downarrow}^{\dagger}\right)+h.c. \\
&+\sum_{\pmb{k}\alpha}F_{\pmb{k}\alpha}c_{\pmb{k}\alpha\uparrow}^{\dagger}d_{-\pmb{k}\alpha\downarrow}+h.c.\\
&+\sum_{\pmb{k}\alpha}\Delta_{\perp}\left( c_{\pmb{k}1\uparrow}^{\dagger}d_{-\pmb{k}2\downarrow}-d_{-\pmb{k}1\downarrow}c_{\pmb{k}2\uparrow}^{\dagger}  \right)+h.c.\\
&=\Psi_{\pmb{k}}^{\dagger}\begin{pmatrix}
    \varepsilon_{\pmb{k}1} & -t_{\perp} & F_{\pmb{k}1} & \Delta_{\perp}\\
    -t_{\perp} & \varepsilon_{\pmb{k}2} & \Delta_{\perp} & F_{\pmb{k}2}\\
    F_{\pmb{k}1}^{*} & \Delta_{\perp}^{*} & -\varepsilon_{-\pmb{k}1} & t_{\perp}\\
    \Delta_{\perp}^{*} & F_{\pmb{k}2}^{*} & t_{\perp} & -\varepsilon_{-\pmb{k}2}
    \end{pmatrix}
\Psi_{\pmb{k}}+const,
\label{MF}
\end{aligned}
\end{equation}
here $\varepsilon_{\pmb{k}\alpha}=-2{t_{0}}(cosk_x+cosk_y)-\mu$, $\Psi_{\pmb{k}}^{\dagger}=\left(c_{\pmb{k}1\uparrow}^{\dagger},c_{\pmb{k}2\uparrow}^{\dagger},d_{-\pmb{k}1\downarrow}^{\dagger},d_{-\pmb{k}2\downarrow}^{\dagger}\right)$, $F_{\pmb{k}\alpha}=\Delta_x cosk_x+\Delta_y cosk_y$. If the pairing symmetry is s-wave, then $F_{\pmb{k}\alpha}=\Delta_{\parallel} (cosk_x+ cosk_y)$. If the pairing symmetry is d-wave, then $F_{\pmb{k}\alpha}=\Delta_{\parallel} (cos k_x- cos k_y)$ and $\Delta_{\perp}=0$. It can be diagonalized by the Bogoliubov transformation: $H_{mf}=\sum_{\pmb{k}}\gamma_{\pmb{k}}^{\dagger}E_{\pmb{k}}\gamma_{\pmb{k}}$, and the eigenvector is $\xi_{\mu\alpha}(\pmb{k})$ ($\mu$ is the layer and $\alpha$ is the band). Now the wavefunction is
\begin{equation}
|FS\rangle=\prod_{\pmb{k},E_{\pmb{k}<0}}\gamma_{\pmb{k}}^{\dagger}|0\rangle,
\end{equation}
and the configuration of electrons is
\begin{equation}
|x\rangle=c_{R_{1}\uparrow}^{\dagger}c_{R_2\uparrow}^{\dagger}\dotsc_{R_{N_e/2}\uparrow}^{\dagger}d_{r_1\downarrow}^{\dagger}d_{r_2\downarrow}^{\dagger}\dots d_{r_{N-N_{e}/2}\downarrow}^{\dagger}|0\rangle.
\end{equation}
The local Hilbert space $\{\uparrow,\downarrow,0\}$ is changing to $\{\uparrow\downarrow_d, 0, \downarrow_d\}$ and there is no $\{\uparrow\}$ configuration corresponding to the exclusion of double occupation. $R_i$ represents the i-site of spin-up electrons and $r_j$ represents the j-site of spin-down hole in the real space. We need to calculate the determinant of the matrix $A$
\begin{equation}
\langle x|P_{G}|FS\rangle=\det(A),
\end{equation}
and 
\begin{equation}
A_{mn}=\xi_{\mu_{m}\alpha_{n}}(\pmb{k})e^{i\pmb{k}_{n}\cdot\pmb{R}_{m}}.
\end{equation}
in the VMC calculation.

\subsection*{Appendix B: Small $\chi_z$ and MF Analysis}
\label{chiz}

\begin{figure}[t]
\centering
\includegraphics[width=1\linewidth]{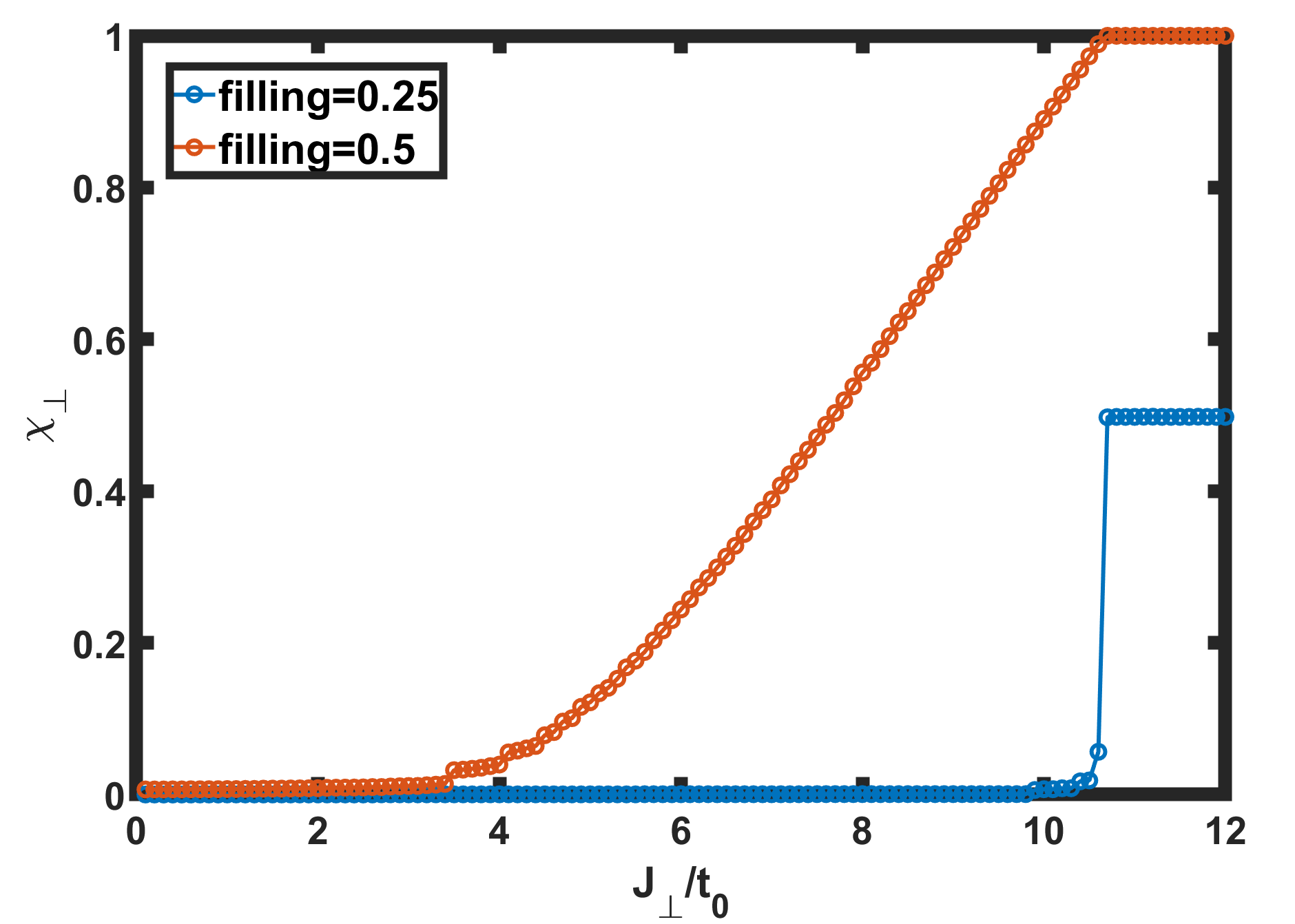}
\caption{MF self-consistent calculation for $\chi_{\perp}$ at half-filling and quarter-filling.}\label{chi}
\end{figure}

In the VMC calculation, we find that $\chi_z$ is very small compared to $\Delta_{\perp}$, so we propose a MF analysis to understand it, which is similar to the Stoner criterion in the itinerant ferromagnetic phase. Only considering the particle-hole channel of $H_{mf}$:
\begin{equation}
\begin{aligned}
H_{mf}&=\sum_{\pmb{k}\alpha\sigma}\varepsilon_{\pmb{k}}c_{\pmb{k}\alpha\sigma}^{\dagger}c_{\pmb{k}\alpha\sigma}-\frac{3}{8}J\chi_{z}\sum_{\pmb{k}\sigma}c_{\pmb{k}1\sigma}^{\dagger}c_{\pmb{k}2\sigma}+h.c.\\&=\sum_{\pmb{k}\sigma}\begin{pmatrix}
c_{\pmb{k}1\sigma}^{\dagger} & c_{\pmb{k}2\sigma}^{\dagger}
\end{pmatrix}
\begin{pmatrix}
\varepsilon_{\pmb{k}} & -\frac{3}{8}J\chi_{z} \\
-\frac{3}{8}J\chi_{z} & \varepsilon_{\pmb{k}}
\end{pmatrix}
\begin{pmatrix}
c_{\pmb{k}1\sigma} \\ c_{\pmb{k}2\sigma} 
\end{pmatrix}\\
&=\sum_{\pmb{k}\sigma}\begin{pmatrix}
\gamma_{\pmb{k}\sigma+}^{\dagger} & \gamma_{\pmb{k\sigma}-}^{\dagger}
\end{pmatrix}\begin{pmatrix}
\epsilon_{\pmb{k}+} & 0  \\
0 & \epsilon_{\pmb{k}-}
\end{pmatrix}
\begin{pmatrix}
\gamma_{\pmb{k}\sigma+} \\ \gamma_{\pmb{k}\sigma-} 
\end{pmatrix},
\end{aligned}
\end{equation}
here $\varepsilon_{\pmb{k}}=-2t_{0}(cosk_x+cosk_y)-\mu$, $c_{\pmb{k}1\sigma}=\frac{1}{\sqrt{2}}(\gamma_{\pmb{k}+}+\gamma_{\pmb{k}-}), c_{\pmb{k}2\sigma}=\frac{1}{\sqrt{2}}(\gamma_{\pmb{k}+}-\gamma_{\pmb{k}-})$ and $\epsilon_{\pmb{k}+}=\varepsilon_{\pmb{k}}-\frac{3}{8}J\chi_{z}, \epsilon_{\pmb{k}-}=\varepsilon_{\pmb{k}}+\frac{3}{8}J\chi_{z}$. So,
\begin{equation}
\begin{aligned}
\langle\chi_{\perp}\rangle&=\frac{1}{N}\sum_{\pmb{k}\sigma}\left( \langle c_{\pmb{k}1\uparrow}^{\dagger}c_{\pmb{k}2\uparrow}\rangle+\langle c_{\pmb{k}1\downarrow}^{\dagger}c_{\pmb{k}2\downarrow}\rangle
\right)\\&=\frac{1}{2N}\sum_{\pmb{k}\sigma}\left( \gamma_{\pmb{k}\sigma+}^{\dagger}\gamma_{\pmb{k}\sigma+}-\gamma_{\pmb{k}\sigma-}^{\dagger}\gamma_{\pmb{k}\sigma-}\right)\\
&=\frac{1}{2N}\left(\sum_{\pmb{k}(\epsilon_{\pmb{k}+}<\mu_c),\sigma}n_{\pmb{k}\sigma+}-\sum_{\pmb{k}(\epsilon_{\pmb{k}-}<\mu_c),\sigma}n_{\pmb{k}\sigma-}\right)
\end{aligned}
\end{equation}
and $\chi_{\perp}\approx\frac{3}{8}J_{\perp}\chi_{\perp}\rho_{\epsilon_{f}}$. This indicates that when the density of the state near the Fermi surface is not large enough, a huge $J_{\perp}$ is required to induce ferromagnetic instability. We adopt a self-consistent calculation, as shown in Fig.~\ref{chi}, the critical $J_{\perp c}\approx 3t_{0}$ for half-filling and $J_{\perp c}\approx 10t_{0}$ for quarter-filling. In our model $J_{\perp}\ll J_{\perp c}$, so we can ignore $\chi_{\perp}$ in our VMC calculation.

\subsection*{Appendix C: Size Effect of VMC Calculation Contrast with SBMF Analysis}
\begin{figure}[h]
\centering
\includegraphics[width=1\linewidth]{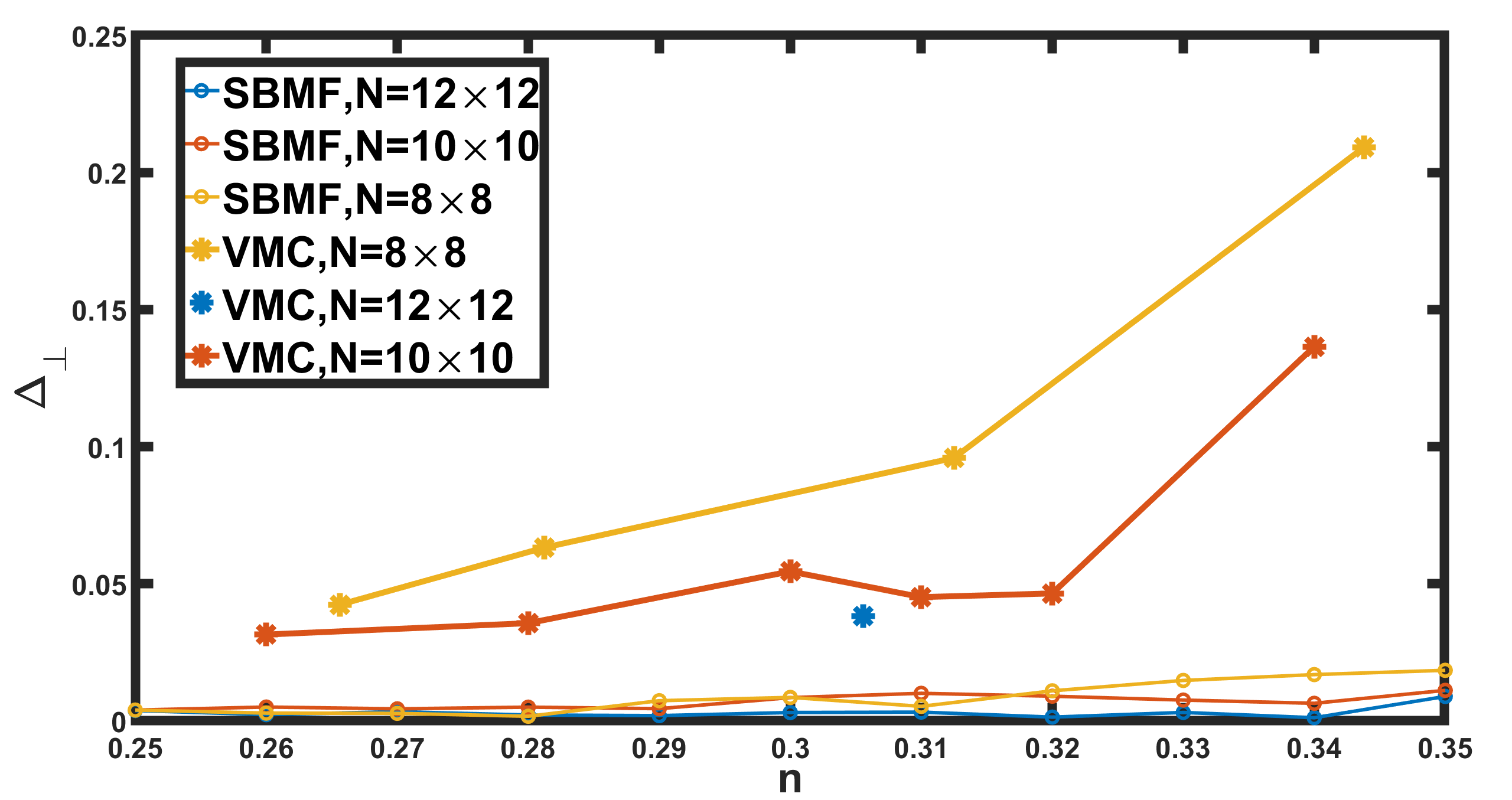}
\caption{Different lattice sizes of $t-J_{\parallel}-J_{\perp}$ model by the VMC calculation and the SBMF theory.}\label{size}
\end{figure}

Considering the finite-size effect of the VMC calculation, we show the results of different lattices in Fig.~\ref{size}, and the results of SBMF theory with the same lattice size. Even with the finite-size effect, the pairing strength of the VMC calculation is larger than that in the SBMF theory, which is in agreement with our conclusion.

\newpage
%\bibliography{reference}

\begin{thebibliography}{121}%
\makeatletter
\providecommand \@ifxundefined [1]{%
 \@ifx{#1\undefined}
}%
\providecommand \@ifnum [1]{%
 \ifnum #1\expandafter \@firstoftwo
 \else \expandafter \@secondoftwo
 \fi
}%
\providecommand \@ifx [1]{%
 \ifx #1\expandafter \@firstoftwo
 \else \expandafter \@secondoftwo
 \fi
}%
\providecommand \natexlab [1]{#1}%
\providecommand \enquote  [1]{``#1''}%
\providecommand \bibnamefont  [1]{#1}%
\providecommand \bibfnamefont [1]{#1}%
\providecommand \citenamefont [1]{#1}%
\providecommand \href@noop [0]{\@secondoftwo}%
\providecommand \href [0]{\begingroup \@sanitize@url \@href}%
\providecommand \@href[1]{\@@startlink{#1}\@@href}%
\providecommand \@@href[1]{\endgroup#1\@@endlink}%
\providecommand \@sanitize@url [0]{\catcode `\\12\catcode `\$12\catcode `\&12\catcode `\#12\catcode `\^12\catcode `\_12\catcode `\%12\relax}%
\providecommand \@@startlink[1]{}%
\providecommand \@@endlink[0]{}%
\providecommand \url  [0]{\begingroup\@sanitize@url \@url }%
\providecommand \@url [1]{\endgroup\@href {#1}{\urlprefix }}%
\providecommand \urlprefix  [0]{URL }%
\providecommand \Eprint [0]{\href }%
\providecommand \doibase [0]{https://doi.org/}%
\providecommand \selectlanguage [0]{\@gobble}%
\providecommand \bibinfo  [0]{\@secondoftwo}%
\providecommand \bibfield  [0]{\@secondoftwo}%
\providecommand \translation [1]{[#1]}%
\providecommand \BibitemOpen [0]{}%
\providecommand \bibitemStop [0]{}%
\providecommand \bibitemNoStop [0]{.\EOS\space}%
\providecommand \EOS [0]{\spacefactor3000\relax}%
\providecommand \BibitemShut  [1]{\csname bibitem#1\endcsname}%
\let\auto@bib@innerbib\@empty
%</preamble>
\bibitem [{\citenamefont {Sun}\ \emph {et~al.}(2023)\citenamefont {Sun}, \citenamefont {Huo}, \citenamefont {Hu}, \citenamefont {Li}, \citenamefont {Liu}, \citenamefont {Han}, \citenamefont {Tang}, \citenamefont {Mao}, \citenamefont {Yang}, \citenamefont {Wang}, \citenamefont {Cheng}, \citenamefont {Yao}, \citenamefont {Zhang},\ and\ \citenamefont {Wang}}]{Wang2023LNO}%
  \BibitemOpen
  \bibfield  {author} {\bibinfo {author} {\bibfnamefont {H.}~\bibnamefont {Sun}}, \bibinfo {author} {\bibfnamefont {M.}~\bibnamefont {Huo}}, \bibinfo {author} {\bibfnamefont {X.}~\bibnamefont {Hu}}, \bibinfo {author} {\bibfnamefont {J.}~\bibnamefont {Li}}, \bibinfo {author} {\bibfnamefont {Z.}~\bibnamefont {Liu}}, \bibinfo {author} {\bibfnamefont {Y.}~\bibnamefont {Han}}, \bibinfo {author} {\bibfnamefont {L.}~\bibnamefont {Tang}}, \bibinfo {author} {\bibfnamefont {Z.}~\bibnamefont {Mao}}, \bibinfo {author} {\bibfnamefont {P.}~\bibnamefont {Yang}}, \bibinfo {author} {\bibfnamefont {B.}~\bibnamefont {Wang}}, \bibinfo {author} {\bibfnamefont {J.}~\bibnamefont {Cheng}}, \bibinfo {author} {\bibfnamefont {D.-X.}\ \bibnamefont {Yao}}, \bibinfo {author} {\bibfnamefont {G.-M.}\ \bibnamefont {Zhang}},\ and\ \bibinfo {author} {\bibfnamefont {M.}~\bibnamefont {Wang}},\ }\bibfield  {title} {\bibinfo {title} {Signatures of superconductivity near 80k in a nickelate under high pressure},\ }\href
  {https://doi.org/10.1038/s41586-023-06408-7} {\bibfield  {journal} {\bibinfo  {journal} {Nature}\ }\textbf {\bibinfo {volume} {621}},\ \bibinfo {pages} {493} (\bibinfo {year} {2023})}\BibitemShut {NoStop}%
\bibitem [{\citenamefont {Zhang}\ \emph {et~al.}(2024{\natexlab{a}})\citenamefont {Zhang}, \citenamefont {Su}, \citenamefont {Huang}, \citenamefont {Shan}, \citenamefont {Sun}, \citenamefont {Huo}, \citenamefont {Ye}, \citenamefont {Zhang}, \citenamefont {Yang}, \citenamefont {Xu}, \citenamefont {Su}, \citenamefont {Li}, \citenamefont {Smidman}, \citenamefont {Wang}, \citenamefont {Jiao},\ and\ \citenamefont {Yuan}}]{YuanHQ2023LNO}%
  \BibitemOpen
  \bibfield  {author} {\bibinfo {author} {\bibfnamefont {Y.}~\bibnamefont {Zhang}}, \bibinfo {author} {\bibfnamefont {D.}~\bibnamefont {Su}}, \bibinfo {author} {\bibfnamefont {Y.}~\bibnamefont {Huang}}, \bibinfo {author} {\bibfnamefont {Z.}~\bibnamefont {Shan}}, \bibinfo {author} {\bibfnamefont {H.}~\bibnamefont {Sun}}, \bibinfo {author} {\bibfnamefont {M.}~\bibnamefont {Huo}}, \bibinfo {author} {\bibfnamefont {K.}~\bibnamefont {Ye}}, \bibinfo {author} {\bibfnamefont {J.}~\bibnamefont {Zhang}}, \bibinfo {author} {\bibfnamefont {Z.}~\bibnamefont {Yang}}, \bibinfo {author} {\bibfnamefont {Y.}~\bibnamefont {Xu}}, \bibinfo {author} {\bibfnamefont {Y.}~\bibnamefont {Su}}, \bibinfo {author} {\bibfnamefont {R.}~\bibnamefont {Li}}, \bibinfo {author} {\bibfnamefont {M.}~\bibnamefont {Smidman}}, \bibinfo {author} {\bibfnamefont {M.}~\bibnamefont {Wang}}, \bibinfo {author} {\bibfnamefont {L.}~\bibnamefont {Jiao}},\ and\ \bibinfo {author} {\bibfnamefont {H.}~\bibnamefont {Yuan}},\ }\bibfield  {title} {\bibinfo {title}
  {High-temperature superconductivity with zero resistance and strange-metal behaviour in {L}a$_3${N}i$_2${O}$_{7-\delta}$},\ }\href {https://doi.org/10.1038/s41567-024-02515-y} {\bibfield  {journal} {\bibinfo  {journal} {Nat. Phys.}\ }\textbf {\bibinfo {volume} {20}},\ \bibinfo {pages} {1269} (\bibinfo {year} {2024}{\natexlab{a}})}\BibitemShut {NoStop}%
\bibitem [{\citenamefont {Hou}\ \emph {et~al.}(2023)\citenamefont {Hou}, \citenamefont {Yang}, \citenamefont {Liu}, \citenamefont {Li}, \citenamefont {Shan}, \citenamefont {Ma}, \citenamefont {Wang}, \citenamefont {Wang}, \citenamefont {Guo}, \citenamefont {Sun}, \citenamefont {Uwatoko}, \citenamefont {Wang}, \citenamefont {Zhang}, \citenamefont {Wang},\ and\ \citenamefont {Cheng}}]{Wang2023LNOb}%
  \BibitemOpen
  \bibfield  {author} {\bibinfo {author} {\bibfnamefont {J.}~\bibnamefont {Hou}}, \bibinfo {author} {\bibfnamefont {P.-T.}\ \bibnamefont {Yang}}, \bibinfo {author} {\bibfnamefont {Z.-Y.}\ \bibnamefont {Liu}}, \bibinfo {author} {\bibfnamefont {J.-Y.}\ \bibnamefont {Li}}, \bibinfo {author} {\bibfnamefont {P.-F.}\ \bibnamefont {Shan}}, \bibinfo {author} {\bibfnamefont {L.}~\bibnamefont {Ma}}, \bibinfo {author} {\bibfnamefont {G.}~\bibnamefont {Wang}}, \bibinfo {author} {\bibfnamefont {N.-N.}\ \bibnamefont {Wang}}, \bibinfo {author} {\bibfnamefont {H.-Z.}\ \bibnamefont {Guo}}, \bibinfo {author} {\bibfnamefont {J.-P.}\ \bibnamefont {Sun}}, \bibinfo {author} {\bibfnamefont {Y.}~\bibnamefont {Uwatoko}}, \bibinfo {author} {\bibfnamefont {M.}~\bibnamefont {Wang}}, \bibinfo {author} {\bibfnamefont {G.-M.}\ \bibnamefont {Zhang}}, \bibinfo {author} {\bibfnamefont {B.-S.}\ \bibnamefont {Wang}},\ and\ \bibinfo {author} {\bibfnamefont {J.-G.}\ \bibnamefont {Cheng}},\ }\bibfield  {title} {\bibinfo {title} {Emergence of
  high-temperature superconducting phase in pressurized {L}a$_{3}${N}i$_{2}${O}$_7$ crystals},\ }\href {https://doi.org/10.1088/0256-307X/40/11/117302} {\bibfield  {journal} {\bibinfo  {journal} {Chin. Phys. Lett.}\ }\textbf {\bibinfo {volume} {40}},\ \bibinfo {eid} {117302} (\bibinfo {year} {2023})}\BibitemShut {NoStop}%
\bibitem [{\citenamefont {Wang}\ \emph {et~al.}(2024{\natexlab{a}})\citenamefont {Wang}, \citenamefont {Wang}, \citenamefont {Shen}, \citenamefont {Hou}, \citenamefont {Ma}, \citenamefont {Shi}, \citenamefont {Ren}, \citenamefont {Gu}, \citenamefont {Ma}, \citenamefont {Yang}, \citenamefont {Liu}, \citenamefont {Guo}, \citenamefont {Sun}, \citenamefont {Zhang}, \citenamefont {Calder}, \citenamefont {Yan}, \citenamefont {Wang}, \citenamefont {Uwatoko},\ and\ \citenamefont {Cheng}}]{wang2023LNOpoly}%
  \BibitemOpen
  \bibfield  {author} {\bibinfo {author} {\bibfnamefont {G.}~\bibnamefont {Wang}}, \bibinfo {author} {\bibfnamefont {N.~N.}\ \bibnamefont {Wang}}, \bibinfo {author} {\bibfnamefont {X.~L.}\ \bibnamefont {Shen}}, \bibinfo {author} {\bibfnamefont {J.}~\bibnamefont {Hou}}, \bibinfo {author} {\bibfnamefont {L.}~\bibnamefont {Ma}}, \bibinfo {author} {\bibfnamefont {L.~F.}\ \bibnamefont {Shi}}, \bibinfo {author} {\bibfnamefont {Z.~A.}\ \bibnamefont {Ren}}, \bibinfo {author} {\bibfnamefont {Y.~D.}\ \bibnamefont {Gu}}, \bibinfo {author} {\bibfnamefont {H.~M.}\ \bibnamefont {Ma}}, \bibinfo {author} {\bibfnamefont {P.~T.}\ \bibnamefont {Yang}}, \bibinfo {author} {\bibfnamefont {Z.~Y.}\ \bibnamefont {Liu}}, \bibinfo {author} {\bibfnamefont {H.~Z.}\ \bibnamefont {Guo}}, \bibinfo {author} {\bibfnamefont {J.~P.}\ \bibnamefont {Sun}}, \bibinfo {author} {\bibfnamefont {G.~M.}\ \bibnamefont {Zhang}}, \bibinfo {author} {\bibfnamefont {S.}~\bibnamefont {Calder}}, \bibinfo {author} {\bibfnamefont {J.-Q.}\ \bibnamefont {Yan}},
  \bibinfo {author} {\bibfnamefont {B.~S.}\ \bibnamefont {Wang}}, \bibinfo {author} {\bibfnamefont {Y.}~\bibnamefont {Uwatoko}},\ and\ \bibinfo {author} {\bibfnamefont {J.-G.}\ \bibnamefont {Cheng}},\ }\bibfield  {title} {\bibinfo {title} {Pressure-induced superconductivity in polycrystalline {L}a$_3${N}i$_2${O}$_7$},\ }\href {https://doi.org/10.1103/PhysRevX.14.011040} {\bibfield  {journal} {\bibinfo  {journal} {Phys. Rev. X}\ }\textbf {\bibinfo {volume} {14}},\ \bibinfo {pages} {011040} (\bibinfo {year} {2024}{\natexlab{a}})}\BibitemShut {NoStop}%
\bibitem [{\citenamefont {Zhou}\ \emph {et~al.}(2025{\natexlab{a}})\citenamefont {Zhou}, \citenamefont {Guo}, \citenamefont {Cai}, \citenamefont {Sun}, \citenamefont {Li}, \citenamefont {Zhao}, \citenamefont {Wang}, \citenamefont {Han}, \citenamefont {Chen}, \citenamefont {Chen}, \citenamefont {Wu}, \citenamefont {Ding}, \citenamefont {Xiang}, \citenamefont {Mao},\ and\ \citenamefont {Sun}}]{10.1063/5.0247684}%
  \BibitemOpen
  \bibfield  {author} {\bibinfo {author} {\bibfnamefont {Y.}~\bibnamefont {Zhou}}, \bibinfo {author} {\bibfnamefont {J.}~\bibnamefont {Guo}}, \bibinfo {author} {\bibfnamefont {S.}~\bibnamefont {Cai}}, \bibinfo {author} {\bibfnamefont {H.}~\bibnamefont {Sun}}, \bibinfo {author} {\bibfnamefont {C.}~\bibnamefont {Li}}, \bibinfo {author} {\bibfnamefont {J.}~\bibnamefont {Zhao}}, \bibinfo {author} {\bibfnamefont {P.}~\bibnamefont {Wang}}, \bibinfo {author} {\bibfnamefont {J.}~\bibnamefont {Han}}, \bibinfo {author} {\bibfnamefont {X.}~\bibnamefont {Chen}}, \bibinfo {author} {\bibfnamefont {Y.}~\bibnamefont {Chen}}, \bibinfo {author} {\bibfnamefont {Q.}~\bibnamefont {Wu}}, \bibinfo {author} {\bibfnamefont {Y.}~\bibnamefont {Ding}}, \bibinfo {author} {\bibfnamefont {T.}~\bibnamefont {Xiang}}, \bibinfo {author} {\bibfnamefont {H.-k.}\ \bibnamefont {Mao}},\ and\ \bibinfo {author} {\bibfnamefont {L.}~\bibnamefont {Sun}},\ }\bibfield  {title} {\bibinfo {title} {Investigations of key issues on the reproducibility of
  high-{T}c superconductivity emerging from compressed {L}a$_3${N}i$_2${O}$_7$},\ }\href {https://doi.org/10.1063/5.0247684} {\bibfield  {journal} {\bibinfo  {journal} {Matter and Radiation at Extremes}\ }\textbf {\bibinfo {volume} {10}},\ \bibinfo {pages} {027801} (\bibinfo {year} {2025}{\natexlab{a}})}\BibitemShut {NoStop}%
\bibitem [{\citenamefont {Zhang}\ \emph {et~al.}(2024{\natexlab{b}})\citenamefont {Zhang}, \citenamefont {Pei}, \citenamefont {Wang}, \citenamefont {Zhao}, \citenamefont {Li}, \citenamefont {Cao}, \citenamefont {Zhu}, \citenamefont {Wu},\ and\ \citenamefont {Qi}}]{zhang2023pressure}%
  \BibitemOpen
  \bibfield  {author} {\bibinfo {author} {\bibfnamefont {M.}~\bibnamefont {Zhang}}, \bibinfo {author} {\bibfnamefont {C.}~\bibnamefont {Pei}}, \bibinfo {author} {\bibfnamefont {Q.}~\bibnamefont {Wang}}, \bibinfo {author} {\bibfnamefont {Y.}~\bibnamefont {Zhao}}, \bibinfo {author} {\bibfnamefont {C.}~\bibnamefont {Li}}, \bibinfo {author} {\bibfnamefont {W.}~\bibnamefont {Cao}}, \bibinfo {author} {\bibfnamefont {S.}~\bibnamefont {Zhu}}, \bibinfo {author} {\bibfnamefont {J.}~\bibnamefont {Wu}},\ and\ \bibinfo {author} {\bibfnamefont {Y.}~\bibnamefont {Qi}},\ }\bibfield  {title} {\bibinfo {title} {Effects of pressure and doping on ruddlesden-popper phases {L}a$_{n+1}${N}i$_n${O}$_{3n+1}$},\ }\href {https://www.sciencedirect.com/science/article/pii/S1005030223009829} {\bibfield  {journal} {\bibinfo  {journal} {J. Mater. Sci. Technol.}\ }\textbf {\bibinfo {volume} {185}},\ \bibinfo {pages} {147} (\bibinfo {year} {2024}{\natexlab{b}})}\BibitemShut {NoStop}%
\bibitem [{\citenamefont {Puphal}\ \emph {et~al.}(2024)\citenamefont {Puphal}, \citenamefont {Reiss}, \citenamefont {Enderlein}, \citenamefont {Wu}, \citenamefont {Khaliullin}, \citenamefont {Sundaramurthy}, \citenamefont {Priessnitz}, \citenamefont {Knauft}, \citenamefont {Suthar}, \citenamefont {Richter}, \citenamefont {Isobe}, \citenamefont {van Aken}, \citenamefont {Takagi}, \citenamefont {Keimer}, \citenamefont {Suyolcu}, \citenamefont {Wehinger}, \citenamefont {Hansmann},\ and\ \citenamefont {Hepting}}]{puphal2024unconven}%
  \BibitemOpen
  \bibfield  {author} {\bibinfo {author} {\bibfnamefont {P.}~\bibnamefont {Puphal}}, \bibinfo {author} {\bibfnamefont {P.}~\bibnamefont {Reiss}}, \bibinfo {author} {\bibfnamefont {N.}~\bibnamefont {Enderlein}}, \bibinfo {author} {\bibfnamefont {Y.-M.}\ \bibnamefont {Wu}}, \bibinfo {author} {\bibfnamefont {G.}~\bibnamefont {Khaliullin}}, \bibinfo {author} {\bibfnamefont {V.}~\bibnamefont {Sundaramurthy}}, \bibinfo {author} {\bibfnamefont {T.}~\bibnamefont {Priessnitz}}, \bibinfo {author} {\bibfnamefont {M.}~\bibnamefont {Knauft}}, \bibinfo {author} {\bibfnamefont {A.}~\bibnamefont {Suthar}}, \bibinfo {author} {\bibfnamefont {L.}~\bibnamefont {Richter}}, \bibinfo {author} {\bibfnamefont {M.}~\bibnamefont {Isobe}}, \bibinfo {author} {\bibfnamefont {P.~A.}\ \bibnamefont {van Aken}}, \bibinfo {author} {\bibfnamefont {H.}~\bibnamefont {Takagi}}, \bibinfo {author} {\bibfnamefont {B.}~\bibnamefont {Keimer}}, \bibinfo {author} {\bibfnamefont {Y.~E.}\ \bibnamefont {Suyolcu}}, \bibinfo {author} {\bibfnamefont
  {B.}~\bibnamefont {Wehinger}}, \bibinfo {author} {\bibfnamefont {P.}~\bibnamefont {Hansmann}},\ and\ \bibinfo {author} {\bibfnamefont {M.}~\bibnamefont {Hepting}},\ }\bibfield  {title} {\bibinfo {title} {Unconventional crystal structure of the high-pressure superconductor {L}a$_{3}${N}i$_{2}${O}$_{7}$},\ }\href {https://doi.org/10.1103/PhysRevLett.133.146002} {\bibfield  {journal} {\bibinfo  {journal} {Phys. Rev. Lett.}\ }\textbf {\bibinfo {volume} {133}},\ \bibinfo {pages} {146002} (\bibinfo {year} {2024})}\BibitemShut {NoStop}%
\bibitem [{\citenamefont {Wang}\ \emph {et~al.}(2024{\natexlab{b}})\citenamefont {Wang}, \citenamefont {Li}, \citenamefont {Xie}, \citenamefont {Liu}, \citenamefont {Sun}, \citenamefont {Huang}, \citenamefont {Gao}, \citenamefont {Nakagawa}, \citenamefont {Fu}, \citenamefont {Dong}, \citenamefont {Cao}, \citenamefont {Yu}, \citenamefont {Kawaguchi}, \citenamefont {Kadobayashi}, \citenamefont {Wang}, \citenamefont {Jin}, \citenamefont {kwang Mao},\ and\ \citenamefont {Liu}}]{wang2023structure}%
  \BibitemOpen
  \bibfield  {author} {\bibinfo {author} {\bibfnamefont {L.}~\bibnamefont {Wang}}, \bibinfo {author} {\bibfnamefont {Y.}~\bibnamefont {Li}}, \bibinfo {author} {\bibfnamefont {S.}~\bibnamefont {Xie}}, \bibinfo {author} {\bibfnamefont {F.}~\bibnamefont {Liu}}, \bibinfo {author} {\bibfnamefont {H.}~\bibnamefont {Sun}}, \bibinfo {author} {\bibfnamefont {C.}~\bibnamefont {Huang}}, \bibinfo {author} {\bibfnamefont {Y.}~\bibnamefont {Gao}}, \bibinfo {author} {\bibfnamefont {T.}~\bibnamefont {Nakagawa}}, \bibinfo {author} {\bibfnamefont {B.}~\bibnamefont {Fu}}, \bibinfo {author} {\bibfnamefont {B.}~\bibnamefont {Dong}}, \bibinfo {author} {\bibfnamefont {Z.}~\bibnamefont {Cao}}, \bibinfo {author} {\bibfnamefont {R.}~\bibnamefont {Yu}}, \bibinfo {author} {\bibfnamefont {S.~I.}\ \bibnamefont {Kawaguchi}}, \bibinfo {author} {\bibfnamefont {H.}~\bibnamefont {Kadobayashi}}, \bibinfo {author} {\bibfnamefont {M.}~\bibnamefont {Wang}}, \bibinfo {author} {\bibfnamefont {C.}~\bibnamefont {Jin}}, \bibinfo {author} {\bibfnamefont
  {H.}~\bibnamefont {kwang Mao}},\ and\ \bibinfo {author} {\bibfnamefont {H.}~\bibnamefont {Liu}},\ }\bibfield  {title} {\bibinfo {title} {Structure responsible for the superconducting state in {L}a$_3${N}i$_2${O}$_7$ at low temperature and high pressure conditions},\ }\href {https://doi.org/10.1021/jacs.3c13094} {\bibfield  {journal} {\bibinfo  {journal} {Journal of the American Chemical Society}\ }\textbf {\bibinfo {volume} {146}},\ \bibinfo {pages} {7506} (\bibinfo {year} {2024}{\natexlab{b}})}\BibitemShut {NoStop}%
\bibitem [{\citenamefont {Li}\ \emph {et~al.}(2025{\natexlab{a}})\citenamefont {Li}, \citenamefont {Peng}, \citenamefont {Ma}, \citenamefont {Zhang}, \citenamefont {Xing}, \citenamefont {Huang}, \citenamefont {Huang}, \citenamefont {Huo}, \citenamefont {Hu}, \citenamefont {Dong}, \citenamefont {Chen}, \citenamefont {Xie}, \citenamefont {Dong}, \citenamefont {Sun}, \citenamefont {Zeng}, \citenamefont {Mao},\ and\ \citenamefont {Wang}}]{li2024pressure}%
  \BibitemOpen
  \bibfield  {author} {\bibinfo {author} {\bibfnamefont {J.}~\bibnamefont {Li}}, \bibinfo {author} {\bibfnamefont {D.}~\bibnamefont {Peng}}, \bibinfo {author} {\bibfnamefont {P.}~\bibnamefont {Ma}}, \bibinfo {author} {\bibfnamefont {H.}~\bibnamefont {Zhang}}, \bibinfo {author} {\bibfnamefont {Z.}~\bibnamefont {Xing}}, \bibinfo {author} {\bibfnamefont {X.}~\bibnamefont {Huang}}, \bibinfo {author} {\bibfnamefont {C.}~\bibnamefont {Huang}}, \bibinfo {author} {\bibfnamefont {M.}~\bibnamefont {Huo}}, \bibinfo {author} {\bibfnamefont {D.}~\bibnamefont {Hu}}, \bibinfo {author} {\bibfnamefont {Z.}~\bibnamefont {Dong}}, \bibinfo {author} {\bibfnamefont {X.}~\bibnamefont {Chen}}, \bibinfo {author} {\bibfnamefont {T.}~\bibnamefont {Xie}}, \bibinfo {author} {\bibfnamefont {H.}~\bibnamefont {Dong}}, \bibinfo {author} {\bibfnamefont {H.}~\bibnamefont {Sun}}, \bibinfo {author} {\bibfnamefont {Q.}~\bibnamefont {Zeng}}, \bibinfo {author} {\bibfnamefont {H.-k.}\ \bibnamefont {Mao}},\ and\ \bibinfo {author} {\bibfnamefont
  {M.}~\bibnamefont {Wang}},\ }\bibfield  {title} {\bibinfo {title} {Identification of superconductivity in bilayer nickelate {L}a$_3${N}i$_2${O}$_7$ under high pressure up to 100 {G}pa},\ }\href {https://doi.org/10.1093/nsr/nwaf220} {\bibfield  {journal} {\bibinfo  {journal} {National Science Review}\ ,\ \bibinfo {pages} {nwaf220}} (\bibinfo {year} {2025}{\natexlab{a}})}\BibitemShut {NoStop}%
\bibitem [{\citenamefont {Dong}\ \emph {et~al.}(2024)\citenamefont {Dong}, \citenamefont {Huo}, \citenamefont {Li}, \citenamefont {Li}, \citenamefont {Li}, \citenamefont {Sun}, \citenamefont {Gu}, \citenamefont {Lu}, \citenamefont {Wang}, \citenamefont {Wang},\ and\ \citenamefont {Chen}}]{Dong2024vis}%
  \BibitemOpen
  \bibfield  {author} {\bibinfo {author} {\bibfnamefont {Z.}~\bibnamefont {Dong}}, \bibinfo {author} {\bibfnamefont {M.}~\bibnamefont {Huo}}, \bibinfo {author} {\bibfnamefont {J.}~\bibnamefont {Li}}, \bibinfo {author} {\bibfnamefont {J.}~\bibnamefont {Li}}, \bibinfo {author} {\bibfnamefont {P.}~\bibnamefont {Li}}, \bibinfo {author} {\bibfnamefont {H.}~\bibnamefont {Sun}}, \bibinfo {author} {\bibfnamefont {L.}~\bibnamefont {Gu}}, \bibinfo {author} {\bibfnamefont {Y.}~\bibnamefont {Lu}}, \bibinfo {author} {\bibfnamefont {M.}~\bibnamefont {Wang}}, \bibinfo {author} {\bibfnamefont {Y.}~\bibnamefont {Wang}},\ and\ \bibinfo {author} {\bibfnamefont {Z.}~\bibnamefont {Chen}},\ }\bibfield  {title} {\bibinfo {title} {Visualization of oxygen vacancies and self-doped ligand holes in {L}a$_3${N}i$_2${O}$_{7-\delta}$},\ }\href {https://doi.org/10.1038/s41586-024-07482-1} {\bibfield  {journal} {\bibinfo  {journal} {Nature}\ }\textbf {\bibinfo {volume} {630}},\ \bibinfo {pages} {847} (\bibinfo {year} {2024})}\BibitemShut
  {NoStop}%
\bibitem [{\citenamefont {Xu}\ \emph {et~al.}(2025)\citenamefont {Xu}, \citenamefont {Chen}, \citenamefont {Huo}, \citenamefont {Hu}, \citenamefont {Wang}, \citenamefont {Wu}, \citenamefont {Li}, \citenamefont {Wu}, \citenamefont {Wang}, \citenamefont {Yao}, \citenamefont {Dong},\ and\ \citenamefont {Wang}}]{PhysRevB.111.075140}%
  \BibitemOpen
  \bibfield  {author} {\bibinfo {author} {\bibfnamefont {S.}~\bibnamefont {Xu}}, \bibinfo {author} {\bibfnamefont {C.-Q.}\ \bibnamefont {Chen}}, \bibinfo {author} {\bibfnamefont {M.}~\bibnamefont {Huo}}, \bibinfo {author} {\bibfnamefont {D.}~\bibnamefont {Hu}}, \bibinfo {author} {\bibfnamefont {H.}~\bibnamefont {Wang}}, \bibinfo {author} {\bibfnamefont {Q.}~\bibnamefont {Wu}}, \bibinfo {author} {\bibfnamefont {R.}~\bibnamefont {Li}}, \bibinfo {author} {\bibfnamefont {D.}~\bibnamefont {Wu}}, \bibinfo {author} {\bibfnamefont {M.}~\bibnamefont {Wang}}, \bibinfo {author} {\bibfnamefont {D.-X.}\ \bibnamefont {Yao}}, \bibinfo {author} {\bibfnamefont {T.}~\bibnamefont {Dong}},\ and\ \bibinfo {author} {\bibfnamefont {N.}~\bibnamefont {Wang}},\ }\bibfield  {title} {\bibinfo {title} {Origin of the density wave instability in trilayer nickelate {L}a$_4${N}i$_3${O}$_{10}$ revealed by optical and ultrafast spectroscopy},\ }\href {https://doi.org/10.1103/PhysRevB.111.075140} {\bibfield  {journal} {\bibinfo  {journal}
  {Phys. Rev. B}\ }\textbf {\bibinfo {volume} {111}},\ \bibinfo {pages} {075140} (\bibinfo {year} {2025})}\BibitemShut {NoStop}%
\bibitem [{\citenamefont {Huang}\ \emph {et~al.}(2024)\citenamefont {Huang}, \citenamefont {Zhang}, \citenamefont {Li}, \citenamefont {Huo}, \citenamefont {Chen}, \citenamefont {Qiu}, \citenamefont {Ma}, \citenamefont {Huang}, \citenamefont {Sun},\ and\ \citenamefont {Wang}}]{huang2024signature}%
  \BibitemOpen
  \bibfield  {author} {\bibinfo {author} {\bibfnamefont {X.}~\bibnamefont {Huang}}, \bibinfo {author} {\bibfnamefont {H.}~\bibnamefont {Zhang}}, \bibinfo {author} {\bibfnamefont {J.}~\bibnamefont {Li}}, \bibinfo {author} {\bibfnamefont {M.}~\bibnamefont {Huo}}, \bibinfo {author} {\bibfnamefont {J.}~\bibnamefont {Chen}}, \bibinfo {author} {\bibfnamefont {Z.}~\bibnamefont {Qiu}}, \bibinfo {author} {\bibfnamefont {P.}~\bibnamefont {Ma}}, \bibinfo {author} {\bibfnamefont {C.}~\bibnamefont {Huang}}, \bibinfo {author} {\bibfnamefont {H.}~\bibnamefont {Sun}},\ and\ \bibinfo {author} {\bibfnamefont {M.}~\bibnamefont {Wang}},\ }\bibfield  {title} {\bibinfo {title} {Signature of superconductivity in pressurized trilayer-nickelate {P}r$_4${N}i$_3${O}$_{10–\delta}$},\ }\href {https://doi.org/10.1088/0256-307X/41/12/127403} {\bibfield  {journal} {\bibinfo  {journal} {Chin. Phys. Lett.}\ }\textbf {\bibinfo {volume} {41}},\ \bibinfo {pages} {127403} (\bibinfo {year} {2024})}\BibitemShut {NoStop}%
\bibitem [{\citenamefont {Du}\ \emph {et~al.}(2024)\citenamefont {Du}, \citenamefont {Li}, \citenamefont {Cao}, \citenamefont {Pei}, \citenamefont {Zhang}, \citenamefont {Zhao}, \citenamefont {Zhai}, \citenamefont {Xu}, \citenamefont {Liu}, \citenamefont {Li} \emph {et~al.}}]{du2024correlated}%
  \BibitemOpen
  \bibfield  {author} {\bibinfo {author} {\bibfnamefont {X.}~\bibnamefont {Du}}, \bibinfo {author} {\bibfnamefont {Y.}~\bibnamefont {Li}}, \bibinfo {author} {\bibfnamefont {Y.}~\bibnamefont {Cao}}, \bibinfo {author} {\bibfnamefont {C.}~\bibnamefont {Pei}}, \bibinfo {author} {\bibfnamefont {M.}~\bibnamefont {Zhang}}, \bibinfo {author} {\bibfnamefont {W.}~\bibnamefont {Zhao}}, \bibinfo {author} {\bibfnamefont {K.}~\bibnamefont {Zhai}}, \bibinfo {author} {\bibfnamefont {R.}~\bibnamefont {Xu}}, \bibinfo {author} {\bibfnamefont {Z.}~\bibnamefont {Liu}}, \bibinfo {author} {\bibfnamefont {Z.}~\bibnamefont {Li}}, \emph {et~al.},\ }\bibfield  {title} {\bibinfo {title} {Correlated electronic structure and density-wave gap in trilayer nickelate {L}a$_4${N}i$_3${O}$_{10}$},\ }\href {https://arxiv.org/abs/2405.19853} {\bibfield  {journal} {\bibinfo  {journal} {arXiv:2405.19853}\ } (\bibinfo {year} {2024})}\BibitemShut {NoStop}%
\bibitem [{\citenamefont {Cui}\ \emph {et~al.}(2024)\citenamefont {Cui}, \citenamefont {Choi}, \citenamefont {Lin}, \citenamefont {Liu}, \citenamefont {Wang}, \citenamefont {Wang}, \citenamefont {Chen}, \citenamefont {Hong}, \citenamefont {Rong}, \citenamefont {Wang}, \citenamefont {Jin}, \citenamefont {Wang}, \citenamefont {Gu}, \citenamefont {Ge}, \citenamefont {Wang}, \citenamefont {Cheng}, \citenamefont {Zhang}, \citenamefont {Si}, \citenamefont {juan Jin},\ and\ \citenamefont {Guo}}]{cui2023strain}%
  \BibitemOpen
  \bibfield  {author} {\bibinfo {author} {\bibfnamefont {T.}~\bibnamefont {Cui}}, \bibinfo {author} {\bibfnamefont {S.}~\bibnamefont {Choi}}, \bibinfo {author} {\bibfnamefont {T.}~\bibnamefont {Lin}}, \bibinfo {author} {\bibfnamefont {C.}~\bibnamefont {Liu}}, \bibinfo {author} {\bibfnamefont {G.}~\bibnamefont {Wang}}, \bibinfo {author} {\bibfnamefont {N.}~\bibnamefont {Wang}}, \bibinfo {author} {\bibfnamefont {S.}~\bibnamefont {Chen}}, \bibinfo {author} {\bibfnamefont {H.}~\bibnamefont {Hong}}, \bibinfo {author} {\bibfnamefont {D.}~\bibnamefont {Rong}}, \bibinfo {author} {\bibfnamefont {Q.}~\bibnamefont {Wang}}, \bibinfo {author} {\bibfnamefont {Q.}~\bibnamefont {Jin}}, \bibinfo {author} {\bibfnamefont {J.-O.}\ \bibnamefont {Wang}}, \bibinfo {author} {\bibfnamefont {L.}~\bibnamefont {Gu}}, \bibinfo {author} {\bibfnamefont {C.}~\bibnamefont {Ge}}, \bibinfo {author} {\bibfnamefont {C.}~\bibnamefont {Wang}}, \bibinfo {author} {\bibfnamefont {J.~G.}\ \bibnamefont {Cheng}}, \bibinfo {author} {\bibfnamefont
  {Q.}~\bibnamefont {Zhang}}, \bibinfo {author} {\bibfnamefont {L.}~\bibnamefont {Si}}, \bibinfo {author} {\bibfnamefont {K.}~\bibnamefont {juan Jin}},\ and\ \bibinfo {author} {\bibfnamefont {E.-J.}\ \bibnamefont {Guo}},\ }\bibfield  {title} {\bibinfo {title} {Strain mediated phase crossover in {R}uddlesden {P}opper nickelates},\ }\href {https://doi.org/10.1038/s43246-024-00478-4} {\bibfield  {journal} {\bibinfo  {journal} {Communications Materials}\ }\textbf {\bibinfo {volume} {5}},\ \bibinfo {pages} {32} (\bibinfo {year} {2024})}\BibitemShut {NoStop}%
\bibitem [{\citenamefont {Li}\ \emph {et~al.}(2024{\natexlab{a}})\citenamefont {Li}, \citenamefont {Wang}, \citenamefont {Pei}, \citenamefont {Zhang}, \citenamefont {Li}, \citenamefont {Guan}, \citenamefont {Amboage}, \citenamefont {Adama}, \citenamefont {Kong}, \citenamefont {Qi},\ and\ \citenamefont {Yang}}]{li2024distinguishing}%
  \BibitemOpen
  \bibfield  {author} {\bibinfo {author} {\bibfnamefont {M.}~\bibnamefont {Li}}, \bibinfo {author} {\bibfnamefont {Y.}~\bibnamefont {Wang}}, \bibinfo {author} {\bibfnamefont {C.}~\bibnamefont {Pei}}, \bibinfo {author} {\bibfnamefont {M.}~\bibnamefont {Zhang}}, \bibinfo {author} {\bibfnamefont {N.}~\bibnamefont {Li}}, \bibinfo {author} {\bibfnamefont {J.}~\bibnamefont {Guan}}, \bibinfo {author} {\bibfnamefont {M.}~\bibnamefont {Amboage}}, \bibinfo {author} {\bibfnamefont {N.-D.}\ \bibnamefont {Adama}}, \bibinfo {author} {\bibfnamefont {Q.}~\bibnamefont {Kong}}, \bibinfo {author} {\bibfnamefont {Y.}~\bibnamefont {Qi}},\ and\ \bibinfo {author} {\bibfnamefont {W.}~\bibnamefont {Yang}},\ }\bibfield  {title} {\bibinfo {title} {Distinguishing electronic band structure of single-layer and bilayer {R}uddlesden-{P}opper nickelates probed by in-situ high pressure {X}-ray absorption near-edge spectroscopy},\ }\href {https://arxiv.org/abs/2410.04230} {\bibfield  {journal} {\bibinfo  {journal} {arXiv:2410.04230}\ }
  (\bibinfo {year} {2024}{\natexlab{a}})}\BibitemShut {NoStop}%
\bibitem [{\citenamefont {Zhou}\ \emph {et~al.}(2024)\citenamefont {Zhou}, \citenamefont {He}, \citenamefont {Ni}, \citenamefont {Huo}, \citenamefont {Hu}, \citenamefont {Zhu}, \citenamefont {Zhang}, \citenamefont {Jiang}, \citenamefont {Zhang}, \citenamefont {Su}, \citenamefont {Jiang}, \citenamefont {Yan}, \citenamefont {Wang}, \citenamefont {Shen}, \citenamefont {Liu}, \citenamefont {Zhao}, \citenamefont {Wang}, \citenamefont {Du},\ and\ \citenamefont {Feng}}]{zhou2024revealing}%
  \BibitemOpen
  \bibfield  {author} {\bibinfo {author} {\bibfnamefont {X.}~\bibnamefont {Zhou}}, \bibinfo {author} {\bibfnamefont {W.}~\bibnamefont {He}}, \bibinfo {author} {\bibfnamefont {K.}~\bibnamefont {Ni}}, \bibinfo {author} {\bibfnamefont {M.}~\bibnamefont {Huo}}, \bibinfo {author} {\bibfnamefont {D.}~\bibnamefont {Hu}}, \bibinfo {author} {\bibfnamefont {Y.}~\bibnamefont {Zhu}}, \bibinfo {author} {\bibfnamefont {E.}~\bibnamefont {Zhang}}, \bibinfo {author} {\bibfnamefont {Z.}~\bibnamefont {Jiang}}, \bibinfo {author} {\bibfnamefont {S.}~\bibnamefont {Zhang}}, \bibinfo {author} {\bibfnamefont {S.}~\bibnamefont {Su}}, \bibinfo {author} {\bibfnamefont {J.}~\bibnamefont {Jiang}}, \bibinfo {author} {\bibfnamefont {Y.}~\bibnamefont {Yan}}, \bibinfo {author} {\bibfnamefont {Y.}~\bibnamefont {Wang}}, \bibinfo {author} {\bibfnamefont {D.}~\bibnamefont {Shen}}, \bibinfo {author} {\bibfnamefont {X.}~\bibnamefont {Liu}}, \bibinfo {author} {\bibfnamefont {J.}~\bibnamefont {Zhao}}, \bibinfo {author} {\bibfnamefont
  {M.}~\bibnamefont {Wang}}, \bibinfo {author} {\bibfnamefont {Z.}~\bibnamefont {Du}},\ and\ \bibinfo {author} {\bibfnamefont {D.}~\bibnamefont {Feng}},\ }\bibfield  {title} {\bibinfo {title} {Revealing nanoscale structural phase separation in {L}a$_{3}${N}i$_{2}${O}$_{7-\delta}$ single crystal via scanning near-field optical microscopy},\ }\href {https://arxiv.org/abs/2410.06602} {\bibfield  {journal} {\bibinfo  {journal} {arXiv:2410.06602}\ } (\bibinfo {year} {2024})}\BibitemShut {NoStop}%
\bibitem [{\citenamefont {Fan}\ \emph {et~al.}(2024)\citenamefont {Fan}, \citenamefont {Luo}, \citenamefont {Huo}, \citenamefont {Wang}, \citenamefont {Li}, \citenamefont {Yang}, \citenamefont {Wang}, \citenamefont {Yao},\ and\ \citenamefont {Wen}}]{fan2024tunn}%
  \BibitemOpen
  \bibfield  {author} {\bibinfo {author} {\bibfnamefont {S.}~\bibnamefont {Fan}}, \bibinfo {author} {\bibfnamefont {Z.}~\bibnamefont {Luo}}, \bibinfo {author} {\bibfnamefont {M.}~\bibnamefont {Huo}}, \bibinfo {author} {\bibfnamefont {Z.}~\bibnamefont {Wang}}, \bibinfo {author} {\bibfnamefont {H.}~\bibnamefont {Li}}, \bibinfo {author} {\bibfnamefont {H.}~\bibnamefont {Yang}}, \bibinfo {author} {\bibfnamefont {M.}~\bibnamefont {Wang}}, \bibinfo {author} {\bibfnamefont {D.-X.}\ \bibnamefont {Yao}},\ and\ \bibinfo {author} {\bibfnamefont {H.-H.}\ \bibnamefont {Wen}},\ }\bibfield  {title} {\bibinfo {title} {Tunneling spectra with gaplike features observed in nickelate {L}a$_{3}${N}i$_{2}${O}$_{7}$ at ambient pressure},\ }\href {https://doi.org/10.1103/PhysRevB.110.134520} {\bibfield  {journal} {\bibinfo  {journal} {Phys. Rev. B}\ }\textbf {\bibinfo {volume} {110}},\ \bibinfo {pages} {134520} (\bibinfo {year} {2024})}\BibitemShut {NoStop}%
\bibitem [{\citenamefont {Li}\ \emph {et~al.}(2024{\natexlab{b}})\citenamefont {Li}, \citenamefont {Chen}, \citenamefont {Huang}, \citenamefont {Han}, \citenamefont {Huo}, \citenamefont {Huang}, \citenamefont {Ma}, \citenamefont {Qiu}, \citenamefont {Chen}, \citenamefont {Hu}, \citenamefont {Chen}, \citenamefont {Xie}, \citenamefont {Shen}, \citenamefont {Sun}, \citenamefont {Yao},\ and\ \citenamefont {Wang}}]{li2024la3}%
  \BibitemOpen
  \bibfield  {author} {\bibinfo {author} {\bibfnamefont {J.}~\bibnamefont {Li}}, \bibinfo {author} {\bibfnamefont {C.-Q.}\ \bibnamefont {Chen}}, \bibinfo {author} {\bibfnamefont {C.}~\bibnamefont {Huang}}, \bibinfo {author} {\bibfnamefont {Y.}~\bibnamefont {Han}}, \bibinfo {author} {\bibfnamefont {M.}~\bibnamefont {Huo}}, \bibinfo {author} {\bibfnamefont {X.}~\bibnamefont {Huang}}, \bibinfo {author} {\bibfnamefont {P.}~\bibnamefont {Ma}}, \bibinfo {author} {\bibfnamefont {Z.}~\bibnamefont {Qiu}}, \bibinfo {author} {\bibfnamefont {J.}~\bibnamefont {Chen}}, \bibinfo {author} {\bibfnamefont {X.}~\bibnamefont {Hu}}, \bibinfo {author} {\bibfnamefont {L.}~\bibnamefont {Chen}}, \bibinfo {author} {\bibfnamefont {T.}~\bibnamefont {Xie}}, \bibinfo {author} {\bibfnamefont {B.}~\bibnamefont {Shen}}, \bibinfo {author} {\bibfnamefont {H.}~\bibnamefont {Sun}}, \bibinfo {author} {\bibfnamefont {D.}~\bibnamefont {Yao}},\ and\ \bibinfo {author} {\bibfnamefont {M.}~\bibnamefont {Wang}},\ }\bibfield  {title} {\bibinfo {title}
  {Structural transition, electric transport, and electronic structures in the compressed trilayer nickelate {L}a$_{4}${N}i$_{3}${O}$_{10}$},\ }\href {https://www.sciengine.com/SCPMA/doi/10.1007/s11433-023-2329-x} {\bibfield  {journal} {\bibinfo  {journal} {Sci. China Phys. Mech. Astron.}\ }\textbf {\bibinfo {volume} {67}},\ \bibinfo {pages} {117403} (\bibinfo {year} {2024}{\natexlab{b}})}\BibitemShut {NoStop}%
\bibitem [{\citenamefont {Zhu}\ \emph {et~al.}(2024)\citenamefont {Zhu}, \citenamefont {Peng}, \citenamefont {Zhang}, \citenamefont {Pan}, \citenamefont {Chen}, \citenamefont {Chen}, \citenamefont {Ren}, \citenamefont {Liu}, \citenamefont {Hao}, \citenamefont {Li} \emph {et~al.}}]{zhu2024superconductivity}%
  \BibitemOpen
  \bibfield  {author} {\bibinfo {author} {\bibfnamefont {Y.}~\bibnamefont {Zhu}}, \bibinfo {author} {\bibfnamefont {D.}~\bibnamefont {Peng}}, \bibinfo {author} {\bibfnamefont {E.}~\bibnamefont {Zhang}}, \bibinfo {author} {\bibfnamefont {B.}~\bibnamefont {Pan}}, \bibinfo {author} {\bibfnamefont {X.}~\bibnamefont {Chen}}, \bibinfo {author} {\bibfnamefont {L.}~\bibnamefont {Chen}}, \bibinfo {author} {\bibfnamefont {H.}~\bibnamefont {Ren}}, \bibinfo {author} {\bibfnamefont {F.}~\bibnamefont {Liu}}, \bibinfo {author} {\bibfnamefont {Y.}~\bibnamefont {Hao}}, \bibinfo {author} {\bibfnamefont {N.}~\bibnamefont {Li}}, \emph {et~al.},\ }\bibfield  {title} {\bibinfo {title} {Superconductivity in pressurized trilayer {L}a$_4${N}i$_3${O}$_{10- \delta}$ single crystals},\ }\href {https://www.nature.com/articles/s41586-024-07553-3} {\bibfield  {journal} {\bibinfo  {journal} {Nature}\ }\textbf {\bibinfo {volume} {631}},\ \bibinfo {pages} {531} (\bibinfo {year} {2024})}\BibitemShut {NoStop}%
\bibitem [{\citenamefont {Li}\ \emph {et~al.}(2024{\natexlab{c}})\citenamefont {Li}, \citenamefont {Zhang}, \citenamefont {Xiang}, \citenamefont {Zhang}, \citenamefont {Zhu},\ and\ \citenamefont {Wen}}]{Li2023trilayer}%
  \BibitemOpen
  \bibfield  {author} {\bibinfo {author} {\bibfnamefont {Q.}~\bibnamefont {Li}}, \bibinfo {author} {\bibfnamefont {Y.-J.}\ \bibnamefont {Zhang}}, \bibinfo {author} {\bibfnamefont {Z.-N.}\ \bibnamefont {Xiang}}, \bibinfo {author} {\bibfnamefont {Y.}~\bibnamefont {Zhang}}, \bibinfo {author} {\bibfnamefont {X.}~\bibnamefont {Zhu}},\ and\ \bibinfo {author} {\bibfnamefont {H.-H.}\ \bibnamefont {Wen}},\ }\bibfield  {title} {\bibinfo {title} {Signature of superconductivity in pressurized {L}a$_4${N}i$_3${O}$_{10}$},\ }\href {https://doi.org/10.1088/0256-307X/41/1/017401} {\bibfield  {journal} {\bibinfo  {journal} {Chin. Phys. Lett.}\ }\textbf {\bibinfo {volume} {41}},\ \bibinfo {pages} {017401} (\bibinfo {year} {2024}{\natexlab{c}})}\BibitemShut {NoStop}%
\bibitem [{\citenamefont {Zhang}\ \emph {et~al.}(2025)\citenamefont {Zhang}, \citenamefont {Pei}, \citenamefont {Peng}, \citenamefont {Du}, \citenamefont {Hu}, \citenamefont {Cao}, \citenamefont {Wang}, \citenamefont {Wu}, \citenamefont {Li}, \citenamefont {Liu}, \citenamefont {Wen}, \citenamefont {Song}, \citenamefont {Zhao}, \citenamefont {Li}, \citenamefont {Cao}, \citenamefont {Zhu}, \citenamefont {Zhang}, \citenamefont {Yu}, \citenamefont {Cheng}, \citenamefont {Zhang}, \citenamefont {Li}, \citenamefont {Zhao}, \citenamefont {Chen}, \citenamefont {Jin}, \citenamefont {Guo}, \citenamefont {Wu}, \citenamefont {Yang}, \citenamefont {Zeng}, \citenamefont {Yan}, \citenamefont {Yang},\ and\ \citenamefont {Qi}}]{zhang2023superconductivity}%
  \BibitemOpen
  \bibfield  {author} {\bibinfo {author} {\bibfnamefont {M.}~\bibnamefont {Zhang}}, \bibinfo {author} {\bibfnamefont {C.}~\bibnamefont {Pei}}, \bibinfo {author} {\bibfnamefont {D.}~\bibnamefont {Peng}}, \bibinfo {author} {\bibfnamefont {X.}~\bibnamefont {Du}}, \bibinfo {author} {\bibfnamefont {W.}~\bibnamefont {Hu}}, \bibinfo {author} {\bibfnamefont {Y.}~\bibnamefont {Cao}}, \bibinfo {author} {\bibfnamefont {Q.}~\bibnamefont {Wang}}, \bibinfo {author} {\bibfnamefont {J.}~\bibnamefont {Wu}}, \bibinfo {author} {\bibfnamefont {Y.}~\bibnamefont {Li}}, \bibinfo {author} {\bibfnamefont {H.}~\bibnamefont {Liu}}, \bibinfo {author} {\bibfnamefont {C.}~\bibnamefont {Wen}}, \bibinfo {author} {\bibfnamefont {J.}~\bibnamefont {Song}}, \bibinfo {author} {\bibfnamefont {Y.}~\bibnamefont {Zhao}}, \bibinfo {author} {\bibfnamefont {C.}~\bibnamefont {Li}}, \bibinfo {author} {\bibfnamefont {W.}~\bibnamefont {Cao}}, \bibinfo {author} {\bibfnamefont {S.}~\bibnamefont {Zhu}}, \bibinfo {author} {\bibfnamefont {Q.}~\bibnamefont
  {Zhang}}, \bibinfo {author} {\bibfnamefont {N.}~\bibnamefont {Yu}}, \bibinfo {author} {\bibfnamefont {P.}~\bibnamefont {Cheng}}, \bibinfo {author} {\bibfnamefont {L.}~\bibnamefont {Zhang}}, \bibinfo {author} {\bibfnamefont {Z.}~\bibnamefont {Li}}, \bibinfo {author} {\bibfnamefont {J.}~\bibnamefont {Zhao}}, \bibinfo {author} {\bibfnamefont {Y.}~\bibnamefont {Chen}}, \bibinfo {author} {\bibfnamefont {C.}~\bibnamefont {Jin}}, \bibinfo {author} {\bibfnamefont {H.}~\bibnamefont {Guo}}, \bibinfo {author} {\bibfnamefont {C.}~\bibnamefont {Wu}}, \bibinfo {author} {\bibfnamefont {F.}~\bibnamefont {Yang}}, \bibinfo {author} {\bibfnamefont {Q.}~\bibnamefont {Zeng}}, \bibinfo {author} {\bibfnamefont {S.}~\bibnamefont {Yan}}, \bibinfo {author} {\bibfnamefont {L.}~\bibnamefont {Yang}},\ and\ \bibinfo {author} {\bibfnamefont {Y.}~\bibnamefont {Qi}},\ }\bibfield  {title} {\bibinfo {title} {Superconductivity in trilayer nickelate {L}a$_{4}${N}i$_{3}${O}$_{10}$ under pressure},\ }\href
  {https://doi.org/10.1103/PhysRevX.15.021005} {\bibfield  {journal} {\bibinfo  {journal} {Phys. Rev. X}\ }\textbf {\bibinfo {volume} {15}},\ \bibinfo {pages} {021005} (\bibinfo {year} {2025})}\BibitemShut {NoStop}%
\bibitem [{\citenamefont {Wang}\ \emph {et~al.}(2024{\natexlab{c}})\citenamefont {Wang}, \citenamefont {Wang}, \citenamefont {Shen}, \citenamefont {Hou}, \citenamefont {Luo}, \citenamefont {Ma}, \citenamefont {Yang}, \citenamefont {Shi}, \citenamefont {Dou}, \citenamefont {Feng}, \citenamefont {Yang}, \citenamefont {Shi}, \citenamefont {Ren}, \citenamefont {Ma}, \citenamefont {Yang}, \citenamefont {Liu}, \citenamefont {Liu}, \citenamefont {Zhang}, \citenamefont {Dong}, \citenamefont {Wang}, \citenamefont {Jiang}, \citenamefont {Hu}, \citenamefont {Calder}, \citenamefont {Yan}, \citenamefont {Sun}, \citenamefont {Wang}, \citenamefont {Zhou}, \citenamefont {Uwatoko},\ and\ \citenamefont {Cheng}}]{wang2024bulk}%
  \BibitemOpen
  \bibfield  {author} {\bibinfo {author} {\bibfnamefont {N.}~\bibnamefont {Wang}}, \bibinfo {author} {\bibfnamefont {G.}~\bibnamefont {Wang}}, \bibinfo {author} {\bibfnamefont {X.}~\bibnamefont {Shen}}, \bibinfo {author} {\bibfnamefont {J.}~\bibnamefont {Hou}}, \bibinfo {author} {\bibfnamefont {J.}~\bibnamefont {Luo}}, \bibinfo {author} {\bibfnamefont {X.}~\bibnamefont {Ma}}, \bibinfo {author} {\bibfnamefont {H.}~\bibnamefont {Yang}}, \bibinfo {author} {\bibfnamefont {L.}~\bibnamefont {Shi}}, \bibinfo {author} {\bibfnamefont {J.}~\bibnamefont {Dou}}, \bibinfo {author} {\bibfnamefont {J.}~\bibnamefont {Feng}}, \bibinfo {author} {\bibfnamefont {J.}~\bibnamefont {Yang}}, \bibinfo {author} {\bibfnamefont {Y.}~\bibnamefont {Shi}}, \bibinfo {author} {\bibfnamefont {Z.}~\bibnamefont {Ren}}, \bibinfo {author} {\bibfnamefont {H.}~\bibnamefont {Ma}}, \bibinfo {author} {\bibfnamefont {P.}~\bibnamefont {Yang}}, \bibinfo {author} {\bibfnamefont {Z.}~\bibnamefont {Liu}}, \bibinfo {author} {\bibfnamefont {Y.}~\bibnamefont
  {Liu}}, \bibinfo {author} {\bibfnamefont {H.}~\bibnamefont {Zhang}}, \bibinfo {author} {\bibfnamefont {X.}~\bibnamefont {Dong}}, \bibinfo {author} {\bibfnamefont {Y.}~\bibnamefont {Wang}}, \bibinfo {author} {\bibfnamefont {K.}~\bibnamefont {Jiang}}, \bibinfo {author} {\bibfnamefont {J.}~\bibnamefont {Hu}}, \bibinfo {author} {\bibfnamefont {S.}~\bibnamefont {Calder}}, \bibinfo {author} {\bibfnamefont {J.}~\bibnamefont {Yan}}, \bibinfo {author} {\bibfnamefont {J.}~\bibnamefont {Sun}}, \bibinfo {author} {\bibfnamefont {B.}~\bibnamefont {Wang}}, \bibinfo {author} {\bibfnamefont {R.}~\bibnamefont {Zhou}}, \bibinfo {author} {\bibfnamefont {Y.}~\bibnamefont {Uwatoko}},\ and\ \bibinfo {author} {\bibfnamefont {J.}~\bibnamefont {Cheng}},\ }\bibfield  {title} {\bibinfo {title} {Bulk high-temperature superconductivity in the high-pressure tetragonal phase of bilayer {L}a$_2${P}r{N}i$_2${O}$_7$},\ }\href {https://doi.org/10.1038/s41586-024-07996-8} {\bibfield  {journal} {\bibinfo  {journal} {Nature}\ }\textbf {\bibinfo
  {volume} {634}},\ \bibinfo {pages} {579} (\bibinfo {year} {2024}{\natexlab{c}})}\BibitemShut {NoStop}%
\bibitem [{\citenamefont {Ko}\ \emph {et~al.}(2025)\citenamefont {Ko}, \citenamefont {Yu}, \citenamefont {Liu}, \citenamefont {Bhatt}, \citenamefont {Li}, \citenamefont {Thampy}, \citenamefont {Kuo}, \citenamefont {Wang}, \citenamefont {Lee}, \citenamefont {Lee}, \citenamefont {Lee}, \citenamefont {Goodge}, \citenamefont {Muller},\ and\ \citenamefont {Hwang}}]{ko2024signatures}%
  \BibitemOpen
  \bibfield  {author} {\bibinfo {author} {\bibfnamefont {E.~K.}\ \bibnamefont {Ko}}, \bibinfo {author} {\bibfnamefont {Y.}~\bibnamefont {Yu}}, \bibinfo {author} {\bibfnamefont {Y.}~\bibnamefont {Liu}}, \bibinfo {author} {\bibfnamefont {L.}~\bibnamefont {Bhatt}}, \bibinfo {author} {\bibfnamefont {J.}~\bibnamefont {Li}}, \bibinfo {author} {\bibfnamefont {V.}~\bibnamefont {Thampy}}, \bibinfo {author} {\bibfnamefont {C.-T.}\ \bibnamefont {Kuo}}, \bibinfo {author} {\bibfnamefont {B.~Y.}\ \bibnamefont {Wang}}, \bibinfo {author} {\bibfnamefont {Y.}~\bibnamefont {Lee}}, \bibinfo {author} {\bibfnamefont {K.}~\bibnamefont {Lee}}, \bibinfo {author} {\bibfnamefont {J.-S.}\ \bibnamefont {Lee}}, \bibinfo {author} {\bibfnamefont {B.~H.}\ \bibnamefont {Goodge}}, \bibinfo {author} {\bibfnamefont {D.~A.}\ \bibnamefont {Muller}},\ and\ \bibinfo {author} {\bibfnamefont {H.~Y.}\ \bibnamefont {Hwang}},\ }\bibfield  {title} {\bibinfo {title} {Signatures of ambient pressure superconductivity in thin film {L}a$_3${N}i$_2${O}$_7$},\
  }\href {https://doi.org/10.1038/s41586-024-08525-3} {\bibfield  {journal} {\bibinfo  {journal} {Nature}\ }\textbf {\bibinfo {volume} {638}},\ \bibinfo {pages} {935} (\bibinfo {year} {2025})}\BibitemShut {NoStop}%
\bibitem [{\citenamefont {Liu}\ \emph {et~al.}(2025{\natexlab{a}})\citenamefont {Liu}, \citenamefont {Ko}, \citenamefont {Tarn}, \citenamefont {Bhatt}, \citenamefont {Li}, \citenamefont {Thampy}, \citenamefont {Goodge}, \citenamefont {Muller}, \citenamefont {Raghu}, \citenamefont {Yu},\ and\ \citenamefont {Hwang}}]{liu2025}%
  \BibitemOpen
  \bibfield  {author} {\bibinfo {author} {\bibfnamefont {Y.}~\bibnamefont {Liu}}, \bibinfo {author} {\bibfnamefont {E.~K.}\ \bibnamefont {Ko}}, \bibinfo {author} {\bibfnamefont {Y.}~\bibnamefont {Tarn}}, \bibinfo {author} {\bibfnamefont {L.}~\bibnamefont {Bhatt}}, \bibinfo {author} {\bibfnamefont {J.}~\bibnamefont {Li}}, \bibinfo {author} {\bibfnamefont {V.}~\bibnamefont {Thampy}}, \bibinfo {author} {\bibfnamefont {B.~H.}\ \bibnamefont {Goodge}}, \bibinfo {author} {\bibfnamefont {D.~A.}\ \bibnamefont {Muller}}, \bibinfo {author} {\bibfnamefont {S.}~\bibnamefont {Raghu}}, \bibinfo {author} {\bibfnamefont {Y.}~\bibnamefont {Yu}},\ and\ \bibinfo {author} {\bibfnamefont {H.~Y.}\ \bibnamefont {Hwang}},\ }\bibfield  {title} {\bibinfo {title} {Superconductivity and normal-state transport in compressively strained {L}a$_2${P}r{N}i$_2${O}$_7$ thin films},\ }\href {https://doi.org/10.1038/s41563-025-02258-y} {\bibfield  {journal} {\bibinfo  {journal} {Nature Materials}\ }\textbf {\bibinfo {volume} {24}},\ \bibinfo
  {pages} {1221} (\bibinfo {year} {2025}{\natexlab{a}})}\BibitemShut {NoStop}%
\bibitem [{\citenamefont {Zhou}\ \emph {et~al.}(2025{\natexlab{b}})\citenamefont {Zhou}, \citenamefont {Lv}, \citenamefont {Wang}, \citenamefont {Nie}, \citenamefont {Chen}, \citenamefont {Li}, \citenamefont {Huang}, \citenamefont {Chen}, \citenamefont {Sun}, \citenamefont {Xue},\ and\ \citenamefont {Chen}}]{Zhou2025}%
  \BibitemOpen
  \bibfield  {author} {\bibinfo {author} {\bibfnamefont {G.}~\bibnamefont {Zhou}}, \bibinfo {author} {\bibfnamefont {W.}~\bibnamefont {Lv}}, \bibinfo {author} {\bibfnamefont {H.}~\bibnamefont {Wang}}, \bibinfo {author} {\bibfnamefont {Z.}~\bibnamefont {Nie}}, \bibinfo {author} {\bibfnamefont {Y.}~\bibnamefont {Chen}}, \bibinfo {author} {\bibfnamefont {Y.}~\bibnamefont {Li}}, \bibinfo {author} {\bibfnamefont {H.}~\bibnamefont {Huang}}, \bibinfo {author} {\bibfnamefont {W.-Q.}\ \bibnamefont {Chen}}, \bibinfo {author} {\bibfnamefont {Y.-J.}\ \bibnamefont {Sun}}, \bibinfo {author} {\bibfnamefont {Q.-K.}\ \bibnamefont {Xue}},\ and\ \bibinfo {author} {\bibfnamefont {Z.}~\bibnamefont {Chen}},\ }\bibfield  {title} {\bibinfo {title} {Ambient-pressure superconductivity onset above 40 {K} in ({L}a,{P}r)$_3${N}i$_2${O}$_7$ films},\ }\href {https://doi.org/10.1038/s41586-025-08755-z} {\bibfield  {journal} {\bibinfo  {journal} {Nature}\ }\textbf {\bibinfo {volume} {640}},\ \bibinfo {pages} {641} (\bibinfo {year}
  {2025}{\natexlab{b}})}\BibitemShut {NoStop}%
\bibitem [{\citenamefont {Yue}\ \emph {et~al.}(2025)\citenamefont {Yue}, \citenamefont {Miao}, \citenamefont {Huang}, \citenamefont {Hua}, \citenamefont {Li}, \citenamefont {Li}, \citenamefont {Zhou}, \citenamefont {Lv}, \citenamefont {Yang}, \citenamefont {Yang}, \citenamefont {Sun}, \citenamefont {Sun}, \citenamefont {Lin}, \citenamefont {Xue}, \citenamefont {Chen},\ and\ \citenamefont {Chen}}]{10.1093/nsr/nwaf253}%
  \BibitemOpen
  \bibfield  {author} {\bibinfo {author} {\bibfnamefont {C.}~\bibnamefont {Yue}}, \bibinfo {author} {\bibfnamefont {J.-J.}\ \bibnamefont {Miao}}, \bibinfo {author} {\bibfnamefont {H.}~\bibnamefont {Huang}}, \bibinfo {author} {\bibfnamefont {Y.}~\bibnamefont {Hua}}, \bibinfo {author} {\bibfnamefont {P.}~\bibnamefont {Li}}, \bibinfo {author} {\bibfnamefont {Y.}~\bibnamefont {Li}}, \bibinfo {author} {\bibfnamefont {G.}~\bibnamefont {Zhou}}, \bibinfo {author} {\bibfnamefont {W.}~\bibnamefont {Lv}}, \bibinfo {author} {\bibfnamefont {Q.}~\bibnamefont {Yang}}, \bibinfo {author} {\bibfnamefont {F.}~\bibnamefont {Yang}}, \bibinfo {author} {\bibfnamefont {H.}~\bibnamefont {Sun}}, \bibinfo {author} {\bibfnamefont {Y.-J.}\ \bibnamefont {Sun}}, \bibinfo {author} {\bibfnamefont {J.}~\bibnamefont {Lin}}, \bibinfo {author} {\bibfnamefont {Q.-K.}\ \bibnamefont {Xue}}, \bibinfo {author} {\bibfnamefont {Z.}~\bibnamefont {Chen}},\ and\ \bibinfo {author} {\bibfnamefont {W.-Q.}\ \bibnamefont {Chen}},\ }\bibfield  {title} {\bibinfo
  {title} {Correlated electronic structures and unconventional superconductivity in bilayer nickelate heterostructures},\ }\href {https://doi.org/10.1093/nsr/nwaf253} {\bibfield  {journal} {\bibinfo  {journal} {National Science Review}\ ,\ \bibinfo {pages} {nwaf253}} (\bibinfo {year} {2025})}\BibitemShut {NoStop}%
\bibitem [{\citenamefont {Bhatt}\ \emph {et~al.}(2025)\citenamefont {Bhatt}, \citenamefont {Jiang}, \citenamefont {Ko}, \citenamefont {Schnitzer}, \citenamefont {Pan}, \citenamefont {Segedin}, \citenamefont {Liu}, \citenamefont {Yu}, \citenamefont {Zhao}, \citenamefont {Morales}, \citenamefont {Brooks}, \citenamefont {Botana}, \citenamefont {Hwang}, \citenamefont {Mundy}, \citenamefont {Muller},\ and\ \citenamefont {Goodge}}]{bhatt2025}%
  \BibitemOpen
  \bibfield  {author} {\bibinfo {author} {\bibfnamefont {L.}~\bibnamefont {Bhatt}}, \bibinfo {author} {\bibfnamefont {A.~Y.}\ \bibnamefont {Jiang}}, \bibinfo {author} {\bibfnamefont {E.~K.}\ \bibnamefont {Ko}}, \bibinfo {author} {\bibfnamefont {N.}~\bibnamefont {Schnitzer}}, \bibinfo {author} {\bibfnamefont {G.~A.}\ \bibnamefont {Pan}}, \bibinfo {author} {\bibfnamefont {D.~F.}\ \bibnamefont {Segedin}}, \bibinfo {author} {\bibfnamefont {Y.}~\bibnamefont {Liu}}, \bibinfo {author} {\bibfnamefont {Y.}~\bibnamefont {Yu}}, \bibinfo {author} {\bibfnamefont {Y.-F.}\ \bibnamefont {Zhao}}, \bibinfo {author} {\bibfnamefont {E.~A.}\ \bibnamefont {Morales}}, \bibinfo {author} {\bibfnamefont {C.~M.}\ \bibnamefont {Brooks}}, \bibinfo {author} {\bibfnamefont {A.~S.}\ \bibnamefont {Botana}}, \bibinfo {author} {\bibfnamefont {H.~Y.}\ \bibnamefont {Hwang}}, \bibinfo {author} {\bibfnamefont {J.~A.}\ \bibnamefont {Mundy}}, \bibinfo {author} {\bibfnamefont {D.~A.}\ \bibnamefont {Muller}},\ and\ \bibinfo {author} {\bibfnamefont
  {B.~H.}\ \bibnamefont {Goodge}},\ }\bibfield  {title} {\bibinfo {title} {Resolving structural origins for superconductivity in strain-engineered {L}a$_3${N}i$_2${O}$_7$ thin films},\ }\href {https://arxiv.org/abs/2501.08204} {\bibfield  {journal} {\bibinfo  {journal} {arXiv:2501.08204}\ } (\bibinfo {year} {2025})}\BibitemShut {NoStop}%
\bibitem [{\citenamefont {Luo}\ \emph {et~al.}(2023)\citenamefont {Luo}, \citenamefont {Hu}, \citenamefont {Wang}, \citenamefont {W\'u},\ and\ \citenamefont {Yao}}]{YaoDX2023}%
  \BibitemOpen
  \bibfield  {author} {\bibinfo {author} {\bibfnamefont {Z.}~\bibnamefont {Luo}}, \bibinfo {author} {\bibfnamefont {X.}~\bibnamefont {Hu}}, \bibinfo {author} {\bibfnamefont {M.}~\bibnamefont {Wang}}, \bibinfo {author} {\bibfnamefont {W.}~\bibnamefont {W\'u}},\ and\ \bibinfo {author} {\bibfnamefont {D.-X.}\ \bibnamefont {Yao}},\ }\bibfield  {title} {\bibinfo {title} {Bilayer two-orbital model of {L}a$_3${N}i$_2${O}$_7$ under pressure},\ }\href {https://doi.org/10.1103/PhysRevLett.131.126001} {\bibfield  {journal} {\bibinfo  {journal} {Phys. Rev. Lett.}\ }\textbf {\bibinfo {volume} {131}},\ \bibinfo {pages} {126001} (\bibinfo {year} {2023})}\BibitemShut {NoStop}%
\bibitem [{\citenamefont {Zhang}\ \emph {et~al.}(2023{\natexlab{a}})\citenamefont {Zhang}, \citenamefont {Lin}, \citenamefont {Moreo},\ and\ \citenamefont {Dagotto}}]{Dagotto2023}%
  \BibitemOpen
  \bibfield  {author} {\bibinfo {author} {\bibfnamefont {Y.}~\bibnamefont {Zhang}}, \bibinfo {author} {\bibfnamefont {L.-F.}\ \bibnamefont {Lin}}, \bibinfo {author} {\bibfnamefont {A.}~\bibnamefont {Moreo}},\ and\ \bibinfo {author} {\bibfnamefont {E.}~\bibnamefont {Dagotto}},\ }\bibfield  {title} {\bibinfo {title} {Electronic structure, dimer physics, orbital-selective behavior, and magnetic tendencies in the bilayer nickelate superconductor {L}a$_3${N}i$_2${O}$_7$ under pressure},\ }\href {https://doi.org/10.1103/PhysRevB.108.L180510} {\bibfield  {journal} {\bibinfo  {journal} {Phys. Rev. B}\ }\textbf {\bibinfo {volume} {108}},\ \bibinfo {pages} {L180510} (\bibinfo {year} {2023}{\natexlab{a}})}\BibitemShut {NoStop}%
\bibitem [{\citenamefont {Rhodes}\ and\ \citenamefont {Wahl}(2024)}]{rhodes2023structural}%
  \BibitemOpen
  \bibfield  {author} {\bibinfo {author} {\bibfnamefont {L.~C.}\ \bibnamefont {Rhodes}}\ and\ \bibinfo {author} {\bibfnamefont {P.}~\bibnamefont {Wahl}},\ }\bibfield  {title} {\bibinfo {title} {Structural routes to stabilize superconducting {L}a$_3${N}i$_2${O}$_7$ at ambient pressure},\ }\href {https://doi.org/10.1103/PhysRevMaterials.8.044801} {\bibfield  {journal} {\bibinfo  {journal} {Phys. Rev. Mater.}\ }\textbf {\bibinfo {volume} {8}},\ \bibinfo {pages} {044801} (\bibinfo {year} {2024})}\BibitemShut {NoStop}%
\bibitem [{\citenamefont {Ouyang}\ \emph {et~al.}(2024{\natexlab{a}})\citenamefont {Ouyang}, \citenamefont {Gao},\ and\ \citenamefont {Lu}}]{Ouyang2024absence}%
  \BibitemOpen
  \bibfield  {author} {\bibinfo {author} {\bibfnamefont {Z.}~\bibnamefont {Ouyang}}, \bibinfo {author} {\bibfnamefont {M.}~\bibnamefont {Gao}},\ and\ \bibinfo {author} {\bibfnamefont {Z.-Y.}\ \bibnamefont {Lu}},\ }\bibfield  {title} {\bibinfo {title} {Absence of electron-phonon coupling superconductivity in the bilayer phase of {L}a$_3${N}i$_2${O}$_7$ under pressure},\ }\href {https://doi.org/10.1038/s41535-024-00689-5} {\bibfield  {journal} {\bibinfo  {journal} {npj Quantum Materials}\ }\textbf {\bibinfo {volume} {9}},\ \bibinfo {pages} {80} (\bibinfo {year} {2024}{\natexlab{a}})}\BibitemShut {NoStop}%
\bibitem [{\citenamefont {Yi}\ \emph {et~al.}(2024)\citenamefont {Yi}, \citenamefont {Meng}, \citenamefont {Li}, \citenamefont {Liao}, \citenamefont {Li}, \citenamefont {You}, \citenamefont {Gu},\ and\ \citenamefont {Su}}]{Yi2024nature}%
  \BibitemOpen
  \bibfield  {author} {\bibinfo {author} {\bibfnamefont {X.-W.}\ \bibnamefont {Yi}}, \bibinfo {author} {\bibfnamefont {Y.}~\bibnamefont {Meng}}, \bibinfo {author} {\bibfnamefont {J.-W.}\ \bibnamefont {Li}}, \bibinfo {author} {\bibfnamefont {Z.-W.}\ \bibnamefont {Liao}}, \bibinfo {author} {\bibfnamefont {W.}~\bibnamefont {Li}}, \bibinfo {author} {\bibfnamefont {J.-Y.}\ \bibnamefont {You}}, \bibinfo {author} {\bibfnamefont {B.}~\bibnamefont {Gu}},\ and\ \bibinfo {author} {\bibfnamefont {G.}~\bibnamefont {Su}},\ }\bibfield  {title} {\bibinfo {title} {Nature of charge density waves and metal-insulator transition in pressurized {L}a$_3${N}i$_2${O}$_7$},\ }\href {https://doi.org/10.1103/PhysRevB.110.L140508} {\bibfield  {journal} {\bibinfo  {journal} {Phys. Rev. B}\ }\textbf {\bibinfo {volume} {110}},\ \bibinfo {pages} {L140508} (\bibinfo {year} {2024})}\BibitemShut {NoStop}%
\bibitem [{\citenamefont {LaBollita}\ \emph {et~al.}(2024{\natexlab{a}})\citenamefont {LaBollita}, \citenamefont {Pardo}, \citenamefont {Norman},\ and\ \citenamefont {Botana}}]{labollita2024}%
  \BibitemOpen
  \bibfield  {author} {\bibinfo {author} {\bibfnamefont {H.}~\bibnamefont {LaBollita}}, \bibinfo {author} {\bibfnamefont {V.}~\bibnamefont {Pardo}}, \bibinfo {author} {\bibfnamefont {M.~R.}\ \bibnamefont {Norman}},\ and\ \bibinfo {author} {\bibfnamefont {A.~S.}\ \bibnamefont {Botana}},\ }\bibfield  {title} {\bibinfo {title} {Electronic structure and magnetic properties of {L}a$_{3}${N}i$_{2}${O}$_{7}$ under pressure: active role of the {N}i-$d_{x^2-y^2}$ orbitals},\ }\href {https://arxiv.org/abs/2309.17279} {\bibfield  {journal} {\bibinfo  {journal} {arXiv:2309.17279}\ } (\bibinfo {year} {2024}{\natexlab{a}})}\BibitemShut {NoStop}%
\bibitem [{\citenamefont {Zhang}\ \emph {et~al.}(2024{\natexlab{c}})\citenamefont {Zhang}, \citenamefont {Xu},\ and\ \citenamefont {Xiang}}]{zhang2024emergent}%
  \BibitemOpen
  \bibfield  {author} {\bibinfo {author} {\bibfnamefont {B.}~\bibnamefont {Zhang}}, \bibinfo {author} {\bibfnamefont {C.}~\bibnamefont {Xu}},\ and\ \bibinfo {author} {\bibfnamefont {H.}~\bibnamefont {Xiang}},\ }\bibfield  {title} {\bibinfo {title} {Emergent spin-charge-orbital order in superconductor {L}a$_3${N}i$_2${O}$_7$},\ }\href {https://arxiv.org/abs/2407.18473} {\bibfield  {journal} {\bibinfo  {journal} {arXiv:2407.18473}\ } (\bibinfo {year} {2024}{\natexlab{c}})}\BibitemShut {NoStop}%
\bibitem [{\citenamefont {LaBollita}\ \emph {et~al.}(2024{\natexlab{b}})\citenamefont {LaBollita}, \citenamefont {Pardo}, \citenamefont {Norman},\ and\ \citenamefont {Botana}}]{labollita2024assessing}%
  \BibitemOpen
  \bibfield  {author} {\bibinfo {author} {\bibfnamefont {H.}~\bibnamefont {LaBollita}}, \bibinfo {author} {\bibfnamefont {V.}~\bibnamefont {Pardo}}, \bibinfo {author} {\bibfnamefont {M.~R.}\ \bibnamefont {Norman}},\ and\ \bibinfo {author} {\bibfnamefont {A.~S.}\ \bibnamefont {Botana}},\ }\bibfield  {title} {\bibinfo {title} {Assessing spin-density wave formation in {L}a$_3${N}i$_2${O}$_7$ from electronic structure calculations},\ }\href {https://doi.org/10.1103/PhysRevMaterials.8.L111801} {\bibfield  {journal} {\bibinfo  {journal} {Phys. Rev. Mater.}\ }\textbf {\bibinfo {volume} {8}},\ \bibinfo {pages} {L111801} (\bibinfo {year} {2024}{\natexlab{b}})}\BibitemShut {NoStop}%
\bibitem [{\citenamefont {Wang}\ \emph {et~al.}(2024{\natexlab{d}})\citenamefont {Wang}, \citenamefont {Jiang}, \citenamefont {Wang}, \citenamefont {Zhang},\ and\ \citenamefont {Hu}}]{wang2024electronic}%
  \BibitemOpen
  \bibfield  {author} {\bibinfo {author} {\bibfnamefont {Y.}~\bibnamefont {Wang}}, \bibinfo {author} {\bibfnamefont {K.}~\bibnamefont {Jiang}}, \bibinfo {author} {\bibfnamefont {Z.}~\bibnamefont {Wang}}, \bibinfo {author} {\bibfnamefont {F.-C.}\ \bibnamefont {Zhang}},\ and\ \bibinfo {author} {\bibfnamefont {J.}~\bibnamefont {Hu}},\ }\bibfield  {title} {\bibinfo {title} {Electronic and magnetic structures of bilayer {L}a$_3${N}i$_2${O}$_7$ at ambient pressure},\ }\href {https://doi.org/10.1103/PhysRevB.110.205122} {\bibfield  {journal} {\bibinfo  {journal} {Phys. Rev. B}\ }\textbf {\bibinfo {volume} {110}},\ \bibinfo {pages} {205122} (\bibinfo {year} {2024}{\natexlab{d}})}\BibitemShut {NoStop}%
\bibitem [{\citenamefont {Chen}\ \emph {et~al.}(2025)\citenamefont {Chen}, \citenamefont {Jiang}, \citenamefont {Li}, \citenamefont {Zhong},\ and\ \citenamefont {Lu}}]{PhysRevB.111.014515}%
  \BibitemOpen
  \bibfield  {author} {\bibinfo {author} {\bibfnamefont {X.}~\bibnamefont {Chen}}, \bibinfo {author} {\bibfnamefont {P.}~\bibnamefont {Jiang}}, \bibinfo {author} {\bibfnamefont {J.}~\bibnamefont {Li}}, \bibinfo {author} {\bibfnamefont {Z.}~\bibnamefont {Zhong}},\ and\ \bibinfo {author} {\bibfnamefont {Y.}~\bibnamefont {Lu}},\ }\bibfield  {title} {\bibinfo {title} {Charge and spin instabilities in superconducting {L}a$_3${N}i$_2${O}$_7$},\ }\href {https://doi.org/10.1103/PhysRevB.111.014515} {\bibfield  {journal} {\bibinfo  {journal} {Phys. Rev. B}\ }\textbf {\bibinfo {volume} {111}},\ \bibinfo {pages} {014515} (\bibinfo {year} {2025})}\BibitemShut {NoStop}%
\bibitem [{\citenamefont {Geisler}\ \emph {et~al.}(2024)\citenamefont {Geisler}, \citenamefont {Fanfarillo}, \citenamefont {Hamlin}, \citenamefont {Stewart}, \citenamefont {Hennig},\ and\ \citenamefont {Hirschfeld}}]{geisler2024optical}%
  \BibitemOpen
  \bibfield  {author} {\bibinfo {author} {\bibfnamefont {B.}~\bibnamefont {Geisler}}, \bibinfo {author} {\bibfnamefont {L.}~\bibnamefont {Fanfarillo}}, \bibinfo {author} {\bibfnamefont {J.~J.}\ \bibnamefont {Hamlin}}, \bibinfo {author} {\bibfnamefont {G.~R.}\ \bibnamefont {Stewart}}, \bibinfo {author} {\bibfnamefont {R.~G.}\ \bibnamefont {Hennig}},\ and\ \bibinfo {author} {\bibfnamefont {P.~J.}\ \bibnamefont {Hirschfeld}},\ }\bibfield  {title} {\bibinfo {title} {Optical properties and electronic correlations in {L}a$_3${N}i$_2${O}$_{7-\delta}$ bilayer nickelates under high pressure},\ }\href {https://doi.org/10.1038/s41535-024-00690-y} {\bibfield  {journal} {\bibinfo  {journal} {npj Quantum Materials}\ }\textbf {\bibinfo {volume} {9}},\ \bibinfo {pages} {89} (\bibinfo {year} {2024})}\BibitemShut {NoStop}%
\bibitem [{\citenamefont {Li}\ \emph {et~al.}(2025{\natexlab{b}})\citenamefont {Li}, \citenamefont {Cao}, \citenamefont {Liu}, \citenamefont {Peng}, \citenamefont {Lin}, \citenamefont {Pei}, \citenamefont {Zhang}, \citenamefont {Wu}, \citenamefont {Du}, \citenamefont {Zhao}, \citenamefont {Zhai}, \citenamefont {Zhang}, \citenamefont {Zhao}, \citenamefont {Lin}, \citenamefont {Tan}, \citenamefont {Qi}, \citenamefont {Li}, \citenamefont {Guo}, \citenamefont {Yang},\ and\ \citenamefont {Yang}}]{LI2025180}%
  \BibitemOpen
  \bibfield  {author} {\bibinfo {author} {\bibfnamefont {Y.}~\bibnamefont {Li}}, \bibinfo {author} {\bibfnamefont {Y.}~\bibnamefont {Cao}}, \bibinfo {author} {\bibfnamefont {L.}~\bibnamefont {Liu}}, \bibinfo {author} {\bibfnamefont {P.}~\bibnamefont {Peng}}, \bibinfo {author} {\bibfnamefont {H.}~\bibnamefont {Lin}}, \bibinfo {author} {\bibfnamefont {C.}~\bibnamefont {Pei}}, \bibinfo {author} {\bibfnamefont {M.}~\bibnamefont {Zhang}}, \bibinfo {author} {\bibfnamefont {H.}~\bibnamefont {Wu}}, \bibinfo {author} {\bibfnamefont {X.}~\bibnamefont {Du}}, \bibinfo {author} {\bibfnamefont {W.}~\bibnamefont {Zhao}}, \bibinfo {author} {\bibfnamefont {K.}~\bibnamefont {Zhai}}, \bibinfo {author} {\bibfnamefont {X.}~\bibnamefont {Zhang}}, \bibinfo {author} {\bibfnamefont {J.}~\bibnamefont {Zhao}}, \bibinfo {author} {\bibfnamefont {M.}~\bibnamefont {Lin}}, \bibinfo {author} {\bibfnamefont {P.}~\bibnamefont {Tan}}, \bibinfo {author} {\bibfnamefont {Y.}~\bibnamefont {Qi}}, \bibinfo {author} {\bibfnamefont {G.}~\bibnamefont
  {Li}}, \bibinfo {author} {\bibfnamefont {H.}~\bibnamefont {Guo}}, \bibinfo {author} {\bibfnamefont {L.}~\bibnamefont {Yang}},\ and\ \bibinfo {author} {\bibfnamefont {L.}~\bibnamefont {Yang}},\ }\bibfield  {title} {\bibinfo {title} {Distinct ultrafast dynamics of bilayer and trilayer nickelate superconductors regarding the density-wave-like transitions},\ }\href {https://doi.org/https://doi.org/10.1016/j.scib.2024.10.011} {\bibfield  {journal} {\bibinfo  {journal} {Science Bulletin}\ }\textbf {\bibinfo {volume} {70}},\ \bibinfo {pages} {180} (\bibinfo {year} {2025}{\natexlab{b}})}\BibitemShut {NoStop}%
\bibitem [{\citenamefont {You}\ \emph {et~al.}(2025)\citenamefont {You}, \citenamefont {Zhu}, \citenamefont {Del~Ben}, \citenamefont {Chen},\ and\ \citenamefont {Li}}]{You2025}%
  \BibitemOpen
  \bibfield  {author} {\bibinfo {author} {\bibfnamefont {J.-Y.}\ \bibnamefont {You}}, \bibinfo {author} {\bibfnamefont {Z.}~\bibnamefont {Zhu}}, \bibinfo {author} {\bibfnamefont {M.}~\bibnamefont {Del~Ben}}, \bibinfo {author} {\bibfnamefont {W.}~\bibnamefont {Chen}},\ and\ \bibinfo {author} {\bibfnamefont {Z.}~\bibnamefont {Li}},\ }\bibfield  {title} {\bibinfo {title} {Unlikelihood of a phonon mechanism for the high-temperature superconductivity in la3ni2o7},\ }\href {https://doi.org/10.1038/s41524-024-01483-4} {\bibfield  {journal} {\bibinfo  {journal} {npj Computational Materials}\ }\textbf {\bibinfo {volume} {11}},\ \bibinfo {pages} {3} (\bibinfo {year} {2025})}\BibitemShut {NoStop}%
\bibitem [{\citenamefont {Khasanov}\ \emph {et~al.}(2025)\citenamefont {Khasanov}, \citenamefont {Hicken}, \citenamefont {Gawryluk}, \citenamefont {Sazgari}, \citenamefont {Plokhikh}, \citenamefont {Sorel}, \citenamefont {Bartkowiak}, \citenamefont {B{\"o}tzel}, \citenamefont {Lechermann}, \citenamefont {Eremin}, \citenamefont {Luetkens},\ and\ \citenamefont {Guguchia}}]{Khasanov2025}%
  \BibitemOpen
  \bibfield  {author} {\bibinfo {author} {\bibfnamefont {R.}~\bibnamefont {Khasanov}}, \bibinfo {author} {\bibfnamefont {T.~J.}\ \bibnamefont {Hicken}}, \bibinfo {author} {\bibfnamefont {D.~J.}\ \bibnamefont {Gawryluk}}, \bibinfo {author} {\bibfnamefont {V.}~\bibnamefont {Sazgari}}, \bibinfo {author} {\bibfnamefont {I.}~\bibnamefont {Plokhikh}}, \bibinfo {author} {\bibfnamefont {L.~P.}\ \bibnamefont {Sorel}}, \bibinfo {author} {\bibfnamefont {M.}~\bibnamefont {Bartkowiak}}, \bibinfo {author} {\bibfnamefont {S.}~\bibnamefont {B{\"o}tzel}}, \bibinfo {author} {\bibfnamefont {F.}~\bibnamefont {Lechermann}}, \bibinfo {author} {\bibfnamefont {I.~M.}\ \bibnamefont {Eremin}}, \bibinfo {author} {\bibfnamefont {H.}~\bibnamefont {Luetkens}},\ and\ \bibinfo {author} {\bibfnamefont {Z.}~\bibnamefont {Guguchia}},\ }\bibfield  {title} {\bibinfo {title} {Pressure-enhanced splitting of density wave transitions in {L}a$_3${N}i$_2${O}$_{7-\delta}$},\ }\href {https://doi.org/10.1038/s41567-024-02754-z} {\bibfield  {journal}
  {\bibinfo  {journal} {Nature Physics}\ }\textbf {\bibinfo {volume} {21}},\ \bibinfo {pages} {430} (\bibinfo {year} {2025})}\BibitemShut {NoStop}%
\bibitem [{\citenamefont {Chen}\ \emph {et~al.}(2024{\natexlab{a}})\citenamefont {Chen}, \citenamefont {Liu}, \citenamefont {Jiao}, \citenamefont {Zou}, \citenamefont {Jiang}, \citenamefont {Li}, \citenamefont {Luo}, \citenamefont {Wu}, \citenamefont {Zhang}, \citenamefont {Guo} \emph {et~al.}}]{chen2024evidence}%
  \BibitemOpen
  \bibfield  {author} {\bibinfo {author} {\bibfnamefont {K.}~\bibnamefont {Chen}}, \bibinfo {author} {\bibfnamefont {X.}~\bibnamefont {Liu}}, \bibinfo {author} {\bibfnamefont {J.}~\bibnamefont {Jiao}}, \bibinfo {author} {\bibfnamefont {M.}~\bibnamefont {Zou}}, \bibinfo {author} {\bibfnamefont {C.}~\bibnamefont {Jiang}}, \bibinfo {author} {\bibfnamefont {X.}~\bibnamefont {Li}}, \bibinfo {author} {\bibfnamefont {Y.}~\bibnamefont {Luo}}, \bibinfo {author} {\bibfnamefont {Q.}~\bibnamefont {Wu}}, \bibinfo {author} {\bibfnamefont {N.}~\bibnamefont {Zhang}}, \bibinfo {author} {\bibfnamefont {Y.}~\bibnamefont {Guo}}, \emph {et~al.},\ }\bibfield  {title} {\bibinfo {title} {Evidence of spin density waves in {L}a$_3${N}i$_2${O}$_{7-\delta}$},\ }\href {https://journals.aps.org/prl/abstract/10.1103/PhysRevLett.132.256503} {\bibfield  {journal} {\bibinfo  {journal} {Phys. Rev. Lett.}\ }\textbf {\bibinfo {volume} {132}},\ \bibinfo {pages} {256503} (\bibinfo {year} {2024}{\natexlab{a}})}\BibitemShut {NoStop}%
\bibitem [{\citenamefont {Kakoi}\ \emph {et~al.}(2024{\natexlab{a}})\citenamefont {Kakoi}, \citenamefont {Oi}, \citenamefont {Ohshita}, \citenamefont {Yashima}, \citenamefont {Kuroki}, \citenamefont {Kato}, \citenamefont {Takahashi}, \citenamefont {Ishiwata}, \citenamefont {Adachi}, \citenamefont {Hatada}, \citenamefont {Uda},\ and\ \citenamefont {Mukuda}}]{Kakoi2024}%
  \BibitemOpen
  \bibfield  {author} {\bibinfo {author} {\bibfnamefont {M.}~\bibnamefont {Kakoi}}, \bibinfo {author} {\bibfnamefont {T.}~\bibnamefont {Oi}}, \bibinfo {author} {\bibfnamefont {Y.}~\bibnamefont {Ohshita}}, \bibinfo {author} {\bibfnamefont {M.}~\bibnamefont {Yashima}}, \bibinfo {author} {\bibfnamefont {K.}~\bibnamefont {Kuroki}}, \bibinfo {author} {\bibfnamefont {T.}~\bibnamefont {Kato}}, \bibinfo {author} {\bibfnamefont {H.}~\bibnamefont {Takahashi}}, \bibinfo {author} {\bibfnamefont {S.}~\bibnamefont {Ishiwata}}, \bibinfo {author} {\bibfnamefont {Y.}~\bibnamefont {Adachi}}, \bibinfo {author} {\bibfnamefont {N.}~\bibnamefont {Hatada}}, \bibinfo {author} {\bibfnamefont {T.}~\bibnamefont {Uda}},\ and\ \bibinfo {author} {\bibfnamefont {H.}~\bibnamefont {Mukuda}},\ }\bibfield  {title} {\bibinfo {title} {Multiband metallic ground state in multilayered nickelates {L}a$_3${N}i$_2${O}$_7$ and {L}a$_4${N}i$_3${O}$_{10}$ probed by $^{139}${L}a-{NMR} at ambient pressure},\ }\href {https://doi.org/10.7566/JPSJ.93.053702}
  {\bibfield  {journal} {\bibinfo  {journal} {J. Phys. Soc. Jpn.}\ }\textbf {\bibinfo {volume} {93}},\ \bibinfo {pages} {053702} (\bibinfo {year} {2024}{\natexlab{a}})}\BibitemShut {NoStop}%
\bibitem [{\citenamefont {Ren}\ \emph {et~al.}(2025)\citenamefont {Ren}, \citenamefont {Sutarto}, \citenamefont {Wu}, \citenamefont {Zhang}, \citenamefont {Huang}, \citenamefont {Xiang}, \citenamefont {Hu}, \citenamefont {Comin}, \citenamefont {Zhou},\ and\ \citenamefont {Zhu}}]{Ren2025}%
  \BibitemOpen
  \bibfield  {author} {\bibinfo {author} {\bibfnamefont {X.}~\bibnamefont {Ren}}, \bibinfo {author} {\bibfnamefont {R.}~\bibnamefont {Sutarto}}, \bibinfo {author} {\bibfnamefont {X.}~\bibnamefont {Wu}}, \bibinfo {author} {\bibfnamefont {J.}~\bibnamefont {Zhang}}, \bibinfo {author} {\bibfnamefont {H.}~\bibnamefont {Huang}}, \bibinfo {author} {\bibfnamefont {T.}~\bibnamefont {Xiang}}, \bibinfo {author} {\bibfnamefont {J.}~\bibnamefont {Hu}}, \bibinfo {author} {\bibfnamefont {R.}~\bibnamefont {Comin}}, \bibinfo {author} {\bibfnamefont {X.}~\bibnamefont {Zhou}},\ and\ \bibinfo {author} {\bibfnamefont {Z.}~\bibnamefont {Zhu}},\ }\bibfield  {title} {\bibinfo {title} {Resolving the electronic ground state of {L}a$_3${N}i$_2${O}$_{7-\delta}$ films},\ }\href {https://doi.org/10.1038/s42005-025-01971-z} {\bibfield  {journal} {\bibinfo  {journal} {Communications Physics}\ }\textbf {\bibinfo {volume} {8}},\ \bibinfo {pages} {52} (\bibinfo {year} {2025})}\BibitemShut {NoStop}%
\bibitem [{\citenamefont {Zhao}\ \emph {et~al.}(2025)\citenamefont {Zhao}, \citenamefont {Zhou}, \citenamefont {Huo}, \citenamefont {Wang}, \citenamefont {Nie}, \citenamefont {Yang}, \citenamefont {Ying}, \citenamefont {Wang}, \citenamefont {Wu},\ and\ \citenamefont {Chen}}]{dan2024spin}%
  \BibitemOpen
  \bibfield  {author} {\bibinfo {author} {\bibfnamefont {D.}~\bibnamefont {Zhao}}, \bibinfo {author} {\bibfnamefont {Y.}~\bibnamefont {Zhou}}, \bibinfo {author} {\bibfnamefont {M.}~\bibnamefont {Huo}}, \bibinfo {author} {\bibfnamefont {Y.}~\bibnamefont {Wang}}, \bibinfo {author} {\bibfnamefont {L.}~\bibnamefont {Nie}}, \bibinfo {author} {\bibfnamefont {Y.}~\bibnamefont {Yang}}, \bibinfo {author} {\bibfnamefont {J.}~\bibnamefont {Ying}}, \bibinfo {author} {\bibfnamefont {M.}~\bibnamefont {Wang}}, \bibinfo {author} {\bibfnamefont {T.}~\bibnamefont {Wu}},\ and\ \bibinfo {author} {\bibfnamefont {X.}~\bibnamefont {Chen}},\ }\bibfield  {title} {\bibinfo {title} {Pressure-enhanced spin-density-wave transition in double-layer nickelate {L}a$_3${N}i$_2${O}$_{7\mathscr{-\delta}}$},\ }\href {https://doi.org/https://doi.org/10.1016/j.scib.2025.02.019} {\bibfield  {journal} {\bibinfo  {journal} {Science Bulletin}\ }\textbf {\bibinfo {volume} {70}},\ \bibinfo {pages} {1239} (\bibinfo {year} {2025})}\BibitemShut {NoStop}%
\bibitem [{\citenamefont {Chen}\ \emph {et~al.}(2024{\natexlab{b}})\citenamefont {Chen}, \citenamefont {Choi}, \citenamefont {Jiang}, \citenamefont {Mei}, \citenamefont {Jiang}, \citenamefont {Li}, \citenamefont {Agrestini}, \citenamefont {Garcia-Fernandez}, \citenamefont {Huang}, \citenamefont {Sun}, \citenamefont {Shen}, \citenamefont {Wang}, \citenamefont {Hu}, \citenamefont {Lu}, \citenamefont {Zhou},\ and\ \citenamefont {Feng}}]{chen2024electronic}%
  \BibitemOpen
  \bibfield  {author} {\bibinfo {author} {\bibfnamefont {X.}~\bibnamefont {Chen}}, \bibinfo {author} {\bibfnamefont {J.}~\bibnamefont {Choi}}, \bibinfo {author} {\bibfnamefont {Z.}~\bibnamefont {Jiang}}, \bibinfo {author} {\bibfnamefont {J.}~\bibnamefont {Mei}}, \bibinfo {author} {\bibfnamefont {K.}~\bibnamefont {Jiang}}, \bibinfo {author} {\bibfnamefont {J.}~\bibnamefont {Li}}, \bibinfo {author} {\bibfnamefont {S.}~\bibnamefont {Agrestini}}, \bibinfo {author} {\bibfnamefont {M.}~\bibnamefont {Garcia-Fernandez}}, \bibinfo {author} {\bibfnamefont {X.}~\bibnamefont {Huang}}, \bibinfo {author} {\bibfnamefont {H.}~\bibnamefont {Sun}}, \bibinfo {author} {\bibfnamefont {D.}~\bibnamefont {Shen}}, \bibinfo {author} {\bibfnamefont {M.}~\bibnamefont {Wang}}, \bibinfo {author} {\bibfnamefont {J.}~\bibnamefont {Hu}}, \bibinfo {author} {\bibfnamefont {Y.}~\bibnamefont {Lu}}, \bibinfo {author} {\bibfnamefont {K.-J.}\ \bibnamefont {Zhou}},\ and\ \bibinfo {author} {\bibfnamefont {D.}~\bibnamefont {Feng}},\ }\bibfield
  {title} {\bibinfo {title} {Electronic and magnetic excitations in {L}a$_3${N}i$_2${O}$_7$},\ }\href {https://doi.org/10.1038/s41467-024-53863-5} {\bibfield  {journal} {\bibinfo  {journal} {Nature Communications}\ }\textbf {\bibinfo {volume} {15}},\ \bibinfo {pages} {9597} (\bibinfo {year} {2024}{\natexlab{b}})}\BibitemShut {NoStop}%
\bibitem [{\citenamefont {Liu}\ \emph {et~al.}(2023{\natexlab{a}})\citenamefont {Liu}, \citenamefont {Sun}, \citenamefont {Huo}, \citenamefont {Ma}, \citenamefont {Ji}, \citenamefont {Yi}, \citenamefont {Li}, \citenamefont {Liu}, \citenamefont {Yu}, \citenamefont {Zhang}, \citenamefont {Chen}, \citenamefont {Liang}, \citenamefont {Dong}, \citenamefont {Guo}, \citenamefont {Zhong}, \citenamefont {Shen}, \citenamefont {Li},\ and\ \citenamefont {Wang}}]{Wang2022LNO}%
  \BibitemOpen
  \bibfield  {author} {\bibinfo {author} {\bibfnamefont {Z.}~\bibnamefont {Liu}}, \bibinfo {author} {\bibfnamefont {H.}~\bibnamefont {Sun}}, \bibinfo {author} {\bibfnamefont {M.}~\bibnamefont {Huo}}, \bibinfo {author} {\bibfnamefont {X.}~\bibnamefont {Ma}}, \bibinfo {author} {\bibfnamefont {Y.}~\bibnamefont {Ji}}, \bibinfo {author} {\bibfnamefont {E.}~\bibnamefont {Yi}}, \bibinfo {author} {\bibfnamefont {L.}~\bibnamefont {Li}}, \bibinfo {author} {\bibfnamefont {H.}~\bibnamefont {Liu}}, \bibinfo {author} {\bibfnamefont {J.}~\bibnamefont {Yu}}, \bibinfo {author} {\bibfnamefont {Z.}~\bibnamefont {Zhang}}, \bibinfo {author} {\bibfnamefont {Z.}~\bibnamefont {Chen}}, \bibinfo {author} {\bibfnamefont {F.}~\bibnamefont {Liang}}, \bibinfo {author} {\bibfnamefont {H.}~\bibnamefont {Dong}}, \bibinfo {author} {\bibfnamefont {H.}~\bibnamefont {Guo}}, \bibinfo {author} {\bibfnamefont {D.}~\bibnamefont {Zhong}}, \bibinfo {author} {\bibfnamefont {B.}~\bibnamefont {Shen}}, \bibinfo {author} {\bibfnamefont {S.}~\bibnamefont
  {Li}},\ and\ \bibinfo {author} {\bibfnamefont {M.}~\bibnamefont {Wang}},\ }\bibfield  {title} {\bibinfo {title} {Evidence for charge and spin density waves in single crystals of {L}a$_3${N}i$_2${O}$_7$ and {L}a$_3${N}i$_2${O}$_6$},\ }\href {https://link.springer.com/article/10.1007/s11433-022-1962-4} {\bibfield  {journal} {\bibinfo  {journal} {Sci. China-Phys. Mech. Astron.}\ }\textbf {\bibinfo {volume} {66}},\ \bibinfo {pages} {217411} (\bibinfo {year} {2023}{\natexlab{a}})}\BibitemShut {NoStop}%
\bibitem [{\citenamefont {Tsuei}\ and\ \citenamefont {Kirtley}(2000)}]{tsuei2000pairing}%
  \BibitemOpen
  \bibfield  {author} {\bibinfo {author} {\bibfnamefont {C.}~\bibnamefont {Tsuei}}\ and\ \bibinfo {author} {\bibfnamefont {J.}~\bibnamefont {Kirtley}},\ }\bibfield  {title} {\bibinfo {title} {Pairing symmetry in cuprate superconductors},\ }\href {https://journals.aps.org/rmp/abstract/10.1103/RevModPhys.72.969} {\bibfield  {journal} {\bibinfo  {journal} {Rev. Mod. Phys.}\ }\textbf {\bibinfo {volume} {72}},\ \bibinfo {pages} {969} (\bibinfo {year} {2000})}\BibitemShut {NoStop}%
\bibitem [{\citenamefont {Lee}\ \emph {et~al.}(2006)\citenamefont {Lee}, \citenamefont {Nagaosa},\ and\ \citenamefont {Wen}}]{lee2006doping}%
  \BibitemOpen
  \bibfield  {author} {\bibinfo {author} {\bibfnamefont {P.~A.}\ \bibnamefont {Lee}}, \bibinfo {author} {\bibfnamefont {N.}~\bibnamefont {Nagaosa}},\ and\ \bibinfo {author} {\bibfnamefont {X.-G.}\ \bibnamefont {Wen}},\ }\bibfield  {title} {\bibinfo {title} {Doping a mott insulator: Physics of high-temperature superconductivity},\ }\href {https://journals.aps.org/rmp/abstract/10.1103/RevModPhys.78.17} {\bibfield  {journal} {\bibinfo  {journal} {Rev. Mod. Phys.}\ }\textbf {\bibinfo {volume} {78}},\ \bibinfo {pages} {17} (\bibinfo {year} {2006})}\BibitemShut {NoStop}%
\bibitem [{\citenamefont {Xiang}\ and\ \citenamefont {Wu}(2022)}]{xiang2022d}%
  \BibitemOpen
  \bibfield  {author} {\bibinfo {author} {\bibfnamefont {T.}~\bibnamefont {Xiang}}\ and\ \bibinfo {author} {\bibfnamefont {C.}~\bibnamefont {Wu}},\ }\href@noop {} {\emph {\bibinfo {title} {D-wave Superconductivity}}}\ (\bibinfo  {publisher} {Cambridge University Press},\ \bibinfo {year} {2022})\BibitemShut {NoStop}%
\bibitem [{\citenamefont {Fernandes}\ \emph {et~al.}(2022)\citenamefont {Fernandes}, \citenamefont {Coldea}, \citenamefont {Ding}, \citenamefont {Fisher}, \citenamefont {Hirschfeld},\ and\ \citenamefont {Kotliar}}]{Fernandes2022}%
  \BibitemOpen
  \bibfield  {author} {\bibinfo {author} {\bibfnamefont {R.~M.}\ \bibnamefont {Fernandes}}, \bibinfo {author} {\bibfnamefont {A.~I.}\ \bibnamefont {Coldea}}, \bibinfo {author} {\bibfnamefont {H.}~\bibnamefont {Ding}}, \bibinfo {author} {\bibfnamefont {I.~R.}\ \bibnamefont {Fisher}}, \bibinfo {author} {\bibfnamefont {P.~J.}\ \bibnamefont {Hirschfeld}},\ and\ \bibinfo {author} {\bibfnamefont {G.}~\bibnamefont {Kotliar}},\ }\bibfield  {title} {\bibinfo {title} {Iron pnictides and chalcogenides: a new paradigm for superconductivity},\ }\href {https://doi.org/10.1038/s41586-021-04073-2} {\bibfield  {journal} {\bibinfo  {journal} {Nature}\ }\textbf {\bibinfo {volume} {601}},\ \bibinfo {pages} {35} (\bibinfo {year} {2022})}\BibitemShut {NoStop}%
\bibitem [{\citenamefont {Lu}\ \emph {et~al.}(2024{\natexlab{a}})\citenamefont {Lu}, \citenamefont {Pan}, \citenamefont {Yang},\ and\ \citenamefont {Wu}}]{lu2023bilayertJ}%
  \BibitemOpen
  \bibfield  {author} {\bibinfo {author} {\bibfnamefont {C.}~\bibnamefont {Lu}}, \bibinfo {author} {\bibfnamefont {Z.}~\bibnamefont {Pan}}, \bibinfo {author} {\bibfnamefont {F.}~\bibnamefont {Yang}},\ and\ \bibinfo {author} {\bibfnamefont {C.}~\bibnamefont {Wu}},\ }\bibfield  {title} {\bibinfo {title} {Interlayer-coupling-driven high-temperature superconductivity in {L}a$_3${N}i$_2${O}$_7$ under pressure},\ }\href {https://doi.org/10.1103/PhysRevLett.132.146002} {\bibfield  {journal} {\bibinfo  {journal} {Phys. Rev. Lett.}\ }\textbf {\bibinfo {volume} {132}},\ \bibinfo {pages} {146002} (\bibinfo {year} {2024}{\natexlab{a}})}\BibitemShut {NoStop}%
\bibitem [{\citenamefont {Oh}\ and\ \citenamefont {Zhang}(2023)}]{oh2023type2}%
  \BibitemOpen
  \bibfield  {author} {\bibinfo {author} {\bibfnamefont {H.}~\bibnamefont {Oh}}\ and\ \bibinfo {author} {\bibfnamefont {Y.-H.}\ \bibnamefont {Zhang}},\ }\bibfield  {title} {\bibinfo {title} {Type-{II} $t$-${J}$ model and shared superexchange coupling from hund's rule in superconducting {L}a$_3${N}i$_2${O}$_7$},\ }\href {https://doi.org/10.1103/PhysRevB.108.174511} {\bibfield  {journal} {\bibinfo  {journal} {Phys. Rev. B}\ }\textbf {\bibinfo {volume} {108}},\ \bibinfo {pages} {174511} (\bibinfo {year} {2023})}\BibitemShut {NoStop}%
\bibitem [{\citenamefont {Liao}\ \emph {et~al.}(2023)\citenamefont {Liao}, \citenamefont {Chen}, \citenamefont {Duan}, \citenamefont {Wang}, \citenamefont {Liu}, \citenamefont {Yu},\ and\ \citenamefont {Si}}]{liao2023electron}%
  \BibitemOpen
  \bibfield  {author} {\bibinfo {author} {\bibfnamefont {Z.}~\bibnamefont {Liao}}, \bibinfo {author} {\bibfnamefont {L.}~\bibnamefont {Chen}}, \bibinfo {author} {\bibfnamefont {G.}~\bibnamefont {Duan}}, \bibinfo {author} {\bibfnamefont {Y.}~\bibnamefont {Wang}}, \bibinfo {author} {\bibfnamefont {C.}~\bibnamefont {Liu}}, \bibinfo {author} {\bibfnamefont {R.}~\bibnamefont {Yu}},\ and\ \bibinfo {author} {\bibfnamefont {Q.}~\bibnamefont {Si}},\ }\bibfield  {title} {\bibinfo {title} {Electron correlations and superconductivity in {L}a$_{3}${N}i$_{2}${O}$_{7}$ under pressure tuning},\ }\href {https://doi.org/10.1103/PhysRevB.108.214522} {\bibfield  {journal} {\bibinfo  {journal} {Phys. Rev. B}\ }\textbf {\bibinfo {volume} {108}},\ \bibinfo {pages} {214522} (\bibinfo {year} {2023})}\BibitemShut {NoStop}%
\bibitem [{\citenamefont {Yang}\ \emph {et~al.}(2023{\natexlab{a}})\citenamefont {Yang}, \citenamefont {Zhang},\ and\ \citenamefont {Zhang}}]{Yi_Feng2023}%
  \BibitemOpen
  \bibfield  {author} {\bibinfo {author} {\bibfnamefont {Y.-F.}\ \bibnamefont {Yang}}, \bibinfo {author} {\bibfnamefont {G.-M.}\ \bibnamefont {Zhang}},\ and\ \bibinfo {author} {\bibfnamefont {F.-C.}\ \bibnamefont {Zhang}},\ }\bibfield  {title} {\bibinfo {title} {Interlayer valence bonds and two-component theory for high-${T}_{c}$ superconductivity of {L}a$_3${N}i$_2${O}$_7$ under pressure},\ }\href {https://doi.org/10.1103/PhysRevB.108.L201108} {\bibfield  {journal} {\bibinfo  {journal} {Phys. Rev. B}\ }\textbf {\bibinfo {volume} {108}},\ \bibinfo {pages} {L201108} (\bibinfo {year} {2023}{\natexlab{a}})}\BibitemShut {NoStop}%
\bibitem [{\citenamefont {Jiang}\ \emph {et~al.}(2023)\citenamefont {Jiang}, \citenamefont {Wang},\ and\ \citenamefont {Zhang}}]{jiang2023high}%
  \BibitemOpen
  \bibfield  {author} {\bibinfo {author} {\bibfnamefont {K.}~\bibnamefont {Jiang}}, \bibinfo {author} {\bibfnamefont {Z.}~\bibnamefont {Wang}},\ and\ \bibinfo {author} {\bibfnamefont {F.-C.}\ \bibnamefont {Zhang}},\ }\bibfield  {title} {\bibinfo {title} {High temperature superconductivity in {L}a$_3${N}i$_2${O}$_7$},\ }\href {https://iopscience.iop.org/article/10.1088/0256-307X/41/1/017402} {\bibfield  {journal} {\bibinfo  {journal} {Chin. Phys. Lett.}\ } (\bibinfo {year} {2023})}\BibitemShut {NoStop}%
\bibitem [{\citenamefont {Huang}\ \emph {et~al.}(2023)\citenamefont {Huang}, \citenamefont {Wang},\ and\ \citenamefont {Zhou}}]{huang2023impurity}%
  \BibitemOpen
  \bibfield  {author} {\bibinfo {author} {\bibfnamefont {J.}~\bibnamefont {Huang}}, \bibinfo {author} {\bibfnamefont {Z.~D.}\ \bibnamefont {Wang}},\ and\ \bibinfo {author} {\bibfnamefont {T.}~\bibnamefont {Zhou}},\ }\bibfield  {title} {\bibinfo {title} {Impurity and vortex states in the bilayer high-temperature superconductor {L}a$_3${N}i$_2${O}$_7$},\ }\href {https://doi.org/10.1103/PhysRevB.108.174501} {\bibfield  {journal} {\bibinfo  {journal} {Phys. Rev. B}\ }\textbf {\bibinfo {volume} {108}},\ \bibinfo {pages} {174501} (\bibinfo {year} {2023})}\BibitemShut {NoStop}%
\bibitem [{\citenamefont {Lu}\ \emph {et~al.}(2023)\citenamefont {Lu}, \citenamefont {Li}, \citenamefont {Zeng}, \citenamefont {Hou}, \citenamefont {Wang}, \citenamefont {Yang},\ and\ \citenamefont {You}}]{lu2023sc}%
  \BibitemOpen
  \bibfield  {author} {\bibinfo {author} {\bibfnamefont {D.-C.}\ \bibnamefont {Lu}}, \bibinfo {author} {\bibfnamefont {M.}~\bibnamefont {Li}}, \bibinfo {author} {\bibfnamefont {Z.-Y.}\ \bibnamefont {Zeng}}, \bibinfo {author} {\bibfnamefont {W.}~\bibnamefont {Hou}}, \bibinfo {author} {\bibfnamefont {J.}~\bibnamefont {Wang}}, \bibinfo {author} {\bibfnamefont {F.}~\bibnamefont {Yang}},\ and\ \bibinfo {author} {\bibfnamefont {Y.-Z.}\ \bibnamefont {You}},\ }\bibfield  {title} {\bibinfo {title} {Superconductivity from doping symmetric mass generation insulators: Application to {L}a$_3${N}i$_2${O}$_7$ under pressure},\ }\href {https://arxiv.org/abs/2308.11195} {\bibfield  {journal} {\bibinfo  {journal} {arXiv:2308.11195}\ } (\bibinfo {year} {2023})}\BibitemShut {NoStop}%
\bibitem [{\citenamefont {Luo}\ \emph {et~al.}(2024)\citenamefont {Luo}, \citenamefont {Lv}, \citenamefont {Wang}, \citenamefont {W{\'u}},\ and\ \citenamefont {Yao}}]{luo2023high}%
  \BibitemOpen
  \bibfield  {author} {\bibinfo {author} {\bibfnamefont {Z.}~\bibnamefont {Luo}}, \bibinfo {author} {\bibfnamefont {B.}~\bibnamefont {Lv}}, \bibinfo {author} {\bibfnamefont {M.}~\bibnamefont {Wang}}, \bibinfo {author} {\bibfnamefont {W.}~\bibnamefont {W{\'u}}},\ and\ \bibinfo {author} {\bibfnamefont {D.-X.}\ \bibnamefont {Yao}},\ }\bibfield  {title} {\bibinfo {title} {High-{T}$_c$ superconductivity in {L}a$_3${N}i$_2${O}$_7$ based on the bilayer two-orbital t-{J} model},\ }\href {https://doi.org/10.1038/s41535-024-00668-w} {\bibfield  {journal} {\bibinfo  {journal} {npj Quantum Materials}\ }\textbf {\bibinfo {volume} {9}},\ \bibinfo {pages} {61} (\bibinfo {year} {2024})}\BibitemShut {NoStop}%
\bibitem [{\citenamefont {Pan}\ \emph {et~al.}(2024)\citenamefont {Pan}, \citenamefont {Lu}, \citenamefont {Yang},\ and\ \citenamefont {Wu}}]{pan2023rno}%
  \BibitemOpen
  \bibfield  {author} {\bibinfo {author} {\bibfnamefont {Z.}~\bibnamefont {Pan}}, \bibinfo {author} {\bibfnamefont {C.}~\bibnamefont {Lu}}, \bibinfo {author} {\bibfnamefont {F.}~\bibnamefont {Yang}},\ and\ \bibinfo {author} {\bibfnamefont {C.}~\bibnamefont {Wu}},\ }\bibfield  {title} {\bibinfo {title} {Effect of rare-earth element substitution in superconducting {R}$_3${N}i$_2${O}$_7$ under pressure},\ }\href {https://doi.org/10.1088/0256-307X/41/8/087401} {\bibfield  {journal} {\bibinfo  {journal} {Chinese Physics Letters}\ }\textbf {\bibinfo {volume} {41}},\ \bibinfo {eid} {087401} (\bibinfo {year} {2024})}\BibitemShut {NoStop}%
\bibitem [{\citenamefont {Yang}(2024)}]{yang2024decom}%
  \BibitemOpen
  \bibfield  {author} {\bibinfo {author} {\bibfnamefont {Y.-F.}\ \bibnamefont {Yang}},\ }\bibfield  {title} {\bibinfo {title} {Decomposition of multilayer superconductivity with interlayer pairing},\ }\href {https://doi.org/10.1103/PhysRevB.110.104507} {\bibfield  {journal} {\bibinfo  {journal} {Phys. Rev. B}\ }\textbf {\bibinfo {volume} {110}},\ \bibinfo {pages} {104507} (\bibinfo {year} {2024})}\BibitemShut {NoStop}%
\bibitem [{\citenamefont {Lu}\ \emph {et~al.}(2024{\natexlab{b}})\citenamefont {Lu}, \citenamefont {Pan}, \citenamefont {Yang},\ and\ \citenamefont {Wu}}]{Lu2024interplay}%
  \BibitemOpen
  \bibfield  {author} {\bibinfo {author} {\bibfnamefont {C.}~\bibnamefont {Lu}}, \bibinfo {author} {\bibfnamefont {Z.}~\bibnamefont {Pan}}, \bibinfo {author} {\bibfnamefont {F.}~\bibnamefont {Yang}},\ and\ \bibinfo {author} {\bibfnamefont {C.}~\bibnamefont {Wu}},\ }\bibfield  {title} {\bibinfo {title} {Interplay of two ${E}_{g}$ orbitals in superconducting {L}a$_{3}${N}i$_{2}${O}$_{7}$ under pressure},\ }\href {https://doi.org/10.1103/PhysRevB.110.094509} {\bibfield  {journal} {\bibinfo  {journal} {Phys. Rev. B}\ }\textbf {\bibinfo {volume} {110}},\ \bibinfo {pages} {094509} (\bibinfo {year} {2024}{\natexlab{b}})}\BibitemShut {NoStop}%
\bibitem [{\citenamefont {Wu}\ \emph {et~al.}(2024)\citenamefont {Wu}, \citenamefont {Yang},\ and\ \citenamefont {Zhang}}]{wu2024deconfined}%
  \BibitemOpen
  \bibfield  {author} {\bibinfo {author} {\bibfnamefont {X.}~\bibnamefont {Wu}}, \bibinfo {author} {\bibfnamefont {H.}~\bibnamefont {Yang}},\ and\ \bibinfo {author} {\bibfnamefont {Y.-H.}\ \bibnamefont {Zhang}},\ }\bibfield  {title} {\bibinfo {title} {Deconfined fermi liquid to fermi liquid transition and superconducting instability},\ }\href {https://doi.org/10.1103/PhysRevB.110.125122} {\bibfield  {journal} {\bibinfo  {journal} {Phys. Rev. B}\ }\textbf {\bibinfo {volume} {110}},\ \bibinfo {pages} {125122} (\bibinfo {year} {2024})}\BibitemShut {NoStop}%
\bibitem [{\citenamefont {Duan}\ \emph {et~al.}(2025)\citenamefont {Duan}, \citenamefont {Liao}, \citenamefont {Chen}, \citenamefont {Wang}, \citenamefont {Yu},\ and\ \citenamefont {Si}}]{duan2025}%
  \BibitemOpen
  \bibfield  {author} {\bibinfo {author} {\bibfnamefont {G.}~\bibnamefont {Duan}}, \bibinfo {author} {\bibfnamefont {Z.}~\bibnamefont {Liao}}, \bibinfo {author} {\bibfnamefont {L.}~\bibnamefont {Chen}}, \bibinfo {author} {\bibfnamefont {Y.}~\bibnamefont {Wang}}, \bibinfo {author} {\bibfnamefont {R.}~\bibnamefont {Yu}},\ and\ \bibinfo {author} {\bibfnamefont {Q.}~\bibnamefont {Si}},\ }\bibfield  {title} {\bibinfo {title} {Orbital-selective correlation effects and superconducting pairing symmetry in a multiorbital t-{J} model for bilayer nickelates},\ }\href {https://arxiv.org/abs/2502.09195} {\bibfield  {journal} {\bibinfo  {journal} {arXiv:2502.09195}\ } (\bibinfo {year} {2025})}\BibitemShut {NoStop}%
\bibitem [{\citenamefont {Liu}\ \emph {et~al.}(2023{\natexlab{b}})\citenamefont {Liu}, \citenamefont {Mei}, \citenamefont {Ye}, \citenamefont {Chen},\ and\ \citenamefont {Yang}}]{YangF2023}%
  \BibitemOpen
  \bibfield  {author} {\bibinfo {author} {\bibfnamefont {Y.-B.}\ \bibnamefont {Liu}}, \bibinfo {author} {\bibfnamefont {J.-W.}\ \bibnamefont {Mei}}, \bibinfo {author} {\bibfnamefont {F.}~\bibnamefont {Ye}}, \bibinfo {author} {\bibfnamefont {W.-Q.}\ \bibnamefont {Chen}},\ and\ \bibinfo {author} {\bibfnamefont {F.}~\bibnamefont {Yang}},\ }\bibfield  {title} {\bibinfo {title} {s$^{\pm}$-wave pairing and the destructive role of apical-oxygen deficiencies in {L}a$_3${N}i$_2${O}$_7$ under pressure},\ }\href {https://doi.org/10.1103/PhysRevLett.131.236002} {\bibfield  {journal} {\bibinfo  {journal} {Phys. Rev. Lett.}\ }\textbf {\bibinfo {volume} {131}},\ \bibinfo {pages} {236002} (\bibinfo {year} {2023}{\natexlab{b}})}\BibitemShut {NoStop}%
\bibitem [{\citenamefont {Zhang}\ \emph {et~al.}(2024{\natexlab{d}})\citenamefont {Zhang}, \citenamefont {Lin}, \citenamefont {Moreo}, \citenamefont {Maier},\ and\ \citenamefont {Dagotto}}]{zhang2023structural}%
  \BibitemOpen
  \bibfield  {author} {\bibinfo {author} {\bibfnamefont {Y.}~\bibnamefont {Zhang}}, \bibinfo {author} {\bibfnamefont {L.-F.}\ \bibnamefont {Lin}}, \bibinfo {author} {\bibfnamefont {A.}~\bibnamefont {Moreo}}, \bibinfo {author} {\bibfnamefont {T.~A.}\ \bibnamefont {Maier}},\ and\ \bibinfo {author} {\bibfnamefont {E.}~\bibnamefont {Dagotto}},\ }\bibfield  {title} {\bibinfo {title} {Structural phase transition, $s_{\pm}$-wave pairing, and magnetic stripe order in bilayered superconductor {L}a$_3${N}i$_2${O}$_7$ under pressure},\ }\href {https://www.nature.com/articles/s41467-024-46622-z} {\bibfield  {journal} {\bibinfo  {journal} {Nat. Commun.}\ }\textbf {\bibinfo {volume} {15}},\ \bibinfo {pages} {2470} (\bibinfo {year} {2024}{\natexlab{d}})}\BibitemShut {NoStop}%
\bibitem [{\citenamefont {Zhang}\ \emph {et~al.}(2024{\natexlab{e}})\citenamefont {Zhang}, \citenamefont {Lin}, \citenamefont {Moreo}, \citenamefont {Maier},\ and\ \citenamefont {Dagotto}}]{zhang2024electronic}%
  \BibitemOpen
  \bibfield  {author} {\bibinfo {author} {\bibfnamefont {Y.}~\bibnamefont {Zhang}}, \bibinfo {author} {\bibfnamefont {L.-F.}\ \bibnamefont {Lin}}, \bibinfo {author} {\bibfnamefont {A.}~\bibnamefont {Moreo}}, \bibinfo {author} {\bibfnamefont {T.~A.}\ \bibnamefont {Maier}},\ and\ \bibinfo {author} {\bibfnamefont {E.}~\bibnamefont {Dagotto}},\ }\bibfield  {title} {\bibinfo {title} {Electronic structure, self-doping, and superconducting instability in the alternating single-layer trilayer stacking nickelates {L}a$_{3}${N}i$_{2}${O}$_{7}$},\ }\href {https://doi.org/10.1103/PhysRevB.110.L060510} {\bibfield  {journal} {\bibinfo  {journal} {Phys. Rev. B}\ }\textbf {\bibinfo {volume} {110}},\ \bibinfo {pages} {L060510} (\bibinfo {year} {2024}{\natexlab{e}})}\BibitemShut {NoStop}%
\bibitem [{\citenamefont {Zhang}\ \emph {et~al.}(2024{\natexlab{f}})\citenamefont {Zhang}, \citenamefont {Lin}, \citenamefont {Moreo}, \citenamefont {Maier},\ and\ \citenamefont {Dagotto}}]{zhang2024prediction}%
  \BibitemOpen
  \bibfield  {author} {\bibinfo {author} {\bibfnamefont {Y.}~\bibnamefont {Zhang}}, \bibinfo {author} {\bibfnamefont {L.-F.}\ \bibnamefont {Lin}}, \bibinfo {author} {\bibfnamefont {A.}~\bibnamefont {Moreo}}, \bibinfo {author} {\bibfnamefont {T.~A.}\ \bibnamefont {Maier}},\ and\ \bibinfo {author} {\bibfnamefont {E.}~\bibnamefont {Dagotto}},\ }\bibfield  {title} {\bibinfo {title} {Prediction of s$^{\pm}$-wave superconductivity enhanced by electronic doping in trilayer nickelates {L}a$_4${N}i$_3${O}$_{10}$ under pressure},\ }\href {https://journals.aps.org/prl/abstract/10.1103/PhysRevLett.133.136001} {\bibfield  {journal} {\bibinfo  {journal} {Phys. Rev. Lett.}\ }\textbf {\bibinfo {volume} {133}},\ \bibinfo {pages} {136001} (\bibinfo {year} {2024}{\natexlab{f}})}\BibitemShut {NoStop}%
\bibitem [{\citenamefont {Zhang}\ \emph {et~al.}(2024{\natexlab{g}})\citenamefont {Zhang}, \citenamefont {Sun}, \citenamefont {Liu}, \citenamefont {Liu}, \citenamefont {Chen},\ and\ \citenamefont {Yang}}]{zhang2024s}%
  \BibitemOpen
  \bibfield  {author} {\bibinfo {author} {\bibfnamefont {M.}~\bibnamefont {Zhang}}, \bibinfo {author} {\bibfnamefont {H.}~\bibnamefont {Sun}}, \bibinfo {author} {\bibfnamefont {Y.-B.}\ \bibnamefont {Liu}}, \bibinfo {author} {\bibfnamefont {Q.}~\bibnamefont {Liu}}, \bibinfo {author} {\bibfnamefont {W.-Q.}\ \bibnamefont {Chen}},\ and\ \bibinfo {author} {\bibfnamefont {F.}~\bibnamefont {Yang}},\ }\bibfield  {title} {\bibinfo {title} {The $s^{\pm}$-wave superconductivity in the pressurized {L}a$_{4}${N}i$_{3}${O}$_{10}$},\ }\href {https://journals.aps.org/prb/abstract/10.1103/PhysRevB.110.L180501} {\bibfield  {journal} {\bibinfo  {journal} {Phys. Rev. B}\ }\textbf {\bibinfo {volume} {110}},\ \bibinfo {pages} {L180501} (\bibinfo {year} {2024}{\natexlab{g}})}\BibitemShut {NoStop}%
\bibitem [{\citenamefont {Zhang}\ \emph {et~al.}(2023{\natexlab{b}})\citenamefont {Zhang}, \citenamefont {Lin}, \citenamefont {Moreo}, \citenamefont {Maier},\ and\ \citenamefont {Dagotto}}]{zhang2023trends}%
  \BibitemOpen
  \bibfield  {author} {\bibinfo {author} {\bibfnamefont {Y.}~\bibnamefont {Zhang}}, \bibinfo {author} {\bibfnamefont {L.-F.}\ \bibnamefont {Lin}}, \bibinfo {author} {\bibfnamefont {A.}~\bibnamefont {Moreo}}, \bibinfo {author} {\bibfnamefont {T.~A.}\ \bibnamefont {Maier}},\ and\ \bibinfo {author} {\bibfnamefont {E.}~\bibnamefont {Dagotto}},\ }\bibfield  {title} {\bibinfo {title} {Trends in electronic structures and $s_{\pm}$-wave pairing for the rare-earth series in bilayer nickelate superconductor ${R}_3${N}i$_2${O}$_7$},\ }\href {https://doi.org/10.1103/PhysRevB.108.165141} {\bibfield  {journal} {\bibinfo  {journal} {Phys. Rev. B}\ }\textbf {\bibinfo {volume} {108}},\ \bibinfo {pages} {165141} (\bibinfo {year} {2023}{\natexlab{b}})}\BibitemShut {NoStop}%
\bibitem [{\citenamefont {Chen}\ \emph {et~al.}(2024{\natexlab{c}})\citenamefont {Chen}, \citenamefont {Luo}, \citenamefont {Wang}, \citenamefont {W\'u},\ and\ \citenamefont {Yao}}]{chen2024tri}%
  \BibitemOpen
  \bibfield  {author} {\bibinfo {author} {\bibfnamefont {C.-Q.}\ \bibnamefont {Chen}}, \bibinfo {author} {\bibfnamefont {Z.}~\bibnamefont {Luo}}, \bibinfo {author} {\bibfnamefont {M.}~\bibnamefont {Wang}}, \bibinfo {author} {\bibfnamefont {W.}~\bibnamefont {W\'u}},\ and\ \bibinfo {author} {\bibfnamefont {D.-X.}\ \bibnamefont {Yao}},\ }\bibfield  {title} {\bibinfo {title} {Trilayer multiorbital models of {L}a$_{4}${N}i$_{3}${O}$_{10}$},\ }\href {https://doi.org/10.1103/PhysRevB.110.014503} {\bibfield  {journal} {\bibinfo  {journal} {Phys. Rev. B}\ }\textbf {\bibinfo {volume} {110}},\ \bibinfo {pages} {014503} (\bibinfo {year} {2024}{\natexlab{c}})}\BibitemShut {NoStop}%
\bibitem [{\citenamefont {Zhang}\ \emph {et~al.}(2024{\natexlab{h}})\citenamefont {Zhang}, \citenamefont {Lin}, \citenamefont {Moreo}, \citenamefont {Maier},\ and\ \citenamefont {Dagotto}}]{zhang2023la3ni2o6}%
  \BibitemOpen
  \bibfield  {author} {\bibinfo {author} {\bibfnamefont {Y.}~\bibnamefont {Zhang}}, \bibinfo {author} {\bibfnamefont {L.-F.}\ \bibnamefont {Lin}}, \bibinfo {author} {\bibfnamefont {A.}~\bibnamefont {Moreo}}, \bibinfo {author} {\bibfnamefont {T.~A.}\ \bibnamefont {Maier}},\ and\ \bibinfo {author} {\bibfnamefont {E.}~\bibnamefont {Dagotto}},\ }\bibfield  {title} {\bibinfo {title} {Electronic structure, magnetic correlations, and superconducting pairing in the reduced ruddlesden-popper bilayer {L}a$_3${N}i$_2${O}$_6$ under pressure: Different role of $d_{3z^2-r^2}$ orbital compared with {L}a$_3${N}i$_2${O}$_7$},\ }\href {https://doi.org/10.1103/PhysRevB.109.045151} {\bibfield  {journal} {\bibinfo  {journal} {Phys. Rev. B}\ }\textbf {\bibinfo {volume} {109}},\ \bibinfo {pages} {045151} (\bibinfo {year} {2024}{\natexlab{h}})}\BibitemShut {NoStop}%
\bibitem [{\citenamefont {Xia}\ \emph {et~al.}(2025)\citenamefont {Xia}, \citenamefont {Liu}, \citenamefont {Zhou},\ and\ \citenamefont {Chen}}]{Xia2025}%
  \BibitemOpen
  \bibfield  {author} {\bibinfo {author} {\bibfnamefont {C.}~\bibnamefont {Xia}}, \bibinfo {author} {\bibfnamefont {H.}~\bibnamefont {Liu}}, \bibinfo {author} {\bibfnamefont {S.}~\bibnamefont {Zhou}},\ and\ \bibinfo {author} {\bibfnamefont {H.}~\bibnamefont {Chen}},\ }\bibfield  {title} {\bibinfo {title} {Sensitive dependence of pairing symmetry on {N}i-eg crystal field splitting in the nickelate superconductor {L}a$_3${N}i$_2${O}$_7$},\ }\href {https://doi.org/10.1038/s41467-025-56206-0} {\bibfield  {journal} {\bibinfo  {journal} {Nature Communications}\ }\textbf {\bibinfo {volume} {16}},\ \bibinfo {pages} {1054} (\bibinfo {year} {2025})}\BibitemShut {NoStop}%
\bibitem [{\citenamefont {B\"otzel}\ \emph {et~al.}(2024)\citenamefont {B\"otzel}, \citenamefont {Lechermann}, \citenamefont {Gondolf},\ and\ \citenamefont {Eremin}}]{botzel2024theory}%
  \BibitemOpen
  \bibfield  {author} {\bibinfo {author} {\bibfnamefont {S.}~\bibnamefont {B\"otzel}}, \bibinfo {author} {\bibfnamefont {F.}~\bibnamefont {Lechermann}}, \bibinfo {author} {\bibfnamefont {J.}~\bibnamefont {Gondolf}},\ and\ \bibinfo {author} {\bibfnamefont {I.~M.}\ \bibnamefont {Eremin}},\ }\bibfield  {title} {\bibinfo {title} {Theory of magnetic excitations in multilayer nickelate superconductor {L}a$_3${N}i$_2${O}$_7$},\ }\href {https://doi.org/10.1103/PhysRevB.109.L180502} {\bibfield  {journal} {\bibinfo  {journal} {Phys. Rev. B}\ }\textbf {\bibinfo {volume} {109}},\ \bibinfo {pages} {L180502} (\bibinfo {year} {2024})}\BibitemShut {NoStop}%
\bibitem [{\citenamefont {Liu}\ \emph {et~al.}(2025{\natexlab{b}})\citenamefont {Liu}, \citenamefont {Sun}, \citenamefont {Zhang}, \citenamefont {Liu}, \citenamefont {Chen},\ and\ \citenamefont {Yang}}]{Yubo_Liu2024}%
  \BibitemOpen
  \bibfield  {author} {\bibinfo {author} {\bibfnamefont {Y.-B.}\ \bibnamefont {Liu}}, \bibinfo {author} {\bibfnamefont {H.}~\bibnamefont {Sun}}, \bibinfo {author} {\bibfnamefont {M.}~\bibnamefont {Zhang}}, \bibinfo {author} {\bibfnamefont {Q.}~\bibnamefont {Liu}}, \bibinfo {author} {\bibfnamefont {W.-Q.}\ \bibnamefont {Chen}},\ and\ \bibinfo {author} {\bibfnamefont {F.}~\bibnamefont {Yang}},\ }\bibfield  {title} {\bibinfo {title} {Origin of the diagonal double-stripe spin density wave and potential superconductivity in bulk {L}a$_{3}${N}i$_{2}${O}$_{7}$ at ambient pressure},\ }\href {https://doi.org/10.1103/24f4-349n} {\bibfield  {journal} {\bibinfo  {journal} {Phys. Rev. B}\ }\textbf {\bibinfo {volume} {112}},\ \bibinfo {pages} {014510} (\bibinfo {year} {2025}{\natexlab{b}})}\BibitemShut {NoStop}%
\bibitem [{\citenamefont {Lin}\ \emph {et~al.}(2024)\citenamefont {Lin}, \citenamefont {Zhang}, \citenamefont {Kaushal}, \citenamefont {Alvarez}, \citenamefont {Maier}, \citenamefont {Moreo},\ and\ \citenamefont {Dagotto}}]{lin2024magnetic}%
  \BibitemOpen
  \bibfield  {author} {\bibinfo {author} {\bibfnamefont {L.-F.}\ \bibnamefont {Lin}}, \bibinfo {author} {\bibfnamefont {Y.}~\bibnamefont {Zhang}}, \bibinfo {author} {\bibfnamefont {N.}~\bibnamefont {Kaushal}}, \bibinfo {author} {\bibfnamefont {G.}~\bibnamefont {Alvarez}}, \bibinfo {author} {\bibfnamefont {T.~A.}\ \bibnamefont {Maier}}, \bibinfo {author} {\bibfnamefont {A.}~\bibnamefont {Moreo}},\ and\ \bibinfo {author} {\bibfnamefont {E.}~\bibnamefont {Dagotto}},\ }\bibfield  {title} {\bibinfo {title} {Magnetic phase diagram of a two-orbital model for bilayer nickelates with varying doping},\ }\href {https://doi.org/10.1103/PhysRevB.110.195135} {\bibfield  {journal} {\bibinfo  {journal} {Phys. Rev. B}\ }\textbf {\bibinfo {volume} {110}},\ \bibinfo {pages} {195135} (\bibinfo {year} {2024})}\BibitemShut {NoStop}%
\bibitem [{\citenamefont {Yang}\ \emph {et~al.}(2023{\natexlab{b}})\citenamefont {Yang}, \citenamefont {Wang},\ and\ \citenamefont {Wang}}]{WangQH2023}%
  \BibitemOpen
  \bibfield  {author} {\bibinfo {author} {\bibfnamefont {Q.-G.}\ \bibnamefont {Yang}}, \bibinfo {author} {\bibfnamefont {D.}~\bibnamefont {Wang}},\ and\ \bibinfo {author} {\bibfnamefont {Q.-H.}\ \bibnamefont {Wang}},\ }\bibfield  {title} {\bibinfo {title} {Possible $s_{\pm}$-wave superconductivity in {L}a$_3${N}i$_2${O}$_7$},\ }\href {https://doi.org/10.1103/PhysRevB.108.L140505} {\bibfield  {journal} {\bibinfo  {journal} {Phys. Rev. B}\ }\textbf {\bibinfo {volume} {108}},\ \bibinfo {pages} {L140505} (\bibinfo {year} {2023}{\natexlab{b}})}\BibitemShut {NoStop}%
\bibitem [{\citenamefont {Gu}\ \emph {et~al.}(2025)\citenamefont {Gu}, \citenamefont {Le}, \citenamefont {Yang}, \citenamefont {Wu},\ and\ \citenamefont {Hu}}]{HuJP2023}%
  \BibitemOpen
  \bibfield  {author} {\bibinfo {author} {\bibfnamefont {Y.}~\bibnamefont {Gu}}, \bibinfo {author} {\bibfnamefont {C.}~\bibnamefont {Le}}, \bibinfo {author} {\bibfnamefont {Z.}~\bibnamefont {Yang}}, \bibinfo {author} {\bibfnamefont {X.}~\bibnamefont {Wu}},\ and\ \bibinfo {author} {\bibfnamefont {J.}~\bibnamefont {Hu}},\ }\bibfield  {title} {\bibinfo {title} {Effective model and pairing tendency in the bilayer {N}i-based superconductor {L}a$_3${N}i$_2${O}$_7$},\ }\href {https://doi.org/10.1103/PhysRevB.111.174506} {\bibfield  {journal} {\bibinfo  {journal} {Phys. Rev. B}\ }\textbf {\bibinfo {volume} {111}},\ \bibinfo {pages} {174506} (\bibinfo {year} {2025})}\BibitemShut {NoStop}%
\bibitem [{\citenamefont {Jiang}\ \emph {et~al.}(2025)\citenamefont {Jiang}, \citenamefont {Cao}, \citenamefont {Yang}, \citenamefont {Lu},\ and\ \citenamefont {Wang}}]{PhysRevLett.134.076001}%
  \BibitemOpen
  \bibfield  {author} {\bibinfo {author} {\bibfnamefont {K.-Y.}\ \bibnamefont {Jiang}}, \bibinfo {author} {\bibfnamefont {Y.-H.}\ \bibnamefont {Cao}}, \bibinfo {author} {\bibfnamefont {Q.-G.}\ \bibnamefont {Yang}}, \bibinfo {author} {\bibfnamefont {H.-Y.}\ \bibnamefont {Lu}},\ and\ \bibinfo {author} {\bibfnamefont {Q.-H.}\ \bibnamefont {Wang}},\ }\bibfield  {title} {\bibinfo {title} {Theory of pressure dependence of superconductivity in bilayer nickelate {L}a$_{3}${N}i$_{2}${O}$_{7}$},\ }\href {https://doi.org/10.1103/PhysRevLett.134.076001} {\bibfield  {journal} {\bibinfo  {journal} {Phys. Rev. Lett.}\ }\textbf {\bibinfo {volume} {134}},\ \bibinfo {pages} {076001} (\bibinfo {year} {2025})}\BibitemShut {NoStop}%
\bibitem [{\citenamefont {Yang}\ \emph {et~al.}(2024{\natexlab{a}})\citenamefont {Yang}, \citenamefont {Jiang}, \citenamefont {Wang}, \citenamefont {Lu},\ and\ \citenamefont {Wang}}]{Yang2024effective}%
  \BibitemOpen
  \bibfield  {author} {\bibinfo {author} {\bibfnamefont {Q.-G.}\ \bibnamefont {Yang}}, \bibinfo {author} {\bibfnamefont {K.-Y.}\ \bibnamefont {Jiang}}, \bibinfo {author} {\bibfnamefont {D.}~\bibnamefont {Wang}}, \bibinfo {author} {\bibfnamefont {H.-Y.}\ \bibnamefont {Lu}},\ and\ \bibinfo {author} {\bibfnamefont {Q.-H.}\ \bibnamefont {Wang}},\ }\bibfield  {title} {\bibinfo {title} {Effective model and ${s}_{\ifmmode\pm\else\textpm\fi{}}$-wave superconductivity in trilayer nickelate {L}a$_4${N}i$_3${O}$_{10}$},\ }\href {https://doi.org/10.1103/PhysRevB.109.L220506} {\bibfield  {journal} {\bibinfo  {journal} {Phys. Rev. B}\ }\textbf {\bibinfo {volume} {109}},\ \bibinfo {pages} {L220506} (\bibinfo {year} {2024}{\natexlab{a}})}\BibitemShut {NoStop}%
\bibitem [{\citenamefont {Sakakibara}\ \emph {et~al.}(2024{\natexlab{a}})\citenamefont {Sakakibara}, \citenamefont {Kitamine}, \citenamefont {Ochi},\ and\ \citenamefont {Kuroki}}]{Kuroki2023}%
  \BibitemOpen
  \bibfield  {author} {\bibinfo {author} {\bibfnamefont {H.}~\bibnamefont {Sakakibara}}, \bibinfo {author} {\bibfnamefont {N.}~\bibnamefont {Kitamine}}, \bibinfo {author} {\bibfnamefont {M.}~\bibnamefont {Ochi}},\ and\ \bibinfo {author} {\bibfnamefont {K.}~\bibnamefont {Kuroki}},\ }\bibfield  {title} {\bibinfo {title} {Possible high ${T}_{c}$ superconductivity in {L}a$_3${N}i$_2${O}$_7$ under high pressure through manifestation of a nearly half-filled bilayer {H}ubbard model},\ }\href {https://doi.org/10.1103/PhysRevLett.132.106002} {\bibfield  {journal} {\bibinfo  {journal} {Phys. Rev. Lett.}\ }\textbf {\bibinfo {volume} {132}},\ \bibinfo {pages} {106002} (\bibinfo {year} {2024}{\natexlab{a}})}\BibitemShut {NoStop}%
\bibitem [{\citenamefont {Sakakibara}\ \emph {et~al.}(2024{\natexlab{b}})\citenamefont {Sakakibara}, \citenamefont {Ochi}, \citenamefont {Nagata}, \citenamefont {Ueki}, \citenamefont {Sakurai}, \citenamefont {Matsumoto}, \citenamefont {Terashima}, \citenamefont {Hirose}, \citenamefont {Ohta}, \citenamefont {Kato}, \citenamefont {Takano},\ and\ \citenamefont {Kuroki}}]{sakakibara2023La4Ni3O10}%
  \BibitemOpen
  \bibfield  {author} {\bibinfo {author} {\bibfnamefont {H.}~\bibnamefont {Sakakibara}}, \bibinfo {author} {\bibfnamefont {M.}~\bibnamefont {Ochi}}, \bibinfo {author} {\bibfnamefont {H.}~\bibnamefont {Nagata}}, \bibinfo {author} {\bibfnamefont {Y.}~\bibnamefont {Ueki}}, \bibinfo {author} {\bibfnamefont {H.}~\bibnamefont {Sakurai}}, \bibinfo {author} {\bibfnamefont {R.}~\bibnamefont {Matsumoto}}, \bibinfo {author} {\bibfnamefont {K.}~\bibnamefont {Terashima}}, \bibinfo {author} {\bibfnamefont {K.}~\bibnamefont {Hirose}}, \bibinfo {author} {\bibfnamefont {H.}~\bibnamefont {Ohta}}, \bibinfo {author} {\bibfnamefont {M.}~\bibnamefont {Kato}}, \bibinfo {author} {\bibfnamefont {Y.}~\bibnamefont {Takano}},\ and\ \bibinfo {author} {\bibfnamefont {K.}~\bibnamefont {Kuroki}},\ }\bibfield  {title} {\bibinfo {title} {Theoretical analysis on the possibility of superconductivity in the trilayer ruddlesden-popper nickelate {L}a$_4${N}i$_3${O}$_{10}$ under pressure and its experimental examination: Comparison with
  {L}a$_3${N}i$_2${O}$_7$},\ }\href {https://doi.org/10.1103/PhysRevB.109.144511} {\bibfield  {journal} {\bibinfo  {journal} {Phys. Rev. B}\ }\textbf {\bibinfo {volume} {109}},\ \bibinfo {pages} {144511} (\bibinfo {year} {2024}{\natexlab{b}})}\BibitemShut {NoStop}%
\bibitem [{\citenamefont {Heier}\ \emph {et~al.}(2024)\citenamefont {Heier}, \citenamefont {Park},\ and\ \citenamefont {Savrasov}}]{heier2023competing}%
  \BibitemOpen
  \bibfield  {author} {\bibinfo {author} {\bibfnamefont {G.}~\bibnamefont {Heier}}, \bibinfo {author} {\bibfnamefont {K.}~\bibnamefont {Park}},\ and\ \bibinfo {author} {\bibfnamefont {S.~Y.}\ \bibnamefont {Savrasov}},\ }\bibfield  {title} {\bibinfo {title} {Competing ${d}_{xy}$ and ${s}_{\pm}$ pairing symmetries in superconducting {L}a$_3${N}i$_2${O}$_7$: $\mathrm{LDA}+\mathrm{FLEX}$ calculations},\ }\href {https://doi.org/10.1103/PhysRevB.109.104508} {\bibfield  {journal} {\bibinfo  {journal} {Phys. Rev. B}\ }\textbf {\bibinfo {volume} {109}},\ \bibinfo {pages} {104508} (\bibinfo {year} {2024})}\BibitemShut {NoStop}%
\bibitem [{\citenamefont {Chen}\ \emph {et~al.}(2024{\natexlab{d}})\citenamefont {Chen}, \citenamefont {Tian}, \citenamefont {Wang}, \citenamefont {He},\ and\ \citenamefont {Lu}}]{PhysRevB.110.235119}%
  \BibitemOpen
  \bibfield  {author} {\bibinfo {author} {\bibfnamefont {Y.}~\bibnamefont {Chen}}, \bibinfo {author} {\bibfnamefont {Y.-H.}\ \bibnamefont {Tian}}, \bibinfo {author} {\bibfnamefont {J.-M.}\ \bibnamefont {Wang}}, \bibinfo {author} {\bibfnamefont {R.-Q.}\ \bibnamefont {He}},\ and\ \bibinfo {author} {\bibfnamefont {Z.-Y.}\ \bibnamefont {Lu}},\ }\bibfield  {title} {\bibinfo {title} {Non-fermi liquid and antiferromagnetic correlations with hole doping in the bilayer two-orbital hubbard model of {L}a$_{3}${N}i$_{2}${O}$_{7}$ at zero temperature},\ }\href {https://doi.org/10.1103/PhysRevB.110.235119} {\bibfield  {journal} {\bibinfo  {journal} {Phys. Rev. B}\ }\textbf {\bibinfo {volume} {110}},\ \bibinfo {pages} {235119} (\bibinfo {year} {2024}{\natexlab{d}})}\BibitemShut {NoStop}%
\bibitem [{\citenamefont {Leonov}(2024)}]{Leonov2024electronicc}%
  \BibitemOpen
  \bibfield  {author} {\bibinfo {author} {\bibfnamefont {I.~V.}\ \bibnamefont {Leonov}},\ }\bibfield  {title} {\bibinfo {title} {Electronic structure and magnetic correlations in the trilayer nickelate superconductor {L}a$_{4}${N}i$_{3}${O}$_{10}$ under pressure},\ }\href {https://doi.org/10.1103/PhysRevB.109.235123} {\bibfield  {journal} {\bibinfo  {journal} {Phys. Rev. B}\ }\textbf {\bibinfo {volume} {109}},\ \bibinfo {pages} {235123} (\bibinfo {year} {2024})}\BibitemShut {NoStop}%
\bibitem [{\citenamefont {Lechermann}\ \emph {et~al.}(2023)\citenamefont {Lechermann}, \citenamefont {Gondolf}, \citenamefont {B\"otzel},\ and\ \citenamefont {Eremin}}]{lechermann2023}%
  \BibitemOpen
  \bibfield  {author} {\bibinfo {author} {\bibfnamefont {F.}~\bibnamefont {Lechermann}}, \bibinfo {author} {\bibfnamefont {J.}~\bibnamefont {Gondolf}}, \bibinfo {author} {\bibfnamefont {S.}~\bibnamefont {B\"otzel}},\ and\ \bibinfo {author} {\bibfnamefont {I.~M.}\ \bibnamefont {Eremin}},\ }\bibfield  {title} {\bibinfo {title} {Electronic correlations and superconducting instability in {L}a$_3${N}i$_2${O}$_7$ under high pressure},\ }\href {https://doi.org/10.1103/PhysRevB.108.L201121} {\bibfield  {journal} {\bibinfo  {journal} {Phys. Rev. B}\ }\textbf {\bibinfo {volume} {108}},\ \bibinfo {pages} {L201121} (\bibinfo {year} {2023})}\BibitemShut {NoStop}%
\bibitem [{\citenamefont {Christiansson}\ \emph {et~al.}(2023)\citenamefont {Christiansson}, \citenamefont {Petocchi},\ and\ \citenamefont {Werner}}]{Werner2023}%
  \BibitemOpen
  \bibfield  {author} {\bibinfo {author} {\bibfnamefont {V.}~\bibnamefont {Christiansson}}, \bibinfo {author} {\bibfnamefont {F.}~\bibnamefont {Petocchi}},\ and\ \bibinfo {author} {\bibfnamefont {P.}~\bibnamefont {Werner}},\ }\bibfield  {title} {\bibinfo {title} {Correlated electronic structure of {L}a$_3${N}i$_2${O}$_7$ under pressure},\ }\href {https://doi.org/10.1103/PhysRevLett.131.206501} {\bibfield  {journal} {\bibinfo  {journal} {Phys. Rev. Lett.}\ }\textbf {\bibinfo {volume} {131}},\ \bibinfo {pages} {206501} (\bibinfo {year} {2023})}\BibitemShut {NoStop}%
\bibitem [{\citenamefont {Shilenko}\ and\ \citenamefont {Leonov}(2023)}]{shilenko2023correlated}%
  \BibitemOpen
  \bibfield  {author} {\bibinfo {author} {\bibfnamefont {D.~A.}\ \bibnamefont {Shilenko}}\ and\ \bibinfo {author} {\bibfnamefont {I.~V.}\ \bibnamefont {Leonov}},\ }\bibfield  {title} {\bibinfo {title} {Correlated electronic structure, orbital-selective behavior, and magnetic correlations in double-layer {L}a$_3${N}i$_2${O}$_7$ under pressure},\ }\href {https://doi.org/10.1103/PhysRevB.108.125105} {\bibfield  {journal} {\bibinfo  {journal} {Phys. Rev. B}\ }\textbf {\bibinfo {volume} {108}},\ \bibinfo {pages} {125105} (\bibinfo {year} {2023})}\BibitemShut {NoStop}%
\bibitem [{\citenamefont {W{\'u}}\ \emph {et~al.}(2024)\citenamefont {W{\'u}}, \citenamefont {Luo}, \citenamefont {Yao},\ and\ \citenamefont {Wang}}]{WuWei2023charge}%
  \BibitemOpen
  \bibfield  {author} {\bibinfo {author} {\bibfnamefont {W.}~\bibnamefont {W{\'u}}}, \bibinfo {author} {\bibfnamefont {Z.}~\bibnamefont {Luo}}, \bibinfo {author} {\bibfnamefont {D.-X.}\ \bibnamefont {Yao}},\ and\ \bibinfo {author} {\bibfnamefont {M.}~\bibnamefont {Wang}},\ }\bibfield  {title} {\bibinfo {title} {Superexchange and charge transfer in the nickelate superconductor {L}a$_3${N}i$_2${O}$_7$ under pressure},\ }\href {https://link.springer.com/article/10.1007/s11433-023-2300-4} {\bibfield  {journal} {\bibinfo  {journal} {Sci. China-Phys. Mech. Astron.}\ }\textbf {\bibinfo {volume} {67}},\ \bibinfo {pages} {117402} (\bibinfo {year} {2024})}\BibitemShut {NoStop}%
\bibitem [{\citenamefont {Cao}\ and\ \citenamefont {Yang}(2024)}]{cao2023flat}%
  \BibitemOpen
  \bibfield  {author} {\bibinfo {author} {\bibfnamefont {Y.}~\bibnamefont {Cao}}\ and\ \bibinfo {author} {\bibfnamefont {Y.-f.}\ \bibnamefont {Yang}},\ }\bibfield  {title} {\bibinfo {title} {Flat bands promoted by hund's rule coupling in the candidate double-layer high-temperature superconductor {L}a$_3${N}i$_2${O}$_7$ under high pressure},\ }\href {https://doi.org/10.1103/PhysRevB.109.L081105} {\bibfield  {journal} {\bibinfo  {journal} {Phys. Rev. B}\ }\textbf {\bibinfo {volume} {109}},\ \bibinfo {pages} {L081105} (\bibinfo {year} {2024})}\BibitemShut {NoStop}%
\bibitem [{\citenamefont {Wang}\ \emph {et~al.}(2024{\natexlab{e}})\citenamefont {Wang}, \citenamefont {Ouyang}, \citenamefont {He},\ and\ \citenamefont {Lu}}]{wang2024non}%
  \BibitemOpen
  \bibfield  {author} {\bibinfo {author} {\bibfnamefont {J.-X.}\ \bibnamefont {Wang}}, \bibinfo {author} {\bibfnamefont {Z.}~\bibnamefont {Ouyang}}, \bibinfo {author} {\bibfnamefont {R.-Q.}\ \bibnamefont {He}},\ and\ \bibinfo {author} {\bibfnamefont {Z.-Y.}\ \bibnamefont {Lu}},\ }\bibfield  {title} {\bibinfo {title} {Non-fermi liquid and hund correlation in {L}a$_{4}${N}i$_{3}${O}$_{10}$ under high pressure},\ }\href {https://doi.org/10.1103/PhysRevB.109.165140} {\bibfield  {journal} {\bibinfo  {journal} {Phys. Rev. B}\ }\textbf {\bibinfo {volume} {109}},\ \bibinfo {pages} {165140} (\bibinfo {year} {2024}{\natexlab{e}})}\BibitemShut {NoStop}%
\bibitem [{\citenamefont {Ryee}\ \emph {et~al.}(2024)\citenamefont {Ryee}, \citenamefont {Witt},\ and\ \citenamefont {Wehling}}]{ryee2024quenched}%
  \BibitemOpen
  \bibfield  {author} {\bibinfo {author} {\bibfnamefont {S.}~\bibnamefont {Ryee}}, \bibinfo {author} {\bibfnamefont {N.}~\bibnamefont {Witt}},\ and\ \bibinfo {author} {\bibfnamefont {T.~O.}\ \bibnamefont {Wehling}},\ }\bibfield  {title} {\bibinfo {title} {Quenched pair breaking by interlayer correlations as a key to superconductivity in {L}a$_{3}${N}i$_{2}${O}$_{7}$},\ }\href {https://doi.org/10.1103/PhysRevLett.133.096002} {\bibfield  {journal} {\bibinfo  {journal} {Phys. Rev. Lett.}\ }\textbf {\bibinfo {volume} {133}},\ \bibinfo {pages} {096002} (\bibinfo {year} {2024})}\BibitemShut {NoStop}%
\bibitem [{\citenamefont {Kumar}\ \emph {et~al.}(2025)\citenamefont {Kumar}, \citenamefont {Melnick},\ and\ \citenamefont {Kotliar}}]{PhysRevResearch.7.L012066}%
  \BibitemOpen
  \bibfield  {author} {\bibinfo {author} {\bibfnamefont {U.}~\bibnamefont {Kumar}}, \bibinfo {author} {\bibfnamefont {C.}~\bibnamefont {Melnick}},\ and\ \bibinfo {author} {\bibfnamefont {G.}~\bibnamefont {Kotliar}},\ }\bibfield  {title} {\bibinfo {title} {Softening of $\mathit{dd}$ excitation in the resonant inelastic x-ray scattering spectra as a signature of hund's coupling in nickelates},\ }\href {https://doi.org/10.1103/PhysRevResearch.7.L012066} {\bibfield  {journal} {\bibinfo  {journal} {Phys. Rev. Res.}\ }\textbf {\bibinfo {volume} {7}},\ \bibinfo {pages} {L012066} (\bibinfo {year} {2025})}\BibitemShut {NoStop}%
\bibitem [{\citenamefont {Ouyang}\ \emph {et~al.}(2024{\natexlab{b}})\citenamefont {Ouyang}, \citenamefont {Wang}, \citenamefont {Wang}, \citenamefont {He}, \citenamefont {Huang},\ and\ \citenamefont {Lu}}]{ouyang2023hund}%
  \BibitemOpen
  \bibfield  {author} {\bibinfo {author} {\bibfnamefont {Z.}~\bibnamefont {Ouyang}}, \bibinfo {author} {\bibfnamefont {J.-M.}\ \bibnamefont {Wang}}, \bibinfo {author} {\bibfnamefont {J.-X.}\ \bibnamefont {Wang}}, \bibinfo {author} {\bibfnamefont {R.-Q.}\ \bibnamefont {He}}, \bibinfo {author} {\bibfnamefont {L.}~\bibnamefont {Huang}},\ and\ \bibinfo {author} {\bibfnamefont {Z.-Y.}\ \bibnamefont {Lu}},\ }\bibfield  {title} {\bibinfo {title} {Hund electronic correlation in {L}a$_3${N}i$_2${O}$_7$ under high pressure},\ }\href {https://doi.org/10.1103/PhysRevB.109.115114} {\bibfield  {journal} {\bibinfo  {journal} {Phys. Rev. B}\ }\textbf {\bibinfo {volume} {109}},\ \bibinfo {pages} {115114} (\bibinfo {year} {2024}{\natexlab{b}})}\BibitemShut {NoStop}%
\bibitem [{\citenamefont {Tian}\ \emph {et~al.}(2024)\citenamefont {Tian}, \citenamefont {Chen}, \citenamefont {Wang}, \citenamefont {He},\ and\ \citenamefont {Lu}}]{tian2023correlation}%
  \BibitemOpen
  \bibfield  {author} {\bibinfo {author} {\bibfnamefont {Y.-H.}\ \bibnamefont {Tian}}, \bibinfo {author} {\bibfnamefont {Y.}~\bibnamefont {Chen}}, \bibinfo {author} {\bibfnamefont {J.-M.}\ \bibnamefont {Wang}}, \bibinfo {author} {\bibfnamefont {R.-Q.}\ \bibnamefont {He}},\ and\ \bibinfo {author} {\bibfnamefont {Z.-Y.}\ \bibnamefont {Lu}},\ }\bibfield  {title} {\bibinfo {title} {Correlation effects and concomitant two-orbital ${s}_{\pm}$-wave superconductivity in {L}a$_3${N}i$_2${O}$_7$ under high pressure},\ }\href {https://doi.org/10.1103/PhysRevB.109.165154} {\bibfield  {journal} {\bibinfo  {journal} {Phys. Rev. B}\ }\textbf {\bibinfo {volume} {109}},\ \bibinfo {pages} {165154} (\bibinfo {year} {2024})}\BibitemShut {NoStop}%
\bibitem [{\citenamefont {Zheng}\ and\ \citenamefont {W\'u}(2025)}]{PhysRevB.111.035108}%
  \BibitemOpen
  \bibfield  {author} {\bibinfo {author} {\bibfnamefont {Y.-Y.}\ \bibnamefont {Zheng}}\ and\ \bibinfo {author} {\bibfnamefont {W.}~\bibnamefont {W\'u}},\ }\bibfield  {title} {\bibinfo {title} {${s}_{\ifmmode\pm\else\textpm\fi{}}$-wave superconductivity in the bilayer two-orbital hubbard model},\ }\href {https://doi.org/10.1103/PhysRevB.111.035108} {\bibfield  {journal} {\bibinfo  {journal} {Phys. Rev. B}\ }\textbf {\bibinfo {volume} {111}},\ \bibinfo {pages} {035108} (\bibinfo {year} {2025})}\BibitemShut {NoStop}%
\bibitem [{\citenamefont {Qin}\ and\ \citenamefont {Yang}(2023)}]{qin2023high}%
  \BibitemOpen
  \bibfield  {author} {\bibinfo {author} {\bibfnamefont {Q.}~\bibnamefont {Qin}}\ and\ \bibinfo {author} {\bibfnamefont {Y.-F.}\ \bibnamefont {Yang}},\ }\bibfield  {title} {\bibinfo {title} {High-${T}_{c}$ superconductivity by mobilizing local spin singlets and possible route to higher ${T}_{c}$ in pressurized {L}a$_3${N}i$_2${O}$_7$},\ }\href {https://doi.org/10.1103/PhysRevB.108.L140504} {\bibfield  {journal} {\bibinfo  {journal} {Phys. Rev. B}\ }\textbf {\bibinfo {volume} {108}},\ \bibinfo {pages} {L140504} (\bibinfo {year} {2023})}\BibitemShut {NoStop}%
\bibitem [{\citenamefont {Chang}\ \emph {et~al.}(2023)\citenamefont {Chang}, \citenamefont {Guo}, \citenamefont {You},\ and\ \citenamefont {Li}}]{chang2023fermisurfacesymmetricmass}%
  \BibitemOpen
  \bibfield  {author} {\bibinfo {author} {\bibfnamefont {W.-X.}\ \bibnamefont {Chang}}, \bibinfo {author} {\bibfnamefont {S.}~\bibnamefont {Guo}}, \bibinfo {author} {\bibfnamefont {Y.-Z.}\ \bibnamefont {You}},\ and\ \bibinfo {author} {\bibfnamefont {Z.-X.}\ \bibnamefont {Li}},\ }\bibfield  {title} {\bibinfo {title} {Fermi surface symmetric mass generation: a quantum monte-carlo study},\ }\href {https://arxiv.org/abs/2311.09970} {\bibfield  {journal} {\bibinfo  {journal} {arXiv:2311.09970}\ } (\bibinfo {year} {2023})}\BibitemShut {NoStop}%
\bibitem [{\citenamefont {Qu}\ \emph {et~al.}(2024)\citenamefont {Qu}, \citenamefont {Qu}, \citenamefont {Chen}, \citenamefont {Wu}, \citenamefont {Yang}, \citenamefont {Li},\ and\ \citenamefont {Su}}]{qu2023bilayer}%
  \BibitemOpen
  \bibfield  {author} {\bibinfo {author} {\bibfnamefont {X.-Z.}\ \bibnamefont {Qu}}, \bibinfo {author} {\bibfnamefont {D.-W.}\ \bibnamefont {Qu}}, \bibinfo {author} {\bibfnamefont {J.}~\bibnamefont {Chen}}, \bibinfo {author} {\bibfnamefont {C.}~\bibnamefont {Wu}}, \bibinfo {author} {\bibfnamefont {F.}~\bibnamefont {Yang}}, \bibinfo {author} {\bibfnamefont {W.}~\bibnamefont {Li}},\ and\ \bibinfo {author} {\bibfnamefont {G.}~\bibnamefont {Su}},\ }\bibfield  {title} {\bibinfo {title} {Bilayer $t$-${J}$-${J}_{\perp}$ model and magnetically mediated pairing in the pressurized nickelate {L}a$_3${N}i$_2${O}$_7$},\ }\href {https://doi.org/10.1103/PhysRevLett.132.036502} {\bibfield  {journal} {\bibinfo  {journal} {Phys. Rev. Lett.}\ }\textbf {\bibinfo {volume} {132}},\ \bibinfo {pages} {036502} (\bibinfo {year} {2024})}\BibitemShut {NoStop}%
\bibitem [{\citenamefont {Shen}\ \emph {et~al.}(2023)\citenamefont {Shen}, \citenamefont {Qin},\ and\ \citenamefont {Zhang}}]{ZhangGM2023DMRG}%
  \BibitemOpen
  \bibfield  {author} {\bibinfo {author} {\bibfnamefont {Y.}~\bibnamefont {Shen}}, \bibinfo {author} {\bibfnamefont {M.}~\bibnamefont {Qin}},\ and\ \bibinfo {author} {\bibfnamefont {G.-M.}\ \bibnamefont {Zhang}},\ }\bibfield  {title} {\bibinfo {title} {Effective bi-layer model hamiltonian and density-matrix renormalization group study for the high-${T}_c$ superconductivity {L}a$_3${N}i$_2${O}$_7$ under high pressure},\ }\href {https://iopscience.iop.org/article/10.1088/0256-307X/40/12/127401} {\bibfield  {journal} {\bibinfo  {journal} {Chin. Phys. Lett.}\ }\textbf {\bibinfo {volume} {40}},\ \bibinfo {pages} {127401} (\bibinfo {year} {2023})}\BibitemShut {NoStop}%
\bibitem [{\citenamefont {Zhang}\ \emph {et~al.}(2024{\natexlab{i}})\citenamefont {Zhang}, \citenamefont {Zhang}, \citenamefont {You},\ and\ \citenamefont {Weng}}]{zhang2023strong}%
  \BibitemOpen
  \bibfield  {author} {\bibinfo {author} {\bibfnamefont {J.-X.}\ \bibnamefont {Zhang}}, \bibinfo {author} {\bibfnamefont {H.-K.}\ \bibnamefont {Zhang}}, \bibinfo {author} {\bibfnamefont {Y.-Z.}\ \bibnamefont {You}},\ and\ \bibinfo {author} {\bibfnamefont {Z.-Y.}\ \bibnamefont {Weng}},\ }\bibfield  {title} {\bibinfo {title} {Strong pairing originated from an emergent $\mathbb{Z}_2$ berry phase in {L}a$_3${N}i$_2${O}$_7$},\ }\href {https://doi.org/10.1103/PhysRevLett.133.126501} {\bibfield  {journal} {\bibinfo  {journal} {Phys. Rev. Lett.}\ }\textbf {\bibinfo {volume} {133}},\ \bibinfo {pages} {126501} (\bibinfo {year} {2024}{\natexlab{i}})}\BibitemShut {NoStop}%
\bibitem [{\citenamefont {Lange}\ \emph {et~al.}(2024{\natexlab{a}})\citenamefont {Lange}, \citenamefont {Homeier}, \citenamefont {Demler}, \citenamefont {Schollw\"ock}, \citenamefont {Bohrdt},\ and\ \citenamefont {Grusdt}}]{lange2023mixedtj}%
  \BibitemOpen
  \bibfield  {author} {\bibinfo {author} {\bibfnamefont {H.}~\bibnamefont {Lange}}, \bibinfo {author} {\bibfnamefont {L.}~\bibnamefont {Homeier}}, \bibinfo {author} {\bibfnamefont {E.}~\bibnamefont {Demler}}, \bibinfo {author} {\bibfnamefont {U.}~\bibnamefont {Schollw\"ock}}, \bibinfo {author} {\bibfnamefont {A.}~\bibnamefont {Bohrdt}},\ and\ \bibinfo {author} {\bibfnamefont {F.}~\bibnamefont {Grusdt}},\ }\bibfield  {title} {\bibinfo {title} {Pairing dome from an emergent feshbach resonance in a strongly repulsive bilayer model},\ }\href {https://doi.org/10.1103/PhysRevB.110.L081113} {\bibfield  {journal} {\bibinfo  {journal} {Phys. Rev. B}\ }\textbf {\bibinfo {volume} {110}},\ \bibinfo {pages} {L081113} (\bibinfo {year} {2024}{\natexlab{a}})}\BibitemShut {NoStop}%
\bibitem [{\citenamefont {Yang}\ \emph {et~al.}(2024{\natexlab{b}})\citenamefont {Yang}, \citenamefont {Oh},\ and\ \citenamefont {Zhang}}]{PhysRevB.110.104517}%
  \BibitemOpen
  \bibfield  {author} {\bibinfo {author} {\bibfnamefont {H.}~\bibnamefont {Yang}}, \bibinfo {author} {\bibfnamefont {H.}~\bibnamefont {Oh}},\ and\ \bibinfo {author} {\bibfnamefont {Y.-H.}\ \bibnamefont {Zhang}},\ }\bibfield  {title} {\bibinfo {title} {Strong pairing from a small fermi surface beyond weak coupling: Application to {L}a$_3${N}i$_2${O}$_7$},\ }\href {https://doi.org/10.1103/PhysRevB.110.104517} {\bibfield  {journal} {\bibinfo  {journal} {Phys. Rev. B}\ }\textbf {\bibinfo {volume} {110}},\ \bibinfo {pages} {104517} (\bibinfo {year} {2024}{\natexlab{b}})}\BibitemShut {NoStop}%
\bibitem [{\citenamefont {Lange}\ \emph {et~al.}(2024{\natexlab{b}})\citenamefont {Lange}, \citenamefont {Homeier}, \citenamefont {Demler}, \citenamefont {Schollw\"ock}, \citenamefont {Grusdt},\ and\ \citenamefont {Bohrdt}}]{lange2023feshbach}%
  \BibitemOpen
  \bibfield  {author} {\bibinfo {author} {\bibfnamefont {H.}~\bibnamefont {Lange}}, \bibinfo {author} {\bibfnamefont {L.}~\bibnamefont {Homeier}}, \bibinfo {author} {\bibfnamefont {E.}~\bibnamefont {Demler}}, \bibinfo {author} {\bibfnamefont {U.}~\bibnamefont {Schollw\"ock}}, \bibinfo {author} {\bibfnamefont {F.}~\bibnamefont {Grusdt}},\ and\ \bibinfo {author} {\bibfnamefont {A.}~\bibnamefont {Bohrdt}},\ }\bibfield  {title} {\bibinfo {title} {Feshbach resonance in a strongly repulsive ladder of mixed dimensionality: A possible scenario for bilayer nickelate superconductors},\ }\href {https://doi.org/10.1103/PhysRevB.109.045127} {\bibfield  {journal} {\bibinfo  {journal} {Phys. Rev. B}\ }\textbf {\bibinfo {volume} {109}},\ \bibinfo {pages} {045127} (\bibinfo {year} {2024}{\natexlab{b}})}\BibitemShut {NoStop}%
\bibitem [{\citenamefont {Kaneko}\ \emph {et~al.}(2024)\citenamefont {Kaneko}, \citenamefont {Sakakibara}, \citenamefont {Ochi},\ and\ \citenamefont {Kuroki}}]{kaneko2023pair}%
  \BibitemOpen
  \bibfield  {author} {\bibinfo {author} {\bibfnamefont {T.}~\bibnamefont {Kaneko}}, \bibinfo {author} {\bibfnamefont {H.}~\bibnamefont {Sakakibara}}, \bibinfo {author} {\bibfnamefont {M.}~\bibnamefont {Ochi}},\ and\ \bibinfo {author} {\bibfnamefont {K.}~\bibnamefont {Kuroki}},\ }\bibfield  {title} {\bibinfo {title} {Pair correlations in the two-orbital hubbard ladder: Implications for superconductivity in the bilayer nickelate {L}a$_3${N}i$_2${O}$_7$},\ }\href {https://doi.org/10.1103/PhysRevB.109.045154} {\bibfield  {journal} {\bibinfo  {journal} {Phys. Rev. B}\ }\textbf {\bibinfo {volume} {109}},\ \bibinfo {pages} {045154} (\bibinfo {year} {2024})}\BibitemShut {NoStop}%
\bibitem [{\citenamefont {Schlömer}\ \emph {et~al.}(2024)\citenamefont {Schlömer}, \citenamefont {Schollwöck}, \citenamefont {Grusdt},\ and\ \citenamefont {Bohrdt}}]{Grusdt2023lno03349}%
  \BibitemOpen
  \bibfield  {author} {\bibinfo {author} {\bibfnamefont {H.}~\bibnamefont {Schlömer}}, \bibinfo {author} {\bibfnamefont {U.}~\bibnamefont {Schollwöck}}, \bibinfo {author} {\bibfnamefont {F.}~\bibnamefont {Grusdt}},\ and\ \bibinfo {author} {\bibfnamefont {A.}~\bibnamefont {Bohrdt}},\ }\bibfield  {title} {\bibinfo {title} {Superconductivity in the pressurized nickelate {L}a$_3${N}i$_2${O}$_7$ in the vicinity of a {BEC}-{BCS} crossover},\ }\href {https://doi.org/10.1038/s42005-024-01854-9} {\bibfield  {journal} {\bibinfo  {journal} {Communications Physics}\ }\textbf {\bibinfo {volume} {7}},\ \bibinfo {pages} {366} (\bibinfo {year} {2024})}\BibitemShut {NoStop}%
\bibitem [{\citenamefont {Qu}\ \emph {et~al.}(2023)\citenamefont {Qu}, \citenamefont {Qu}, \citenamefont {Li},\ and\ \citenamefont {Su}}]{qu2023roles}%
  \BibitemOpen
  \bibfield  {author} {\bibinfo {author} {\bibfnamefont {X.-Z.}\ \bibnamefont {Qu}}, \bibinfo {author} {\bibfnamefont {D.-W.}\ \bibnamefont {Qu}}, \bibinfo {author} {\bibfnamefont {W.}~\bibnamefont {Li}},\ and\ \bibinfo {author} {\bibfnamefont {G.}~\bibnamefont {Su}},\ }\bibfield  {title} {\bibinfo {title} {Roles of hund's rule and hybridization in the two-orbital model for high-${T}_c$ superconductivity in the bilayer nickelate},\ }\href {https://arxiv.org/abs/2311.12769} {\bibfield  {journal} {\bibinfo  {journal} {arXiv:2311.12769}\ } (\bibinfo {year} {2023})}\BibitemShut {NoStop}%
\bibitem [{\citenamefont {Kakoi}\ \emph {et~al.}(2024{\natexlab{b}})\citenamefont {Kakoi}, \citenamefont {Kaneko}, \citenamefont {Sakakibara}, \citenamefont {Ochi},\ and\ \citenamefont {Kuroki}}]{kakoi2023pair}%
  \BibitemOpen
  \bibfield  {author} {\bibinfo {author} {\bibfnamefont {M.}~\bibnamefont {Kakoi}}, \bibinfo {author} {\bibfnamefont {T.}~\bibnamefont {Kaneko}}, \bibinfo {author} {\bibfnamefont {H.}~\bibnamefont {Sakakibara}}, \bibinfo {author} {\bibfnamefont {M.}~\bibnamefont {Ochi}},\ and\ \bibinfo {author} {\bibfnamefont {K.}~\bibnamefont {Kuroki}},\ }\bibfield  {title} {\bibinfo {title} {Pair correlations of the hybridized orbitals in a ladder model for the bilayer nickelate {L}a$_3${N}i$_2${O}$_7$},\ }\href {https://doi.org/10.1103/PhysRevB.109.L201124} {\bibfield  {journal} {\bibinfo  {journal} {Phys. Rev. B}\ }\textbf {\bibinfo {volume} {109}},\ \bibinfo {pages} {L201124} (\bibinfo {year} {2024}{\natexlab{b}})}\BibitemShut {NoStop}%
\bibitem [{\citenamefont {Ji}\ \emph {et~al.}(2025)\citenamefont {Ji}, \citenamefont {Lu}, \citenamefont {Shao}, \citenamefont {Pan}, \citenamefont {Yang},\ and\ \citenamefont {Wu}}]{ji2025}%
  \BibitemOpen
  \bibfield  {author} {\bibinfo {author} {\bibfnamefont {J.-H.}\ \bibnamefont {Ji}}, \bibinfo {author} {\bibfnamefont {C.}~\bibnamefont {Lu}}, \bibinfo {author} {\bibfnamefont {Z.-Y.}\ \bibnamefont {Shao}}, \bibinfo {author} {\bibfnamefont {Z.}~\bibnamefont {Pan}}, \bibinfo {author} {\bibfnamefont {F.}~\bibnamefont {Yang}},\ and\ \bibinfo {author} {\bibfnamefont {C.}~\bibnamefont {Wu}},\ }\bibfield  {title} {\bibinfo {title} {A strong-coupling-limit study on the pairing mechanism in the pressurized {L}a$_3${N}i$_2${O}$_7$},\ }\href {https://arxiv.org/abs/2504.12127} {\bibfield  {journal} {\bibinfo  {journal} {arXiv:2504.12127}\ } (\bibinfo {year} {2025})}\BibitemShut {NoStop}%
\bibitem [{\citenamefont {Chen}\ \emph {et~al.}(2024{\natexlab{e}})\citenamefont {Chen}, \citenamefont {Yang},\ and\ \citenamefont {Li}}]{chen2023iPEPS}%
  \BibitemOpen
  \bibfield  {author} {\bibinfo {author} {\bibfnamefont {J.}~\bibnamefont {Chen}}, \bibinfo {author} {\bibfnamefont {F.}~\bibnamefont {Yang}},\ and\ \bibinfo {author} {\bibfnamefont {W.}~\bibnamefont {Li}},\ }\bibfield  {title} {\bibinfo {title} {Orbital-selective superconductivity in the pressurized bilayer nickelate {L}a$_3${N}i$_2${O}$_7$: An infinite projected entangled-pair state study},\ }\href {https://doi.org/10.1103/PhysRevB.110.L041111} {\bibfield  {journal} {\bibinfo  {journal} {Phys. Rev. B}\ }\textbf {\bibinfo {volume} {110}},\ \bibinfo {pages} {L041111} (\bibinfo {year} {2024}{\natexlab{e}})}\BibitemShut {NoStop}%
\bibitem [{\citenamefont {Yang}\ \emph {et~al.}(2024{\natexlab{c}})\citenamefont {Yang}, \citenamefont {Sun}, \citenamefont {Hu}, \citenamefont {Xie}, \citenamefont {Miao}, \citenamefont {Luo}, \citenamefont {Chen}, \citenamefont {Liang}, \citenamefont {Zhu}, \citenamefont {Qu} \emph {et~al.}}]{yang2024orbital}%
  \BibitemOpen
  \bibfield  {author} {\bibinfo {author} {\bibfnamefont {J.}~\bibnamefont {Yang}}, \bibinfo {author} {\bibfnamefont {H.}~\bibnamefont {Sun}}, \bibinfo {author} {\bibfnamefont {X.}~\bibnamefont {Hu}}, \bibinfo {author} {\bibfnamefont {Y.}~\bibnamefont {Xie}}, \bibinfo {author} {\bibfnamefont {T.}~\bibnamefont {Miao}}, \bibinfo {author} {\bibfnamefont {H.}~\bibnamefont {Luo}}, \bibinfo {author} {\bibfnamefont {H.}~\bibnamefont {Chen}}, \bibinfo {author} {\bibfnamefont {B.}~\bibnamefont {Liang}}, \bibinfo {author} {\bibfnamefont {W.}~\bibnamefont {Zhu}}, \bibinfo {author} {\bibfnamefont {G.}~\bibnamefont {Qu}}, \emph {et~al.},\ }\bibfield  {title} {\bibinfo {title} {Orbital-dependent electron correlation in double-layer nickelate {L}a$_3${N}i$_2${O}$_7$},\ }\href {https://www.nature.com/articles/s41467-024-48701-7} {\bibfield  {journal} {\bibinfo  {journal} {Nat. Commun.}\ }\textbf {\bibinfo {volume} {15}},\ \bibinfo {pages} {4373} (\bibinfo {year} {2024}{\natexlab{c}})}\BibitemShut {NoStop}%
\bibitem [{\citenamefont {Li}\ \emph {et~al.}(2024{\natexlab{d}})\citenamefont {Li}, \citenamefont {Du}, \citenamefont {Cao}, \citenamefont {Pei}, \citenamefont {Zhang}, \citenamefont {Zhao}, \citenamefont {Zhai}, \citenamefont {Xu}, \citenamefont {Liu}, \citenamefont {Li}, \citenamefont {Zhao}, \citenamefont {Li}, \citenamefont {Qi}, \citenamefont {Guo}, \citenamefont {Chen},\ and\ \citenamefont {Yang}}]{Li2024ele}%
  \BibitemOpen
  \bibfield  {author} {\bibinfo {author} {\bibfnamefont {Y.}~\bibnamefont {Li}}, \bibinfo {author} {\bibfnamefont {X.}~\bibnamefont {Du}}, \bibinfo {author} {\bibfnamefont {Y.}~\bibnamefont {Cao}}, \bibinfo {author} {\bibfnamefont {C.}~\bibnamefont {Pei}}, \bibinfo {author} {\bibfnamefont {M.}~\bibnamefont {Zhang}}, \bibinfo {author} {\bibfnamefont {W.}~\bibnamefont {Zhao}}, \bibinfo {author} {\bibfnamefont {K.}~\bibnamefont {Zhai}}, \bibinfo {author} {\bibfnamefont {R.}~\bibnamefont {Xu}}, \bibinfo {author} {\bibfnamefont {Z.}~\bibnamefont {Liu}}, \bibinfo {author} {\bibfnamefont {Z.}~\bibnamefont {Li}}, \bibinfo {author} {\bibfnamefont {J.}~\bibnamefont {Zhao}}, \bibinfo {author} {\bibfnamefont {G.}~\bibnamefont {Li}}, \bibinfo {author} {\bibfnamefont {Y.}~\bibnamefont {Qi}}, \bibinfo {author} {\bibfnamefont {H.}~\bibnamefont {Guo}}, \bibinfo {author} {\bibfnamefont {Y.}~\bibnamefont {Chen}},\ and\ \bibinfo {author} {\bibfnamefont {L.}~\bibnamefont {Yang}},\ }\bibfield  {title} {\bibinfo {title} {Electronic
  correlation and pseudogap-like behavior of high-temperature superconductor {L}a$_3${N}i$_2${O}$_7$},\ }\href {https://doi.org/10.1088/0256-307X/41/8/087402} {\bibfield  {journal} {\bibinfo  {journal} {Chin. Phys. Lett.}\ }\textbf {\bibinfo {volume} {41}},\ \bibinfo {pages} {087402} (\bibinfo {year} {2024}{\natexlab{d}})}\BibitemShut {NoStop}%
\bibitem [{\citenamefont {Liu}\ \emph {et~al.}(2024)\citenamefont {Liu}, \citenamefont {Huo}, \citenamefont {Li}, \citenamefont {Li}, \citenamefont {Liu}, \citenamefont {Dai}, \citenamefont {Zhou}, \citenamefont {Hao}, \citenamefont {Lu}, \citenamefont {Wang},\ and\ \citenamefont {Wen}}]{liu2024electronic}%
  \BibitemOpen
  \bibfield  {author} {\bibinfo {author} {\bibfnamefont {Z.}~\bibnamefont {Liu}}, \bibinfo {author} {\bibfnamefont {M.}~\bibnamefont {Huo}}, \bibinfo {author} {\bibfnamefont {J.}~\bibnamefont {Li}}, \bibinfo {author} {\bibfnamefont {Q.}~\bibnamefont {Li}}, \bibinfo {author} {\bibfnamefont {Y.}~\bibnamefont {Liu}}, \bibinfo {author} {\bibfnamefont {Y.}~\bibnamefont {Dai}}, \bibinfo {author} {\bibfnamefont {X.}~\bibnamefont {Zhou}}, \bibinfo {author} {\bibfnamefont {J.}~\bibnamefont {Hao}}, \bibinfo {author} {\bibfnamefont {Y.}~\bibnamefont {Lu}}, \bibinfo {author} {\bibfnamefont {M.}~\bibnamefont {Wang}},\ and\ \bibinfo {author} {\bibfnamefont {H.-H.}\ \bibnamefont {Wen}},\ }\bibfield  {title} {\bibinfo {title} {Electronic correlations and partial gap in the bilayer nickelate {L}a$_3${N}i$_2${O}$_7$},\ }\href {https://doi.org/10.1038/s41467-024-52001-5} {\bibfield  {journal} {\bibinfo  {journal} {Nat. Commun.}\ }\textbf {\bibinfo {volume} {15}},\ \bibinfo {pages} {7570} (\bibinfo {year} {2024})}\BibitemShut
  {NoStop}%
\bibitem [{\citenamefont {Li}\ \emph {et~al.}(2025{\natexlab{c}})\citenamefont {Li}, \citenamefont {Zhou}, \citenamefont {Lv}, \citenamefont {Li}, \citenamefont {Yue}, \citenamefont {Huang}, \citenamefont {Xu}, \citenamefont {Shen}, \citenamefont {Miao}, \citenamefont {Song}, \citenamefont {Nie}, \citenamefont {Chen}, \citenamefont {Wang}, \citenamefont {Chen}, \citenamefont {Huang}, \citenamefont {Chen}, \citenamefont {Qian}, \citenamefont {Lin}, \citenamefont {He}, \citenamefont {Sun}, \citenamefont {Chen},\ and\ \citenamefont {Xue}}]{10.1093}%
  \BibitemOpen
  \bibfield  {author} {\bibinfo {author} {\bibfnamefont {P.}~\bibnamefont {Li}}, \bibinfo {author} {\bibfnamefont {G.}~\bibnamefont {Zhou}}, \bibinfo {author} {\bibfnamefont {W.}~\bibnamefont {Lv}}, \bibinfo {author} {\bibfnamefont {Y.}~\bibnamefont {Li}}, \bibinfo {author} {\bibfnamefont {C.}~\bibnamefont {Yue}}, \bibinfo {author} {\bibfnamefont {H.}~\bibnamefont {Huang}}, \bibinfo {author} {\bibfnamefont {L.}~\bibnamefont {Xu}}, \bibinfo {author} {\bibfnamefont {J.}~\bibnamefont {Shen}}, \bibinfo {author} {\bibfnamefont {Y.}~\bibnamefont {Miao}}, \bibinfo {author} {\bibfnamefont {W.}~\bibnamefont {Song}}, \bibinfo {author} {\bibfnamefont {Z.}~\bibnamefont {Nie}}, \bibinfo {author} {\bibfnamefont {Y.}~\bibnamefont {Chen}}, \bibinfo {author} {\bibfnamefont {H.}~\bibnamefont {Wang}}, \bibinfo {author} {\bibfnamefont {W.}~\bibnamefont {Chen}}, \bibinfo {author} {\bibfnamefont {Y.}~\bibnamefont {Huang}}, \bibinfo {author} {\bibfnamefont {Z.-H.}\ \bibnamefont {Chen}}, \bibinfo {author} {\bibfnamefont
  {T.}~\bibnamefont {Qian}}, \bibinfo {author} {\bibfnamefont {J.}~\bibnamefont {Lin}}, \bibinfo {author} {\bibfnamefont {J.}~\bibnamefont {He}}, \bibinfo {author} {\bibfnamefont {Y.-J.}\ \bibnamefont {Sun}}, \bibinfo {author} {\bibfnamefont {Z.}~\bibnamefont {Chen}},\ and\ \bibinfo {author} {\bibfnamefont {Q.-K.}\ \bibnamefont {Xue}},\ }\bibfield  {title} {\bibinfo {title} {Angle-resolved photoemission spectroscopy of superconducting ({L}a,{P}r)$_3${N}i$_2${O}$_7$/{S}r{L}a{A}l{O}$_4$ heterostructures},\ }\href {https://doi.org/10.1093/nsr/nwaf205} {\bibfield  {journal} {\bibinfo  {journal} {National Science Review}\ ,\ \bibinfo {pages} {nwaf205}} (\bibinfo {year} {2025}{\natexlab{c}})}\BibitemShut {NoStop}%
\bibitem [{\citenamefont {Wang}\ \emph {et~al.}(2025)\citenamefont {Wang}, \citenamefont {Zhong}, \citenamefont {Abadi}, \citenamefont {Liu}, \citenamefont {Yu}, \citenamefont {Zhang}, \citenamefont {Wu}, \citenamefont {Wang}, \citenamefont {Li}, \citenamefont {Tarn}, \citenamefont {Ko}, \citenamefont {Thampy}, \citenamefont {Hashimoto}, \citenamefont {Lu}, \citenamefont {Lee}, \citenamefont {Devereaux}, \citenamefont {Jia}, \citenamefont {Hwang},\ and\ \citenamefont {Shen}}]{wang2025electronic}%
  \BibitemOpen
  \bibfield  {author} {\bibinfo {author} {\bibfnamefont {B.~Y.}\ \bibnamefont {Wang}}, \bibinfo {author} {\bibfnamefont {Y.}~\bibnamefont {Zhong}}, \bibinfo {author} {\bibfnamefont {S.}~\bibnamefont {Abadi}}, \bibinfo {author} {\bibfnamefont {Y.}~\bibnamefont {Liu}}, \bibinfo {author} {\bibfnamefont {Y.}~\bibnamefont {Yu}}, \bibinfo {author} {\bibfnamefont {X.}~\bibnamefont {Zhang}}, \bibinfo {author} {\bibfnamefont {Y.-M.}\ \bibnamefont {Wu}}, \bibinfo {author} {\bibfnamefont {R.}~\bibnamefont {Wang}}, \bibinfo {author} {\bibfnamefont {J.}~\bibnamefont {Li}}, \bibinfo {author} {\bibfnamefont {Y.}~\bibnamefont {Tarn}}, \bibinfo {author} {\bibfnamefont {E.~K.}\ \bibnamefont {Ko}}, \bibinfo {author} {\bibfnamefont {V.}~\bibnamefont {Thampy}}, \bibinfo {author} {\bibfnamefont {M.}~\bibnamefont {Hashimoto}}, \bibinfo {author} {\bibfnamefont {D.}~\bibnamefont {Lu}}, \bibinfo {author} {\bibfnamefont {Y.~S.}\ \bibnamefont {Lee}}, \bibinfo {author} {\bibfnamefont {T.~P.}\ \bibnamefont {Devereaux}}, \bibinfo {author}
  {\bibfnamefont {C.}~\bibnamefont {Jia}}, \bibinfo {author} {\bibfnamefont {H.~Y.}\ \bibnamefont {Hwang}},\ and\ \bibinfo {author} {\bibfnamefont {Z.-X.}\ \bibnamefont {Shen}},\ }\bibfield  {title} {\bibinfo {title} {Electronic structure of compressively strained thin film {L}a$_2${P}r{N}i$_2${O}$_7$},\ }\href {https://arxiv.org/abs/2504.16372} {\bibfield  {journal} {\bibinfo  {journal} {arXiv:2504.16372}\ } (\bibinfo {year} {2025})}\BibitemShut {NoStop}%
\bibitem [{\citenamefont {Yang}\ \emph {et~al.}(2025)\citenamefont {Yang}, \citenamefont {Lu}, \citenamefont {Wan}, \citenamefont {Chen},\ and\ \citenamefont {Gong}}]{yang2025}%
  \BibitemOpen
  \bibfield  {author} {\bibinfo {author} {\bibfnamefont {Y.}~\bibnamefont {Yang}}, \bibinfo {author} {\bibfnamefont {X.}~\bibnamefont {Lu}}, \bibinfo {author} {\bibfnamefont {Y.}~\bibnamefont {Wan}}, \bibinfo {author} {\bibfnamefont {W.-Q.}\ \bibnamefont {Chen}},\ and\ \bibinfo {author} {\bibfnamefont {S.-S.}\ \bibnamefont {Gong}},\ }\bibfield  {title} {\bibinfo {title} {Evolution from intralayer to interlayer superconductivity in a bilayer $t$-${J}$ model},\ }\href {https://arxiv.org/abs/2507.07545} {\bibfield  {journal} {\bibinfo  {journal} {arXiv:2507.07545}\ } (\bibinfo {year} {2025})}\BibitemShut {NoStop}%
\bibitem [{\citenamefont {Hirthe}\ \emph {et~al.}(2023)\citenamefont {Hirthe}, \citenamefont {Chalopin}, \citenamefont {Bourgund}, \citenamefont {Bojovi{\'{c}}}, \citenamefont {Bohrdt}, \citenamefont {Demler}, \citenamefont {Grusdt}, \citenamefont {Bloch},\ and\ \citenamefont {Hilker}}]{Hirthe2023}%
  \BibitemOpen
  \bibfield  {author} {\bibinfo {author} {\bibfnamefont {S.}~\bibnamefont {Hirthe}}, \bibinfo {author} {\bibfnamefont {T.}~\bibnamefont {Chalopin}}, \bibinfo {author} {\bibfnamefont {D.}~\bibnamefont {Bourgund}}, \bibinfo {author} {\bibfnamefont {P.}~\bibnamefont {Bojovi{\'{c}}}}, \bibinfo {author} {\bibfnamefont {A.}~\bibnamefont {Bohrdt}}, \bibinfo {author} {\bibfnamefont {E.}~\bibnamefont {Demler}}, \bibinfo {author} {\bibfnamefont {F.}~\bibnamefont {Grusdt}}, \bibinfo {author} {\bibfnamefont {I.}~\bibnamefont {Bloch}},\ and\ \bibinfo {author} {\bibfnamefont {T.~A.}\ \bibnamefont {Hilker}},\ }\bibfield  {title} {\bibinfo {title} {Magnetically mediated hole pairing in fermionic ladders of ultracold atoms},\ }\href {https://doi.org/10.1038/s41586-022-05437-y} {\bibfield  {journal} {\bibinfo  {journal} {Nature}\ }\textbf {\bibinfo {volume} {613}},\ \bibinfo {pages} {463} (\bibinfo {year} {2023})}\BibitemShut {NoStop}%
\bibitem [{\citenamefont {Bohrdt}\ \emph {et~al.}(2022)\citenamefont {Bohrdt}, \citenamefont {Homeier}, \citenamefont {Bloch}, \citenamefont {Demler},\ and\ \citenamefont {Grusdt}}]{Bohrdt2022}%
  \BibitemOpen
  \bibfield  {author} {\bibinfo {author} {\bibfnamefont {A.}~\bibnamefont {Bohrdt}}, \bibinfo {author} {\bibfnamefont {L.}~\bibnamefont {Homeier}}, \bibinfo {author} {\bibfnamefont {I.}~\bibnamefont {Bloch}}, \bibinfo {author} {\bibfnamefont {E.}~\bibnamefont {Demler}},\ and\ \bibinfo {author} {\bibfnamefont {F.}~\bibnamefont {Grusdt}},\ }\bibfield  {title} {\bibinfo {title} {Strong pairing in mixed-dimensional bilayer antiferromagnetic mott insulators},\ }\href {https://doi.org/10.1038/s41567-022-01561-8} {\bibfield  {journal} {\bibinfo  {journal} {Nature Physics}\ }\textbf {\bibinfo {volume} {18}},\ \bibinfo {pages} {651} (\bibinfo {year} {2022})}\BibitemShut {NoStop}%
\bibitem [{\citenamefont {Schl\"omer}\ \emph {et~al.}(2024)\citenamefont {Schl\"omer}, \citenamefont {Lange}, \citenamefont {Franz}, \citenamefont {Chalopin}, \citenamefont {Bojovi\ifmmode~\acute{c}\else \'{c}\fi{}}, \citenamefont {Wang}, \citenamefont {Bloch}, \citenamefont {Hilker}, \citenamefont {Grusdt},\ and\ \citenamefont {Bohrdt}}]{PRXQuantum.5.040341}%
  \BibitemOpen
  \bibfield  {author} {\bibinfo {author} {\bibfnamefont {H.}~\bibnamefont {Schl\"omer}}, \bibinfo {author} {\bibfnamefont {H.}~\bibnamefont {Lange}}, \bibinfo {author} {\bibfnamefont {T.}~\bibnamefont {Franz}}, \bibinfo {author} {\bibfnamefont {T.}~\bibnamefont {Chalopin}}, \bibinfo {author} {\bibfnamefont {P.}~\bibnamefont {Bojovi\ifmmode~\acute{c}\else \'{c}\fi{}}}, \bibinfo {author} {\bibfnamefont {S.}~\bibnamefont {Wang}}, \bibinfo {author} {\bibfnamefont {I.}~\bibnamefont {Bloch}}, \bibinfo {author} {\bibfnamefont {T.~A.}\ \bibnamefont {Hilker}}, \bibinfo {author} {\bibfnamefont {F.}~\bibnamefont {Grusdt}},\ and\ \bibinfo {author} {\bibfnamefont {A.}~\bibnamefont {Bohrdt}},\ }\bibfield  {title} {\bibinfo {title} {Local control and mixed dimensions: Exploring high-temperature superconductivity in optical lattices},\ }\href {https://doi.org/10.1103/PRXQuantum.5.040341} {\bibfield  {journal} {\bibinfo  {journal} {PRX Quantum}\ }\textbf {\bibinfo {volume} {5}},\ \bibinfo {pages} {040341} (\bibinfo {year}
  {2024})}\BibitemShut {NoStop}%
\bibitem [{\citenamefont {Shen}\ \emph {et~al.}(2025)\citenamefont {Shen}, \citenamefont {Miao}, \citenamefont {Ou}, \citenamefont {Zhou}, \citenamefont {Chen}, \citenamefont {Luan}, \citenamefont {Sun}, \citenamefont {Feng}, \citenamefont {Yong}, \citenamefont {Li}, \citenamefont {Li}, \citenamefont {Xu}, \citenamefont {Lv}, \citenamefont {Nie}, \citenamefont {Wang}, \citenamefont {Huang}, \citenamefont {Sun}, \citenamefont {Xue}, \citenamefont {Chen},\ and\ \citenamefont {He}}]{shen2025}%
  \BibitemOpen
  \bibfield  {author} {\bibinfo {author} {\bibfnamefont {J.}~\bibnamefont {Shen}}, \bibinfo {author} {\bibfnamefont {Y.}~\bibnamefont {Miao}}, \bibinfo {author} {\bibfnamefont {Z.}~\bibnamefont {Ou}}, \bibinfo {author} {\bibfnamefont {G.}~\bibnamefont {Zhou}}, \bibinfo {author} {\bibfnamefont {Y.}~\bibnamefont {Chen}}, \bibinfo {author} {\bibfnamefont {R.}~\bibnamefont {Luan}}, \bibinfo {author} {\bibfnamefont {H.}~\bibnamefont {Sun}}, \bibinfo {author} {\bibfnamefont {Z.}~\bibnamefont {Feng}}, \bibinfo {author} {\bibfnamefont {X.}~\bibnamefont {Yong}}, \bibinfo {author} {\bibfnamefont {P.}~\bibnamefont {Li}}, \bibinfo {author} {\bibfnamefont {Y.}~\bibnamefont {Li}}, \bibinfo {author} {\bibfnamefont {L.}~\bibnamefont {Xu}}, \bibinfo {author} {\bibfnamefont {W.}~\bibnamefont {Lv}}, \bibinfo {author} {\bibfnamefont {Z.}~\bibnamefont {Nie}}, \bibinfo {author} {\bibfnamefont {H.}~\bibnamefont {Wang}}, \bibinfo {author} {\bibfnamefont {H.}~\bibnamefont {Huang}}, \bibinfo {author} {\bibfnamefont {Y.-J.}\
  \bibnamefont {Sun}}, \bibinfo {author} {\bibfnamefont {Q.-K.}\ \bibnamefont {Xue}}, \bibinfo {author} {\bibfnamefont {Z.}~\bibnamefont {Chen}},\ and\ \bibinfo {author} {\bibfnamefont {J.}~\bibnamefont {He}},\ }\bibfield  {title} {\bibinfo {title} {Anomalous energy gap in superconducting {L}a$_{2.85}${P}r$_{0.15}${N}i$_2${O}$_7$/{S}r{L}a{A}l{O}$_4$ heterostructures},\ }\href {https://arxiv.org/abs/2502.17831} {\bibfield  {journal} {\bibinfo  {journal} {arXiv:2502.17831}\ } (\bibinfo {year} {2025})}\BibitemShut {NoStop}%
\bibitem [{\citenamefont {Fan}\ \emph {et~al.}(2025)\citenamefont {Fan}, \citenamefont {Ou}, \citenamefont {Scholten}, \citenamefont {Li}, \citenamefont {Shang}, \citenamefont {Wang}, \citenamefont {Xu}, \citenamefont {Yang}, \citenamefont {Eremin},\ and\ \citenamefont {Wen}}]{fan2025STM}%
  \BibitemOpen
  \bibfield  {author} {\bibinfo {author} {\bibfnamefont {S.}~\bibnamefont {Fan}}, \bibinfo {author} {\bibfnamefont {M.}~\bibnamefont {Ou}}, \bibinfo {author} {\bibfnamefont {M.}~\bibnamefont {Scholten}}, \bibinfo {author} {\bibfnamefont {Q.}~\bibnamefont {Li}}, \bibinfo {author} {\bibfnamefont {Z.}~\bibnamefont {Shang}}, \bibinfo {author} {\bibfnamefont {Y.}~\bibnamefont {Wang}}, \bibinfo {author} {\bibfnamefont {J.}~\bibnamefont {Xu}}, \bibinfo {author} {\bibfnamefont {H.}~\bibnamefont {Yang}}, \bibinfo {author} {\bibfnamefont {I.~M.}\ \bibnamefont {Eremin}},\ and\ \bibinfo {author} {\bibfnamefont {H.-H.}\ \bibnamefont {Wen}},\ }\bibfield  {title} {\bibinfo {title} {Superconducting gap structure and bosonic mode in {L}a$_{2}${P}r{N}i$_{2}${O}$_{7}$ thin films at ambient pressure},\ }\href {https://arxiv.org/abs/2506.01788} {\bibfield  {journal} {\bibinfo  {journal} {arXiv:2506.01788}\ } (\bibinfo {year} {2025})}\BibitemShut {NoStop}%
\end{thebibliography}

%\clearpage
\onecolumngrid
%\appendix
%\renewcommand{\theequation}{S\arabic{equation}}
%\renewcommand{\thefigure}{A\arabic{figure}}
%\setcounter{equation}{0}
%\setcounter{figure}{0}

\end{document}